\documentclass[11pt]{amsart}

\usepackage[english]{babel}
\usepackage{inputenc} 
 \usepackage{comment} 

\usepackage{tikz}
\usetikzlibrary{shapes.geometric, arrows, arrows.meta, positioning}

\newcommand{\firstfigure}{\begin{tikzpicture}
    \pgfplotsset{
        every axis/.style={
            width=5cm, height=5cm,
            ymin=-3.5, ymax=3.5,
            xmin=0.5, xmax=2.5,
            ytick={-3, -2, -1, 0, 1, 2, 3},
            yticklabels={
                STD, D, SWD, N, SWA, A, STA
            },
            xtick={1, 2},
            xticklabels={$k$, $l$},
            axis line style={black},
            tick style={draw=none},
            enlarge y limits=0.1,
        }
    }

    \begin{scope}[xshift=-7.25cm, yshift=4cm]
        \begin{axis}[yticklabels={}]
            \node[anchor=west] at (axis cs:0.5,3) {Strongly agree};
            \node[anchor=west] at (axis cs:0.5,2) {Agree};
            \node[anchor=west] at (axis cs:0.5,1) {Somewhat agree};
            \node[anchor=west] at (axis cs:0.5,0) {No agree/disagree};
            \node[anchor=west] at (axis cs:0.5,-1) {Somewhat disagree};
            \node[anchor=west] at (axis cs:0.5,-2) {Disagree};
            \node[anchor=west] at (axis cs:0.5,-3) {Strongly disagree};
        \end{axis}
    \end{scope}

    \begin{scope}[xshift=-2cm, yshift=4cm]
            \begin{axis}[title={I. $u_{kl}$, $v_{kl}$ cross semispaces \\ in the opposite direction\\ $\pi = -1$}, title style={yshift=-1ex, align=center}]
                \draw[thick] (axis cs:0.5,0) -- (axis cs:2.5,0);

                \addplot[only marks, mark=*, mark options={fill=white}] coordinates {(1, 3)};
                \addplot[only marks, mark=*, mark options={fill=white}] coordinates {(2, -2)};
                \addplot[dotted, thick, black] coordinates {(1, 3) (2, -2)};

                \addplot[only marks, mark=*, mark options={fill=black}] coordinates {(1,-1)};
                \addplot[only marks, mark=*, mark options={fill=black}] coordinates {(2,1)};
                \addplot[dashed, thick, black] coordinates {(1, -1) (2, 1)};

            \end{axis}
        \end{scope}

        \begin{scope}[xshift=3.25cm, yshift=4cm]
            \begin{axis}[title={II. $u_{kl}$, $v_{kl}$ remain in\\ different semispaces\\ $\pi = -1$}, title style={yshift=-1ex,align=center}]
                \draw[thick] (axis cs:0.5,0) -- (axis cs:2.5,0);

                \addplot[only marks, mark=*, mark options={fill=white}] coordinates {(1, 1)};
                \addplot[only marks, mark=*, mark options={fill=white}] coordinates {(2, 1)};
                \addplot[dotted, thick, black] coordinates {(1, 1) (2, 1)};

                \addplot[only marks, mark=*, mark options={fill=black}] coordinates {(1,-2)};
                \addplot[only marks, mark=*, mark options={fill=black}] coordinates {(2,-3)};
                \addplot[dashed, thick, black] coordinates {(1, -2) (2, -3)};

            \end{axis}
        \end{scope}

        \begin{scope}[xshift=-7.25cm, yshift=-2.5cm]
            \begin{axis}[title={III. either $u_{kl}$ or $v_{kl}$ change\\ semispaces but not the other\\ $\pi = 0$}, title style={yshift=-1ex,align=center}]
                \draw[thick] (axis cs:0.5,0) -- (axis cs:2.5,0);

                \addplot[only marks, mark=*, mark options={fill=white}] coordinates {(1, 1)};
                \addplot[only marks, mark=*, mark options={fill=white}] coordinates {(2, 3)};
                \addplot[dotted, thick, black] coordinates {(1, 1) (2, 3)};

                \addplot[only marks, mark=*, mark options={fill=black}] coordinates {(1,-3)};
                \addplot[only marks, mark=*, mark options={fill=black}] coordinates {(2,1)};
                \addplot[dashed, thick, black] coordinates {(1, -3) (2, 1)};

            \end{axis}
        \end{scope}

        \begin{scope}[xshift=-2cm, yshift=-2.5cm]
            \begin{axis}[title={IV. $u_{kl}$, $v_{kl}$ change semi-\\spaces in the same direction\\ $\pi = 1$}, title style={yshift=-1ex,align=center}]
                \draw[thick] (axis cs:0.5,0) -- (axis cs:2.5,0);

                \addplot[only marks, mark=*, mark options={fill=white}] coordinates {(1, -2)};
                \addplot[only marks, mark=*, mark options={fill=white}] coordinates {(2, 2)};
                \addplot[dotted, thick, black] coordinates {(1, -2) (2, 2)};

                \addplot[only marks, mark=*, mark options={fill=black}] coordinates {(1,-1)};
                \addplot[only marks, mark=*, mark options={fill=black}] coordinates {(2,1)};
                \addplot[dashed, thick, black] coordinates {(1, -1) (2, 1)};

            \end{axis}
        \end{scope}

        \begin{scope}[xshift=3.25cm, yshift=-2.5cm]
            \begin{axis}[title={V. $u_{kl}$, $v_{kl}$ remain in\\ the same semispace\\ $\pi = 1$}, title style={yshift=-1ex,align=center}]
                \draw[thick] (axis cs:0.5,0) -- (axis cs:2.5,0);

                \addplot[only marks, mark=*, mark options={fill=white}] coordinates {(1, 1)};
                \addplot[only marks, mark=*, mark options={fill=white}] coordinates {(2, 1)};
                \addplot[dotted, thick, black] coordinates {(1, 1) (2, 1)};

                \addplot[only marks, mark=*, mark options={fill=black}] coordinates {(1,2)};
                \addplot[only marks, mark=*, mark options={fill=black}] coordinates {(2,3)};
                \addplot[dashed, thick, black] coordinates {(1, 2) (2, 3)};

            \end{axis}
        \end{scope}

        \begin{scope}[xshift=-7.25cm, yshift=-9cm]
            \begin{axis}[title={VI. $u_{kl}$ or $v_{kl}$ holds neutral\\ change and the other\\ changes semispace\\ $\pi = 0$}, title style={yshift=-1ex,align=center}]
                \draw[thick] (axis cs:0.5,0) -- (axis cs:2.5,0);

                \addplot[only marks, mark=*, mark options={fill=white}] coordinates {(1, -2)};
                \addplot[only marks, mark=*, mark options={fill=white}] coordinates {(2, 3)};
                \addplot[dotted, thick, black] coordinates {(1, -2) (2, 3)};

                \addplot[only marks, mark=*, mark options={fill=black}] coordinates {(1,0)};
                \addplot[only marks, mark=*, mark options={fill=black}] coordinates {(2,0)};
                \addplot[dashed, thick, black] coordinates {(1, 0) (2, 0)};

            \end{axis}
        \end{scope}

        \begin{scope}[xshift=-2cm, yshift=-9cm]
            \begin{axis}[title={VII. $u_{kl}$ or $v_{kl}$ holds neutral\\ change and the other\\ remains in the same semispace\\ $\pi = 1$}, title style={yshift=-1ex,align=center}]
                \draw[thick] (axis cs:0.5,0) -- (axis cs:2.5,0);

                \addplot[only marks, mark=*, mark options={fill=white}] coordinates {(1, 0)};
                \addplot[only marks, mark=*, mark options={fill=white}] coordinates {(2, 0)};
                \addplot[dotted, thick, black] coordinates {(1, 0) (2, 0)};

                \addplot[only marks, mark=*, mark options={fill=black}] coordinates {(1,1)};
                \addplot[only marks, mark=*, mark options={fill=black}] coordinates {(2,3)};
                \addplot[dashed, thick, black] coordinates {(1, 1) (2, 3)};

            \end{axis}
        \end{scope}

        \begin{scope}[xshift=3.25cm, yshift=-9cm]
            \begin{axis}[title={VIII. Both $u_{kl}$ and $v_{kl}$ hold\\ neutral positions\\ $\pi = 1$}, title style={yshift=-1ex,align=center}]
                \draw[thick] (axis cs:0.5,0) -- (axis cs:2.5,0);

                \addplot[only marks, mark=*, mark options={fill=white}] coordinates {(1, 0)};
                \addplot[only marks, mark=*, mark options={fill=white}] coordinates {(2, 0)};
                \addplot[dotted, thick, black] coordinates {(1, 0) (2, 0)};

                \addplot[only marks, mark=*, mark options={fill=black}] coordinates {(1,0.2)};
                \addplot[only marks, mark=*, mark options={fill=black}] coordinates {(2,0.2)};
                \addplot[dashed, thick, black] coordinates {(1, 0.2) (2, 0.2)};

            \end{axis}
        \end{scope}
\end{tikzpicture}}

\newcommand{\secondfigure}{ \begin{tikzpicture}
        \begin{axis}[
            width=10cm, height=6cm,
            axis x line=bottom,
            axis y line=left,
            ymin=-3.5, ymax=3.5,
            xmin=-0.4, xmax=2.2,
            ytick={-3, -2, -1, 0, 1, 2, 3},
            yticklabels={
                Strongly disagree,
                Disagree,
                Somewhat disagree,
                Neither agree nor disagree,
                Somewhat agree,
                Agree,
                Strongly agree
            },
            xtick={0, 1, 2},
            xticklabels={
                Q1,
                Q2,
                Q3
            },
            grid = major,
             ymajorgrids=true,
             grid style=dotted,
            axis line style={draw=none},
        ]

            \addplot[only marks, mark=*, mark options={fill=black}] coordinates {(0,1.1)};
            \addplot[only marks, mark=*, mark options={fill=black}] coordinates {(1,1.1)};
            \addplot[only marks, mark=*, mark options={fill=black}] coordinates {(0,1)};
            \addplot[only marks, mark=*, mark options={fill=black}] coordinates {(1,1)};
            \addplot[only marks, mark=*, mark options={fill=black}] coordinates {(2,3)};
            \addplot[only marks, mark=*, mark options={fill=black}] coordinates {(2,1)};
            \addplot[dotted, black, thick] coordinates {(0,1.1) (1,1.1) (2,3)};
            \addplot[dotted, black, thick] coordinates {(0,1) (1,1) (2,1)};
            \node[right] at (axis cs:2,2.5) {C};
            \node[right] at (axis cs:2,0.5) {A};

            \addplot[only marks, mark=*, mark options={fill=lightgray}] coordinates {(0,0.9)};
            \addplot[only marks, mark=*, mark options={fill=lightgray}] coordinates {(1,0.9)};
            \addplot[only marks, mark=*, mark options={fill=lightgray}] coordinates {(2,-1)};
            \addplot[dotted, gray, thick] coordinates {(0,0.9) (1,0.9) (2,-1)};
            \node[right] at (axis cs:2,-1) {B};

            \draw[thick] (axis cs:-0.2,0) -- (axis cs:2.2,0);
            \draw[thick] (axis cs:-0.2,-3.5) -- (axis cs:2.2,-3.5);
            \draw[thick] (axis cs:-0.2,3.5) -- (axis cs:2.2,3.5);
            \draw[thick] (axis cs:-0.2,-3.5) -- (axis cs:-0.2,3.5);
            \draw[thick] (axis cs:2.2,-3.5) -- (axis cs:2.2,3.5);
        \end{axis}

        \node[draw=none, anchor=west] at (9,4.25) {\textbf{•} \textit{Construal 1}:};
        \node[draw=none, anchor=west] at (9,3.75) {classic ideologue};
        \node[draw=none, anchor=west] at (9,3.25) {standard};
        \node[draw=none, anchor=west] at (9,2.75) {left/right};
        \node[draw=none, anchor=west] at (9,2) {\textcolor{lightgray}{•} \textit{Construal 2}:};
        \node[draw=none, anchor=west] at (9,1.5) {spend-averse};
        \node[draw=none, anchor=west] at (9,1) {progressives / };
        \node[draw=none, anchor=west] at (9,0.5) {redistributive};
        \node[draw=none, anchor=west] at (9,0) {conservatives};
    \end{tikzpicture}}

    \newcommand{\thirdfigure}{\begin{tikzpicture}
        \node at (-5.785,6) {\textbf{(1)}};
        \draw[thick] (-8.02,5.5) -- (-3.55,5.5);
        \node at (-5.785,5) {A-B (different construals)};

        \node at (-0.265,6) {\textbf{(2)}};
        \draw[thick] (-2.5,5.5) -- (1.97,5.5);
        \node at (-0.265,5) {A-C (same construal)};

        \node at (5.215,6) {\textbf{(3)}};
        \draw[thick] (2.98,5.5) -- (7.45,5.5);
        \node at (5.215,5) {B-C (different construals)};

        \begin{scope}[xshift=-8cm]
            \begin{axis}[
                width=6cm, height=6cm,
                ymin=0, ymax=1,
                xmin=0.5, xmax=4.5,
                ytick={0, 0.5, 1},
                yticklabels={0, 0.5, 1},
                xtick={1, 2, 3, 4},
                xticklabels={\footnotesize{RCA}, \footnotesize{CCA}, \footnotesize{RRCA}, \footnotesize{BCA}},
                xticklabel style={rotate=0, anchor=north},
                grid=major,
                major grid style={dotted, gray},
                axis line style={black, very thick},
                ymajorgrids=true,
                every axis plot/.append style={thick}
            ]
                \addplot[ybar, draw=black, fill=white] coordinates {(1, 0.56)};
                \addplot[ybar, draw=black, fill=white] coordinates {(3, 0.11)};
                \addplot[ybar, draw=black, fill=gray] coordinates {(4,0.33)}; 
                \node at (axis cs:1,0.65) {5/9};
                \node at (axis cs:2,0.05) {0};
                \node at (axis cs:3,0.15) {1/9};
                \node at (axis cs:4,0.4) {1/3};
            \end{axis}
        \end{scope}

        \begin{scope}[xshift=-2.5cm]
            \begin{axis}[
                width=6cm, height=6cm,
                ymin=0, ymax=1,
                xmin=0.5, xmax=4.5,
                ytick={0, 0.5, 1},
                yticklabels={0, 0.5, 1},
                xtick={1, 2, 3, 4},
                xticklabels={\footnotesize{RCA}, \footnotesize{CCA}, \footnotesize{RRCA}, \footnotesize{BCA}},
                xticklabel style={rotate=0, anchor=north},
                grid=major,
                major grid style={dotted, gray},
                axis line style={black, very thick},
                ymajorgrids=true,
                every axis plot/.append style={thick}
            ]
                \addplot[ybar, draw=black, fill=white] coordinates {(1, 0.56)};
                \addplot[ybar, draw=black, fill=white] coordinates {(3, 0.11)};
                \addplot[ybar, draw=black, fill=gray] coordinates {(4,1)};
                \node at (axis cs:1,0.65) {5/9};
                \node at (axis cs:2,0.05) {0};
                \node at (axis cs:3,0.15) {1/9};
            \end{axis}
        \end{scope}

        \begin{scope}[xshift=3cm]
            \begin{axis}[
                width=6cm, height=6cm,
                ymin=0, ymax=1,
                xmin=0.5, xmax=4.5,
                ytick={0, 0.5, 1},
                yticklabels={0, 0.5, 1},
                xtick={1, 2, 3, 4},
                xticklabels={\footnotesize{RCA}, \footnotesize{CCA}, \footnotesize{RRCA}, \footnotesize{BCA}},
                xticklabel style={rotate=0, anchor=north},
                grid=major,
                major grid style={dotted, gray},
                axis line style={black, very thick},
                ymajorgrids=true,
                every axis plot/.append style={thick}
            ]
                \addplot[ybar, draw=black, fill=white] coordinates {(1, 0.33)};
                \addplot[ybar, draw=black, fill=white] coordinates {(3, 0.11)};
                \addplot[ybar, draw=black, fill=white] coordinates {(2, 1)};
                \addplot[ybar, draw=black, fill=gray] coordinates {(4,0.33)};
                \node at (axis cs:1,0.4) {1/3};
                \node at (axis cs:2,1.05) {1};
                \node at (axis cs:3,0.15) {1/9};
                \node at (axis cs:4,0.4) {1/3};
            \end{axis}
        \end{scope}
    \end{tikzpicture}}

\newcommand{\flowchart}{
\begin{tikzpicture}[
    node distance=3cm,
    every node/.style={font=\small},
    startstop/.style={
        rectangle, rounded corners,
        minimum width=3.5cm,
        minimum height=1cm,
        text centered, draw
    },
    process/.style={
        rectangle,
        minimum width=4cm,
        minimum height=1cm,
        text centered, draw
    },
    decision/.style={
        diamond,
        aspect=2,
        minimum width=3.5cm,
        minimum height=1cm,
        text centered, draw
    },
    arrow/.style={thick,->,>=stealth}
]

\node (input) [startstop, align=center] {Data with $N$ respondents \\ and $Q$ questions};

\node (loop1) [decision, right=of input, xshift=0.5cm, align=center]
{Last respondent pair \\ $(u, v)$?};

\node (loop2) [decision, right=of loop1, xshift=0.5cm, align=center]
{Last question pair \\ $(k, l)$?};

\node (proc_pw) [process, right=of loop2, xshift=0.5cm, align=center]
{Find \textit{pairwise polarity} $\pi(u_{kl}, v_{kl})$ \\ (Figure~\ref{fig:polarity_function})};

\node (proc_mean) [process, right=of proc_pw, xshift=0.5cm, align=center]
{Average \textit{paiwise polarities}};

\node (output) [startstop, right=of proc_mean, xshift=0.5cm]
{Adjacency matrix};

\draw [arrow] (input) -- (loop1);
\draw [arrow] (loop1) -- node[above]{no} (loop2);
\draw [arrow] (loop2) -- node[above]{no} (proc_pw);

\draw [arrow] (proc_pw.north) -- ++(0, 1.5) 
node[midway, left] {next pair} -- ++ (-8.73, 0) -- (loop2);

\draw [arrow] (proc_mean.north) -- ++(0, 2) 
node[midway, left] {next pair} -- ++ (-25.77, 0) -- (loop1);

\draw [arrow] (loop2.south) -- ++ (0, -1)
node[midway, right]{yes} -- ++ (17, 0) -- (proc_mean);

\draw [arrow] (loop1.south) -- ++ (0, -1.5)
node[midway, right]{yes} -- ++ (33.2, 0) -- (output);

\end{tikzpicture}
} 

\usepackage{setspace} 
\usepackage{microtype} 

\usepackage[margin=2.5cm]{geometry} 

\usepackage{algorithm}
\usepackage{algorithmic}

\usepackage{etoolbox} 
\makeatletter
\patchcmd{\@sect}{\uppercasenonmath\@secname}{\MakeUppercase{Appendix: \@secname}}{}{}
\makeatother

\usepackage{graphicx}
\usepackage{caption,subcaption}
\usepackage{pdflscape}
\usepackage{placeins}

\usepackage{pgfplots} 
\pgfplotsset{compat=1.15} 
\usepackage{tikz} 
\usetikzlibrary{arrows} 

\usepackage{amsmath, amssymb, amsfonts, amsthm} 
\usepackage{mathtools} 
\usepackage[mathscr]{eucal} 
\usepackage{amscd} 
\usepackage[all,cmtip]{xy} 

\usepackage{lineno}

\usepackage{footnote}
\usepackage{endnotes}
\makesavenoteenv{tabular}
\makesavenoteenv{table}

\usepackage{array} 
\newcolumntype{L}[1]{>{\raggedright\arraybackslash}p{#1}}

\newcolumntype{R}[1]{>{\raggedleft\arraybackslash}p{#1}}

\usepackage{longtable} 
\usepackage{booktabs} 
\usepackage{multirow} 
\usepackage{rotating} 
\usepackage{makecell} 
\usepackage{arydshln}

\usepackage{multicol} 

\usepackage{verbatim}

\usepackage{hyperref}
\hypersetup{
colorlinks=true,
	linkcolor=blue,
	citecolor=blue,
	urlcolor=blue,
    pdfstartview=FitH 
}

\usepackage{caption} 
\usepackage{csquotes} 

\usepackage[title]{appendix} 
\usepackage{threeparttable}  
\usepackage{tabularx}        

\usepackage{enumitem} 

\usepackage{soul} 
\usepackage[normalem]{ulem} 

\usepackage[authordate,backend=biber,maxbibnames=99]{biblatex-chicago} 
\newtheorem{thm}{Theorem}[section]
\newtheorem*{thm*}{Theorem}
	
\newtheorem*{cor*}{Corollary}

\theoremstyle{definition}
\newtheorem{rema}[thm]{Remark} 
\newtheorem*{definition*}{Definition}    	

\DeclareMathOperator{\trace}{trace}

\DeclareMathOperator{\sign}{sign}

\begin{document}



    \title[Reassessing Construal Clustering via BCA]{
    Finding patterns of meaning: Reassessing Construal Clustering via Bipolar Class Analysis}

   \date{\today}


\begin{abstract}
    Empirical research on \textit{construals}--social affinity groups that share similar patterns of meaning--has advanced significantly in recent years. This progress is largely driven by the development of \textit{Construal Clustering Methods} (CCMs), which group survey respondents into construal clusters based on similarities in their response patterns. We identify key limitations of existing CCMs, which affect their accuracy when applied to the typical structures of available data, and introduce Bipolar Class Analysis (BCA), a CCM designed to address these shortcomings. BCA measures similarity in response shifts between expressions of support and rejection across survey respondents, addressing conceptual and measurement challenges in existing methods. We formally define BCA and demonstrate its advantages through extensive simulation analyses,  where it consistently outperforms existing CCMs in accurately identifying construals. Along the way, we develop a novel data-generation process that approximates more closely how individuals map latent opinions onto observable survey responses, as well as a new metric to evaluate the performance of CCMs. Additionally, we find that applying BCA to previously studied real-world datasets reveals substantively different construal patterns compared to those generated by existing CCMs in prior empirical analyses. Finally, we discuss limitations of BCA and outline directions for future research.
\end{abstract}

    \author[M.~Cuerno]{Manuel Cuerno$^{\ast}$}
    \author[F.~Galaz-García]{Fernando Galaz-García}
    \author[S.~Galaz-García]{Sergio Galaz-García$^{\ast\ast}$}
    \author[T.~Pérez-Izquierdo]{Telmo Pérez-Izquierdo}



    \thanks{The authors were supported by the project ``Charting political ideological landscapes in Europe: Fault lines and opportunities (POL-AXES)'' - Programa Primas y Problemas de la Fundación BBVA 2023.}

    \thanks{$^{\ast}$M. Cuerno has also been supported by the FPI Graduate Research Grant PRE2018-084109, by research grants  
	 MTM2017-‐85934-‐C3-‐2-‐P and PID2021-124195NB-C32
from the Ministerio de Econom\'ia y Competitividad de Espa\~{na} (MINECO), by the research grant PID2024-158664NB-C2 from Miciu/AEI/10.13039/501100011033/FEDER/UE, by the Quantitative Methods and Mathematics Departments of CUNEF University, Madrid, Spain, and by the Department of Mathematical Sciences of Durham University, United Kingdom.}

\thanks{$^{\ast\ast}$S. Galaz-García has been supported by the Juan de la Cierva Research Grant FJC2021-047259-I  from the Ministerio de Ciencia, Innovación y Universidades de España.}


     \address[M.~Cuerno]{Department of Mathematics, CUNEF University, Spain.}
     \email{manuel.mellado@cunef.edu}

     \address[F.~Galaz-García]{Department of Mathematical Sciences, Durham University, United Kingdom.}
     \email{fernando.galaz-garcia@durham.ac.uk} 

     \address[S.~Galaz-García]{Department of Quantitative Methods, CUNEF University, Spain.}
     \email{sergio.galaz@uc3m.es}
    
     \address[T.~Pérez-Izquierdo]{Department of Economic Analysis, University of the Basque Country, Spain.}
     \email{telmo.perezizquierdo@ehu.eus}




    \maketitle

    \setcounter{tocdepth}{1} 
    \tableofcontents


\section{Introduction}

Empirical knowledge on \emph{construals}, or ``meaning structures upon which actors draw to understand social life'' (\cite{dimaggio_etal_2018}), has significantly expanded in recent years. More specifically, a \textit{construal} can be defined as a social affinity group whose members share similar patterns of meaning in interpreting and navigating social life. Consider, for instance, two people answering a political opinion survey consisting of questions on progressive issues with answers ranging from ``strongly agree'' to ``strongly disagree''. The first respondent selects ``strongly agree'' for every question, aligning with progressive views on politics, while the second one chooses ``strongly disagree'', aligning with conservative ones. Despite their opposing political opinions, both respondents likely belong to the same construal because their response patterns are identical, differing only along the progressive--conservative axis.

Empirical knowledge on construals has grown remarkably in recent times. This increase has been driven by Construal Clustering Methods (CCMs) that use community-detection algorithms to group survey respondents based on the similarity of their response patterns. Rather than clustering respondents according to how similar their specific responses are, CCMs cluster respondents according to how similarly their response patterns across questions are. 

\textcolor{black}{This article introduces a new CCM, Bipolar Class Analysis (BCA). Our contribution is structured around three main fronts. First, through mathematical modeling, we isolate the problem presented by bipolar questions and incorporate how human cognition engages with this type of questions, leading to the construction of a \textit{polarity function} that differentiates our method from other CCMs. Second, we evaluate BCA and existent CCMs via a novel data simulation technique based on families of copulae, which more closely captures the concept of construal and its operation in opinion surveys. Third, we present two in-depth reanalyses of datasets previously studied using other CCMs, along with complementary results showing that the construal clusters produced by BCA are substantively different than those generated by existing CCMs.}

CCMs work in two steps. They first compute an adjacency matrix out of pairs of survey respondents. Adjacency between two respondents is meant to capture similarity in response patterns. Second, they apply a community-detection algorithm to identify construal clusters. \textcolor{black}{For this second step, any community-detection algorithm may be combined with any adjacency matrix. Thus, the main difference among CCMs arises in how they compute the adjacency between pairs of survey respondents.}

 We begin by showing that \emph{bipolar survey items} with varying numbers of options are the most widely available data type for construal clustering analysis. Drawing from the literature on survey response (\cite{krosnick_1999, ostrom_1987, ostrom_etal_1992, malhotra_krosnick_2007, malhotra_etal_2009}), we note that an appropriate response adjacency measurement for this type of data requires:
\begin{enumerate}
\item Prioritizing measuring how respondents' answers shift between rejection and support. 
\item Addressing comparability issues in the meanings of answer options across questions, particularly when the number of options differs. 
\item Interpreting responses as discrete choices. 
\end{enumerate}
In this paper, we give a motivating example showing that the adjacency measures used in existing CCMs may not meet these requirements, making them prone to generating invalid construal clusters.
To remedy this, BCA adopts a novel function for measuring respondent adjacency. This measure, which we call \textit{polarity}, captures how similarly respondents’ answers shift along rejection (negative) or support (positive) responses  across questions. BCA constructs construal clusters solely based on the variation of these shifts between pairs of respondents.

We evaluate BCA's performance relative to existing CCMs using a large number of simulated datasets. The data is drawn from a data-generation process (DGP) that incorporates ordered choice and stochastic dependency (\cite{greene_2010, fisher_1997}). This DGP is designed to approximate how respondents translate their opinions into concrete survey answers. We see this as a valuable contribution to CCM evaluation, as previous simulation methods relied on shift-scale transformations of survey responses (see \cite{boutyline_2017, sotoudeh_dimaggio_2023}). The shift-scale approach entirely bypassed the modeling of respondents' thought processes and had certain technical shortcomings---for example, draws outside of the available options were truncated. To evaluate performance, we employ four different measures, including several well-established in the CCM and cluster analysis literature, along with an original measure of construal clustering performance, \textit{correlation dissimilarity} (CDIS). 

    We show that BCA outperforms existing CCMs by conducting experiments that emulate construal structures in the empirical literature. We conduct two experiments. In the first experiment, we generate datasets that simulate the presence of two construals, based on our motivating example. BCA correctly estimates the number of construals in these synthetic datasets in $93\%$ of cases, while the \textcolor{black}{rest of the methods miss in all cases}. In this experiment, BCA's performance measures are one or two orders of magnitude better than those of existing CCMs. In the second experiment, we examine BCA's performance in data structures with between $2$ and $4$ construals---empirical studies mostly report numbers within this range. In this experiment, the gap between BCA and other CCMs narrows. Nevertheless, BCA performs best. We  also reanalyze datasets that have been studied in empirical applications of CCMs. We find that BCA's results are substantially different from those of existing CCMs.

We offer three interrelated contributions in this study. First, for empirical researchers, we introduce a method that produces more accurate construal clusters from answers to bipolar questions, the  dominant data type in  social surveys. Second, for methodological research, we highlight challenges to construal clustering accuracy arising from the cognitive processes underlying answer selection. Additionally, we propose a new modeling strategy tailored to the study of construals using survey data. Third, we expand the methodological toolkit for evaluating CCM performance by introducing a new performance metric.

Our article is organized as follows. Section~\ref{sec:lit_review} presents a literature review of existent CCMs, some variants and empirical studies.  Section~\ref{sec:bipolar_data} discusses the structure and use of bipolar data. Section~\ref{sec:BCA} introduces BCA. Section~\ref{sec:RCACCA} reviews existing CCMs and their limitations. Sections~\ref{sec:simulations} and~\ref{sec:reanalyses} 
present our simulation analysis and examine empirical differences among CCMs using real-world datasets, respectively. Section~\ref{sec:limitations} explores the limitations of BCA and outlines directions for future research, while Section~\ref{sec:conclusion} summarizes the study's results.
Finally, Appendices~\ref{app:polarity}, \ref{app:simulation}, and \ref{app:performance} provide technical details on the methods and tools used in our study, and Appendix \ref{app:reanalyses} presents more complete and additional results that those presented in the main body of the paper.


\section{Related work}{\label{sec:lit_review}}

 \textcolor{black}{The first CCM introduced in the literature was Relational Class Analysis (RCA) (\cite{goldberg_2011}).} This method groups respondents into construal clusters using a function known as \textit{relationality} as a measure of adjacency among respondents. More recently, Boutyline (\citeyear{boutyline_2017}) introduced Correlational Class Analysis (CCA), which uses correlation coefficients instead of relationality as a measure of respondent adjacency. \textcolor{black}{Finally, Sotoudeh and DiMaggio (\citeyear{sotoudeh_dimaggio_2023}) introduced a variant of Goldberg's RCA by modifying the relationality function and using a different community-detection algorithm. We will refer to this method as Recursive Relational Class Analysis (RRCA), in line with Sotoudeh and DiMaggio’s use of the term ``Recursive Relationality'' for their modification of Goldberg’s original relationality in their R implementation.}
 
\textcolor{black}{Although the original paper in which Goldberg introduced RCA dates back to 2011, only CCA and RRCA are strictly tied to that framework. Nevertheless, other classical methods such as Latent Class Analysis (LCA) (\cite{mccutcheon1987latent}) have been compared with CCMs, particularly with RCA (\cite{dimaggio_goldberg_2018}). In addition, CCMs and their underlying philosophy have been used as a baseline for studying cultural schemas using alternative tools, such as the Implicit Association Test (IAT) and the Affect Misattribution Procedure (AMP) (\cite{hunzaker_2019}), as well as Subcoalition Class Analysis (SCA), which, rather than clustering based on ``who agrees on what the argument is about,'' focuses on ``who agrees on the substance of issues'' (\cite[p.~7949]{ganz2025subcoalition}). Finally, CCMs have also inspired approaches to the study of cultural schemas that rely on different types of input data. This is the case of Concept Class Analysis (CoCA), which is tailored to construal clustering using open text (\cite{taylor2020concept}).}

\textcolor{black}{Existing CCMs have led to insights into the number, prevalence, and characteristics of construals across a wide range of topics. These include areas as diverse as political orientations, positions on the penetration of markets into social life, attitudes toward religion and science, and broader cultural orientations (see \cite[p.~1842]{sotoudeh_dimaggio_2023} for an overview; \cite{baldasarri_goldberg_2014, dimaggio_goldberg_2018, daenekindt_2017, brensinger_sotoudeh_2022}). Furthermore, CCMs continue to be widely used across diverse empirical contexts (\cite{bertero2024inequality, batzke2025cognitive, lindner2024stances, van2022support}). Construals have also begun to inform broader public interpretations of opinion in the public sphere, as illustrated by the \textit{New York Times} article ``The 6 Kinds of Republican Voters'' (\cite{NYtimes_republicans}). All these studies illustrate the growing importance of CCMs both as instruments for empirical analysis and targets for methodological refinement.}


\section{Bipolar Data}
\label{sec:bipolar_data}

    Identifying the cultural structures that guide people's social lives using survey responses is not straightforward. Answers to survey questions are simplified approximations rather than precise, high-fidelity representations of the opinions people have on social issues. The survey response literature suggests that these opinions serve as inputs to an answer selection process that is situationally bound and highly sensitive to questionnaire characteristics, including the types of answers that survey questions offer to respondents (\cite{schwarz_sudman_1996}). For example, a question asking to express degrees of support or opposition to migration prompts different answering strategies in respondents than one asking them how many new migrants should be admitted to the U.S. In other words, the types of answers to which survey respondents react influence the cognitive processes respondents use to make their selections, and consequently, the inferences we can draw about their opinions and construals. Against this backdrop, we ask: What kind of answer structure constitutes the typical data environment available for construal analysis?
    
    Research on social surveys defines \textit{bipolar survey items} as a ``linearly ordered set between elements considered as negative and positive categories” (\cite{batyrshin_etal_2017, ostrom_etal_1992, krosnick_1999}). We identify bipolar survey items  as the most common data structure in opinion surveys. Bipolar questions prompt survey takers to choose among options that express magnitudes of rejection or support for statements (corresponding, respectively, to negative and positive categories), or to express relative degrees of preference between two competing options. 
    
    Bipolar questions are the most commonly used type of questions in construal analysis. For example, 54\% of the variables in the databases used by Sotoudeh and DiMaggio (\cite{sotoudeh_dimaggio_2023}) for construal analysis are bipolar; this percentage increases to 97--100\% for analyses focused on the study of cultural tastes, political attitudes, opinions and science and religion, and stances about the role of the market in structuring social interactions (\cite{baldasarri_goldberg_2014, dimaggio_etal_2018, dimaggio_goldberg_2018}). Bipolar survey items are also predominant in well-known social survey instruments. For example, in the most recent codebooks from the American National Election Studies (ANES) and General Social Survey (GSS) cumulative databases, respectively, 70\% and 81.4\%  of the variables measuring public opinion are bipolar. 
    
    Cognitive research on survey item response shows that respondents follow a ``first sign, then magnitude” decision-making sequence when answering bipolar questions. In other words, they begin by deciding whether they identify with a rejection or support stance, and afterwards, they choose the answer that best expresses the intensity of this rejection or support position (\cite{ostrom_etal_1992}). This process has important implications for construal analysis. It suggests that assessing shared cultural outlooks between respondents requires first determining the extent to which their responses align during their transitions between negative and positive answer categories.

    Formally, the space of available options in a \textit{bipolar question} $q$ can be seen as the union of a \textit{positive opinion semispace} $\mathcal{P}_q$ and a \textit{negative opinion semispace} $\mathcal{N}_q$. These semispaces are composed of the answers expressing support or rejection, respectively. Additionally, both semispaces can include a \textit{neutrality} option $n_q$, which expresses a neutral position indicating neither support nor rejection of the statement $q$. 
    
    For example, let $\mathcal{A}_q$ be the \textit{answer space} formed by all the possible answers to question $q$, and let $q$ represent a $7$-point \textit{strongly agree--strongly disagree} question. Then
   \[
    \begin{aligned}
        \mathcal{A}_q=&\{\texttt{strongly disagree, disagree, slightly disagree, neither agree nor disagree,}\\
        &\texttt{slightly agree, agree, strongly agree}\},\\
        \mathcal{P}_q=&\{\texttt{strongly agree, agree, slightly agree, neither agree nor disagree}\} \text{, and }\\
        \mathcal{N}_q=&\{\texttt{strongly disagree, disagree, slightly disagree, neither agree nor disagree}\}.
    \end{aligned}
    \]
   
    We would like to remark on two points regarding bipolar questions. First, neutrality may not always be an explicit option. For example, in an \textit{in favor--against} question, 
    \[
    \mathcal{A}_q = \{\texttt{in favor, against}\}, \quad \mathcal{P}_q = \{\texttt{in favor}\}, \quad \text{and} \quad \mathcal{N}_q = \{\texttt{against}\}.
    \]
    We still consider this as a bipolar question, as it adheres to 
    the original definitions provided by (\cite{ostrom_etal_1992, batyrshin_etal_2017, krosnick_1999}), albeit with implicit neutrality. By contrast, the above $7$-point \textit{strongly agree--strongly disagree} question  above has explicit neutrality, as $n_q = \texttt{Neither agree nor disagree}$ is explicitly included in $\mathcal{A}_q$. This formal representation illustrates how semispaces can capture both explicit and implicit neutrality, as seen in typical survey question designs. Without loss of generality, for the remainder of this paper, we will assume that all questions have explicit neutrality. Note that, although typically bipolar survey items have the same number of negative and positive answer choices, and these are ordered, our framework does not require these assumptions.
    
    Finally, we highlight a key feature of opinion surveys relevant to construal clustering. In these surveys, the answer options presented to respondents are generally not numbers but strings, such as ``agree'' or ``disagree''. In most cases, there is no standard way to map these strings to numeric values. Moreover, the number of  answer choices varies widely across questions, ranging from as few as $2$ for binary \textit{agree--disagree} questions to as many as $98$ in $0$--$97$ ``thermometer'' scores. As a result, the intensity of opinion associated with each string can vary across questions. For example, the ``agreement'' or ``disagreement'' answer options to a bipolar question do not match the options provided by a $5$-point scale question that distinguishes among ``completely'', ``somewhat'', and ``neither agreeing/disagreeing''. As we discuss in the next section, this issue poses challenges to construal clustering methods.

\section{Bipolar Class Analysis}
{\label{sec:BCA}}

        This section introduces Bipolar Class Analysis (BCA), a construal clustering method (CCM) developed specifically for analyzing bipolar survey data. Existing CCMs identify construal clusters from survey data through a two-step procedure:
        \begin{enumerate}
            \item Compute an adjacency matrix.
            \item Apply a clustering algorithm to identify distinct construals.
        \end{enumerate}
        The main difference among CCMs is how they compute the adjacency matrix. Two respondents are considered adjacent if their response patterns exhibit a similar structure. This number ranges from $0$ to $1$, where high values suggest a shared construal. The key innovation of BCA lies in replacing the distance-based adjacency measures of existing CCMs with a polar measure adapted to the bipolar nature of the data. The polar measure obviates the need to represent responses numerically or directly compare responses across survey questions, making BCA particularly suited for datasets with a bipolar structure. 

        \subsection{Step one: Polarity}
        Let $u=(u_1,\dots,u_Q)$ and $v=(v_1,\dots,v_Q)$ be the answers of two respondents to a survey. The \emph{polarity} $P(u,v)$ between these two respondents is defined by
        \begin{equation*}
        P(u,v) = \frac{2}{Q(Q-1)}\sum_{k=1}^{Q-1}\sum_{l=k+1}^Q\pi(u_{kl}, v_{kl}), 
    \end{equation*}
    where $Q$ is the number of survey questions, and $u_{kl}=(u_k, u_l)$, $v_{kl}=(v_k,v_l)$ are the responses of respondents $u$ and $v$ to questions $k$ and $l$, respectively. One may think of $P(u,v)$ as capturing the overall polarity between the two respondents based on their paired answers.


      The \textit{pairwise polarity function} $\pi$ assigns a value of $-1$, $0$, or $1$ to the pairs $u_{kl}$ and $v_{kl}$ of responses from respondents $u$ and $v$. \textcolor{black}{It focuses on whether the two respondents' answers across questions $k$ and $l$ remain in the same opinion semispace or shift to the other. A value of $1$ captures similarity in opinion shift or permanence. By contrast, $-1$ is assigned to opposite direction shifts or permanence in different opinion semispaces. Since the polarity function aims to compare opinion shifts, when either $u_{kl}$ or $v_{kl}$ remains in the same opinion semispace, but the other does not, polarity takes value $0$.} Figure~\ref{fig:polarity_function} illustrates the behavior of $\pi$, including how it evaluates these shifts. 
      
      In summary, $\pi(u_{kl},v_{kl})=-1$ when the responses to a pair of questions move to opposite semispaces (\emph{panels I and II} in Figure \ref{fig:polarity_function}), $\pi(u_{kl},v_{kl})=0$ if the responses change semispaces for one respondent but not for the other (\emph{panels III and VI}), and $\pi(u_{kl},v_{kl}) =1$ if the responses either both shift in the same direction or remain within the same semispace (\emph{panels IV, V, VII, and VIII}). \textcolor{black}{A flowchart detailing the steps of the algorithm  followed to production of the polarity-function is available in  Appendix~\ref{app:polarity} (Figure~\ref{fig:flowchart}).} 

            \begin{figure}
            \firstfigure
            \caption{BCA: polarity function $\pi$.}
                 \caption*{\footnotesize\textit{Notes:} Black dots represent answers of respondent $u$ to to question items $k$ and $l$; white ones represent answers of a hypothetical respondent $v$ for the same question.}
                 \label{fig:polarity_function}
            \end{figure}

        \textcolor{black}{Polarity may resemble more familiar sign-based tools, such as computing correlations between the signs of pairwise differences in responses. However, such transformations typically require embedding responses in a vector space (usually the real line) to enable algebraic operations on vectors, such as computing means. By contrast, polarity is agnostic to these requirements: it relies on comparisons of subspace changes and does not depend on embeddings or vector algebra. This makes polarity more versatile and generally applicable.}


    \subsection{Step two: Partitioning algorithm}

        \textcolor{black}{To cluster respondents sharing the same construal, the literature has employed two modularity partitioning algorithms: Newman's partitioning algorithm (\cite{newman_2004,newman_2006}) and Louvain's partitioning algorithm (\cite{blondel_2011}). Goldberg argued for the adoption of modularity partitioning as an appropriate tool for construal clustering (\citeyear[p.~1409]{goldberg_2011}), and we indeed consider it for the problem of construal clustering. Modularity is based on a comparison between the observed structure of edges in a graph and the structure expected under a random graph. This logic aligns well with construal clustering tasks, in which the goal is to identify robust patterns of association in the network while distinguishing them from random structure. For the interested reader, we recommend the following paper as a comprehensive review of modularity partition algorithms (\cite{bayan2024modularity_review}).}

        \textcolor{black}{These algorithms address the problem from two different perspectives: either by computing the exact modularity of the network or by approximating it through an optimization procedure. Exact modularity computations are not scalable in the size of the dataset. According to (\cite{bayan2024modularity_review}), Bayan's algorithm is the one capable of handling exact modularity computations for the largest networks, with up to $3,000$ nodes. Our simulations deal with sample sizes that are close to that upper limit, thus incorporating exact modularity algorithms to the simulation study was not feasible.}
        
        \textcolor{black}{By contrast, approximate algorithms, the most widely used ones, converge within reasonable computational times (in the order of seconds). We obtain results for three approximate algorithms: Newman (\cite{newman_2004,newman_2006}), Louvain (\cite{blondel_2011}), and Leiden (\cite{traag2019louvain}). These algorithms partition the data based on an adjacency matrix, such as the one provided by any CCM. Other algorithms, as the Paris partition algorithm (\cite{bonald2018hierarchical}), are tailored for distance-based matrices. In our experiments, transformation of the adjacency matrix  into a distance-based matrix led to convergence issues.}
        

\section{Construal Clustering Methods and Their Limitations}
\label{sec:RCACCA}
This section provides an overview of current construal clustering methods (CCMs) and examines their limitations using a concise example.

\subsection{Overview of existing CCMs}
        
        The first adjacency measure introduced in the CCM literature is \textit{relationality}, developed for Relational Class Analysis (RCA) by (\cite{goldberg_2011}). To compute relationality, responses must first be mapped to the $[0, 1]$ interval (\cite{goldberg_2011}, p.~1406). Consider $u= (u_1,\dots,u_Q)$ and $v=(v_1,\dots,v_Q)$ to be the answers of two respondents. We denote by $\tilde{u}$ and $\tilde{v}$ the numerical values where these answers are mapped to. Also, $\tilde{u}_{kl}= (\tilde{u}_k, \tilde{u}_l)$ and $\tilde{v}_{kl}=(\tilde{v}_k,\tilde{v}_l)$ denote the pairs of numerical values corresponding to questions $k$ and $l$. Then, the relationality $R(u,v)$ between $u$ and $v$ is given by
        \begin{equation*}
            R(u,v) = \frac{2}{Q(Q-1)}\sum_{k=1}^{Q-1}\sum_{l=k+1}^Q \lambda(\tilde u_{kl}, \tilde v_{kl})\delta(\tilde u_{kl}, \tilde v_{kl}),
        \end{equation*}
        where
        \begin{align}{\label{eq:delta_goldberg}}
            \delta(\tilde u_{kl}, \tilde v_{kl}) &= 1 - \lvert \lvert \tilde u_{k} - \tilde u_{l} \rvert - \lvert \tilde v_{k} - \tilde v_{l} \rvert\rvert
         \end{align}   
            and 
        \[
        \lambda(\tilde u_{kl}, \tilde v_{kl}) =
        \begin{cases}    
                      1, & \text{if } (\tilde u_{k}-\tilde u_{l})(\tilde v_{k}-\tilde v_{l}) \geq 0  \\
                     -1, & \text{otherwise}.
             \end{cases}
        \]
         According to Goldberg (\citeyear{goldberg_2011}), $\delta(\cdot)$ measures schematic similarity between the pair and $\lambda(\cdot)$ ``is a binary coefficient that changes the sign of the schematic similarity if both distances are in opposite directions" (\cite[p.~1407]{goldberg_2011}). Relationality \textcolor{black}{is a real number that} ranges from $-1$ to $1$. Before building the adjacency matrix, RCA conducts an edge removal process, where relationalities that are small in absolute value are set to zero. This procedure has been criticized by (\cite[p.~366]{boutyline_2017}). After this step, the adjacency matrix is constructed using the absolute values of the remaining nonzero pairwise relationalities. 
            
        Boutyline (\cite{boutyline_2017}) introduced Correlational Class Analysis (CCA) as an alternative to RCA. Before any computation, each respondent's answers are mapped to the real line. Then, CCA computes the Pearson correlation coefficient between the numerical values $\tilde{u}$ and $\tilde{v}$ and uses its absolute value to construct the adjacency matrix. CCA also performs an edge removal process---though different from the one implemented in RCA---by setting statistically insignificant correlations to $0$.

        More recently, Sotoudeh and DiMaggio revisited Goldberg's relationality to develop a method which we call Recursive Relationality Class Analysis (RRCA), based on its R implementation. They introduced several variations to the original RCA (\cite[Appendix A]{sotoudeh_dimaggio_2023}). We consider three to be the most relevant. First, RRCA \textit{binarizes} the difference between each pair of answers. Specifically, they replaced $\delta(\cdot)$ in equation~\eqref{eq:delta_goldberg} with:
        \begin{align*}
            \delta'(\tilde u_{kl}, \tilde v_{kl}) = 1 - \lvert \lvert\sign(\tilde u_{k} - \tilde u_{l})\rvert - \lvert\sign( \tilde v_{k} - \tilde v_{l})\rvert\rvert.
        \end{align*}
        In practice, the role of $\delta'(\cdot)$ is to modify the values taken by $\lambda(\cdot)$ in cases where $\tilde u_k = \tilde u_l$ or $\tilde v_k = \tilde v_l$. Second, instead of working with differences between numerical values of each respondent's answers, RRCA employs a ``second differences" or recursive approach. With this framework, the authors compute adjacency using  difference-in-differences between numerical values of answers. Lastly, rather than taking the absolute value of the modified relationality, they compute its square to construct the adjacency matrix.

       \subsection{Motivating example}
        We illustrate the challenges that RCA, CCA, and RRCA face in clustering respondents into construals with bipolar data using a motivating example from politics. This example uses hypothetical survey responses on divisive political issues in the United States. The goal is to analyze shared patterns of political views among respondents A, B, and C. The survey asks them to select one option from
        \texttt{Strongly Disagree}, \texttt{Disagree}, \texttt{Somewhat Disagree}, \texttt{Neither Agree nor Disagree}, \texttt{Somewhat Agree}, \texttt{Agree}, and \texttt{Strongly Agree} in response to the following statements:
        \begin{enumerate}
          \item[Q1.] \textit{Undocumented migrants should be summarily deported.}
          \item[Q2.] \textit{Gun control should be free of background checks.}
          \item[Q3.] \textit{Medicaid should not be further expanded.}
        \end{enumerate}  
        Figure \ref{fig:example_opinions} plots the answer values for each respondent.

    \begin{figure}
    \centering
    \secondfigure
    \caption{Political opinion answers in a hypothetical survey example.} 
    \caption*{\footnotesize\textit{Notes:} Black dots refer to respondents that rank as ``classic ideologues", gray dots indicate those characterizable as ``alternative ideologues" (see \cite{baldasarri_goldberg_2014}).}
    \label{fig:example_opinions}
\end{figure}

    We follow the nomenclature and empirical findings of Baldassarri and Goldberg (\cite{baldasarri_goldberg_2014}) to characterize the environment as divided into two political opinion construals, each populated by either ``classic" or ``alternative" ideologues. \textit{Classic ideologues} are divided between ``progressives'' --who strongly reject expedient migrant deportation, easy access to guns, and freezes in public health care-- and ``conservatives'', who strongly support all of these policies. Among traditional ideologues, responses across survey questions are positively related. 
            
    In our example, respondent A ranks as a ``classic ideologue'' and can be characterized as a moderate conservative. Her responses are (\texttt{Somewhat Agree}, \texttt{ Somewhat Agree}, \texttt{ Somewhat Agree}). Respondent C is also a conservative within the classic ideologues but expresses stronger support for the last question. Her responses are (\texttt{Somewhat Agree}, \texttt{ Somewhat Agree}, \texttt{ Strongly Agree}).

    In the second political construal, \textit{alternative ideologues}, responses to questions $1$ and $2$ are positively related with one another but each is negatively related with the answers given to question 3. Redistributive conservatives lie at one end of this spectrum of political opposition: they support fast migrant deportation and no changes in gun control policies but object halting Medicaid expansion. At the other end are fiscally conservative progressives, who reject expedient migrant expulsion and easy access to guns but also oppose further expansions of Medicaid. Respondent B's views classify her as a moderate redistributive conservative. Her responses are (\texttt{Somewhat Agree}, \texttt{Somewhat Agree}, \texttt{Somewhat Disagree}).

    \subsection{Limitations observed in existing methods}

    Figure~\ref{fig:example_similarity_scores} plots the adjacency values yielded by RCA, CCA, RRCA, and BCA for each respondent pair. Recall that values closer to one indicate that the two respondents share a common pattern of political thought ---that is, they are likely to belong to the same construal. Out of the four methods, only BCA succeeds in assigning a higher adjacency value to the one pair that belongs to the same construal, $(A,C)$. In what follows, we discuss why RCA, CCA, and RRCA fail to provide a satisfactory answer to the problem in this example. 

    \begin{figure}
            \centering
            \thirdfigure
    \caption{Adjacency measures calculated by different CCMs for respondent pairs.}
    \caption*{\footnotesize\textit{Notes:} Scores for RCA, CCA, and BCA are absolute values. Results for RRCA are squares. Results for RCA and CCA do not include the edge removal process.}
     \label{fig:example_similarity_scores}
    \end{figure}

    The primary reason for these discrepancies is the failure of existing methods to account for the bipolar structure of the data. A key feature of bipolar survey items, as we have noted in Section~\ref{sec:bipolar_data}, is that respondents typically position themselves within one of the opinion semispaces when evaluating each question (\cite{ostrom_etal_1992}). Consequently, the placement of each answer relative to these semispaces is crucial for understanding how respondents organize their opinions. However, none of RCA, CCA, or RRCA effectively incorporate this information.

    RCA computes relationality based solely on differences in numerical answers among pairs of questions. This method does not account for whether the responses fall within the same semispace. As a result, RCA assigns the same value to respondent pairs $(A,B)$ and $(A,C)$. For example, the numerical value of respondent B’s opinion on Medicaid expansion (\texttt{Strongly Agree}) is two points higher than respondent A’s (\texttt{Somewhat Agree}), while the numerical value corresponding to respondent C’s opinion (\texttt{Somewhat Disagree}) is two points lower than respondent A’s. Although these opinions fall in different semispaces, RCA treats their relative positions as equivalent. Similarly, RCA assigns a considerably large relationality to the respondent pair $(B,C)$ because the numerical values of their opinions on Medicaid expansion are equidistant from the numerical values of their respective views on the other two issues.

    In our motivating example, the binarization step taken by RRCA aggravates this issue, as it disregards the magnitude of the differences between numerical answers. Thus, it is expected that pairs $(A, B)$ and $(A, C)$ have the same recursive relationality value. After all, the sign of the difference does not indicate whether the answers fall in different opinion semispaces. Similarly to RCA, the pair $(B,C)$ has a considerably large score. Moreover, since the binarization step shrinks the adjacency score of $(A, B)$ and $(A, C)$, the three adjacencies happen to be equal.

    CCA computes the correlation coefficient between the numerical answers of two respondents. Thus, for CCA, what matters is the deviation of each answer from the mean of the numerical answers. Importantly, this mean may not align with the point where the neutral answer is mapped. As a result, being above or below the mean of the numerical answers does not necessarily correspond to being in distinct opinion semispaces. For example, the mean of respondent B's answers lies somewhere between where \texttt{Somewhat Agree} and \texttt{Neither Agree nor Disagree} are mapped. This approach can lead to misleading results.

    The issue is illustrated by the adjacency values that CCA yields for respondents B and C, which are the largest possible. When viewed as deviations from each respondent’s mean numerical answer, respondent C's answers mirror those of respondent B. As a result, the correlation between the two is $-1$. Since CCA uses the absolute value of this correlation to cluster respondents into construals, B and C may be incorrectly clustered into the same construal despite their contrasting patterns of thought.

    Current methods also face several other limitations. The first arises from the mapping of answers into the real line, typically into the $[0,1]$ interval. When the number of response options varies among questions, this operation may assign different numerical values to identical textual expressions --and identical values to different ones. For instance, on a $5$-point Likert scale item, the response  \texttt{Agree} is usually given a value of $0.75$, whereas on a $3$-point question it is assigned a value of $1$. This inconsistency in the normalization of the response set can lead to distorted measurements of pairwise adjacencies. 
            
    To illustrate this point, suppose one respondent answers (\texttt{Agree, Agree}) to a pair of $3$-point Likert scale questions, while another chooses (\texttt{Disagree, Agree}). Their numerical answers would be $(1,1)$ and $(0,1)$, respectively, yielding a relationality of $0$ between them. However, if these responses were instead derived from $7$-point Likert-scale questions, the answers would be mapped to $(5/6,5/6)$ and $(1/6, 5/6)$, respectively. In this case, the relationality would be $1/3$. 

    Second, CCA runs into trouble when clustering respondents who provide the same answer across all items --that is, those whose responses do not deviate from their mean numerical answer. Boutyline (\cite{boutyline_2017}) proposes two approaches to address this issue: removing these respondents from the sample or, as illustrated in Figure~\ref{fig:example_opinions}, setting CCA’s value between ``constant respondents'' and ``non-constant respondents'' to zero. However, as Boutyline acknowledges (\cite[Appendix D]{boutyline_2017}), this latter approach yields a ``null construal'' comprising all respondents with identical answers across all items.
 
    Based on the limitations above, we make a final pair of observations regarding  adjacency measures underlying current CCMs. First, distance-based adjacency measures may be unreliable because they treat answers as cardinal variables. They assume answers are comparable in magnitude when, at most, they are comparable in order. In a $7$-point Likert-scale question, for instance, individuals choose between ``slightly" or ``strongly" agree based on whether they perceive their opinion intensity to cross a certain threshold. These thresholds may vary across questions. As a result, computing adjacency based on differences in numerical answers can lead to misinterpretations, portraying respondents' positions as either farther apart or closer than they actually are. 
        
    Second, current CCMs are predicated on the assumption that there is a ``universal scale" that allows to compare answers across different questions. \textcolor{black}{This assumption underlies the data preprocessing step, consisting on assigning numerical values to responses. RCA and RRCA are sensitive to preprocessing choices, as they use the intensity of responses in their $\delta(\cdot)$ and $\delta'(\cdot)$ functions. Moreover, RCA implicitly assumes that each item is embedded in $[0,1]$, so that relationality takes values in $[-1,1]$. CCA's results remain the same as long as the alternative numerical embedding is an affine transformation of the original one. However, the correlation coefficient is also sensitive to the preprocessing steps taken for each question.}
    
    \textcolor{black}{The ``universal scale" assumption} presents a significant obstacle, as expressions of support or rejection intensity are often incomparable across questions, particularly when the questions have differing numbers of response options. For instance, simple ``agreement'' or ``disagreement'' in binary bipolar questions cannot be canonically mapped into the answer set of $5$-point Likert-scale questions.

     Addressing all the limitations previously presented, BCA is specifically designed to avoid mapping answers to numerical values and to function independently from a universal scale. Furthermore, we have demonstrated through an example that taking into account the bipolar structure of the data can yield more accurate results. To validate and extend these insights, we perform a simulation analysis that evaluates the performance of CCMs across a diverse range of synthetic datasets.


\section{Simulation Analysis}
\label{sec:simulations}

    \subsection{Simulation algorithm} 
        
        Our modeling strategy to generate synthetic datasets has two distinctive features that aim to capture key elements of the cognitive processes underlying survey responses. The first is \emph{latent ordered choice} (\cite{greene_2010}), which assumes that each respondent has an unobserved latent \textit{position} on a continuum for each survey question. These positions are mapped to specific survey responses selected from a finite set of ordered alternatives. The second feature is \emph{non-linear dependency}, which assumes that the latent positions of respondents across different questions are related through a copula function. This approach generalizes the shift-scale simulation process in the CCM literature (\cite{boutyline_2017}), allowing for a more complex representation of respondent behavior in the data generation process.

        The algorithm to generate the synthetic data consists of three steps: 
        \begin{enumerate}
            \item Generate the number of construals ($K$), their population sizes ($N_1, \dots, N_K$), and the number of survey questions ($Q$). The total sample size is $N=N_1 + \cdots +N_K$.
            \item Generate an $N \times Q$ matrix $X^*$ representing respondents' latent positions. 
            \item Map the latent positions onto an equally sized matrix $X$ of observed responses.
        \end{enumerate} 
        The first step is straightforward, while the second and third steps are elaborated on in the following paragraphs. Further details are provided in Appendix~\ref{app:simulation}.

        \subsubsection{Modeling positions}
        Respondents' \textit{positions} on each survey item are modeled as continuous uniform random variables ranging from $0$ (full disagreement) to $1$ (full agreement).\footnote{Our simulation process is equivalent to considering question-specific distributions (see Appendix~\ref{app:simulation}).} For example, if respondent $i$ has a position of $X_{i1}^*=0.7$ on the first issue, this means that 70\% of the population disagrees more strongly than respondent $i$ on this issue. Positions across different questions are not independent; instead, they are constrained by a dependency structure defined by a construal-specific multivariate distribution $C_k(x^*_1,x^*_2,\dots, x^*_Q)$. Consequently, the statement ``respondent $i$ belongs to construal $k$'' is modeled as ``the positions in $X_i^* = (X_{i1}^*, \dots, X_{iQ}^*)$ are drawn from the multivariate distribution $C_k(\cdot)$.''

        The construal-specific multivariate distribution $C_k(\cdot)$ is a \textit{copula}, as it has uniform marginal distributions (see \cite{nelsen_2006} for an introduction to copulae). A wide variety of copulae exist, allowing for the modeling of diverse dependence structures among the latent positions held by individuals within a construal. For our simulation, we use the Gaussian copula family due to its computational efficiency and ease of interpretation. The shape of the Gaussian copula $C_k(\cdot)$ is determined by its correlation matrix $\Sigma_k$. 
    
        \subsubsection{Modeling responses} Respondents map their latent positions onto observable \textit{responses} by selecting from a finite number of ordered options. A respondent's choice depends on where their position falls relative to a question-specific threshold. For instance, consider a respondent with a position value of $X_{i1}^*=0.7$ for the first survey item. If the thresholds for \texttt{Somewhat Agree} and \texttt{Strongly Agree} on this item are $0.6$ and $0.8$, respectively, the respondent will select \texttt{Somewhat Agree} as their response.

        A question $q$ with $H_q\geq 2$ response options is characterized by $H_q - 1$ thresholds that partition $[0, 1]$ into $H_q$ subintervals. These thresholds are used to model question-specific features. Since positions on a given issue are uniformly distributed, the proportion of the population selecting each option corresponds to the length of the associated subinterval. The subinterval lengths can be adjusted to model statements eliciting varying degrees of consensus. To that end, we introduce a skewness parameter that controls the prevalence of responses in the positive \textcolor{black}{opinion} semispace (agreement) relative to those in the negative \textcolor{black}{opinion} semispace (disagreement). 
        For example, if the skewness parameter is $0.2$, then the share of the population selecting agreement responses is $20$ percentage points greater than that selecting disagreement responses. To introduce additional variability across survey questions, the subinterval lengths are randomized, allowing them to differ from one question to another.
    
    \subsection{Experiments}
        We assess the performance of BCA relative to existing CCMs through two simulation experiments. The first experiment aims to confirm the limitations of existing CCMs, as discussed in Section~\ref{sec:RCACCA}. The second experiment evaluates CCM's performance across a broad range of datasets. In both experiments, we use BCA, RCA, RRCA, and CCA to cluster respondents into construals for each dataset.\footnote{For RCA and CCA, we used the R packages \href{https://CRAN.R-project.org/package=RCA}{\texttt{RCA}} (v2.0) and \href{https://CRAN.R-project.org/package=corclass}{\texttt{corclass}} (v0.2.1), respectively. For RRCA, we used the R implementation available at \href{https://github.com/raminasotoudeh/coping_with_plenitude} {\texttt{https://github.com/raminasotoudeh/coping\textunderscore with\textunderscore plenitude}}.} \textcolor{black}{Results from the simulations using Newman's partitioning algorithm are reported through this section; those from simulations employing Leiden's algorithm are available in Appendix~\ref{app:extra_results}.\footnote{\textcolor{black}{Results obtained using Louvain’s partition algorithm are similar to those obtained with Leiden’s algorithm, although moderately worse for all CCMs. The corresponding tables are available upon request.}} }

        \subsubsection{Experiment 1: Exploring limitations} To investigate the limitations identified in Section~\ref{sec:RCACCA}, we construct synthetic datasets comprising three questions ($Q=3$), each with five response options ($H_q=5$).  
        The data structure follows our motivating example, where respondents are members of one of two construals. 
        The dependence structure for each construal is determined by the following correlation matrices:
        \begin{equation} \label{eq:bwors_construals}
            \Sigma_1 = \begin{pmatrix}
                1 & 1 & 0.7 \\
                1 & 1 & 0.7 \\
                0.7 & 0.7 & 1
            \end{pmatrix}
            \text{ and }
            \Sigma_2 = \begin{pmatrix}
                1 & 1 & -0.7 \\
                1 & 1 & -0.7 \\
                -0.7 & -0.7 & 1
            \end{pmatrix}.            
        \end{equation}
        The number of respondents in each construal and the skewness of responses for each question are randomized, as detailed in Table~\ref{tab:experiments}. We generate $1000$ independent datasets with these characteristics. 

        \subsubsection{Experiment $2$: General performance} In this experiment, we randomize all parameters to generate a diverse array of datasets with varying structural characteristics. The number of construals is sampled from $\{2, 3, 4, 5, 6 \}$, with each option assigned equal probability. Importantly, the dependence structure for each construal is generated randomly: the matrices $\Sigma_k$ are positive definite and have random entries (see Appendix~\ref{app:simulation} for the details). The ranges for the remaining parameters are provided in Table~\ref{tab:experiments}, with each option in the specified ranges assigned equal probability. We generate $5000$ independent datasets with these characteristics.

    \begin{table}[h!]
    \centering

    \caption{Description of the experiment parameters.}
    \label{tab:experiments}

        \begin{tabular}{@{}rll@{}}
        \toprule
        \addlinespace[0.4em]
                             & Experiment 1: Exploring limitations           & Experiment 2: General performance \\
                             \addlinespace[0.2em]
                             \midrule
                             \addlinespace[1.0em]
        Number of questions  & Fixed $Q=3$                                   & Random from $10$ to $20$           \\
        Construal population & Random from $200$ to $400$                    & Random from $200$ to $400$        \\
        Number of construals & Fixed $K=2$                                   & Random from $2$ to $6$            \\
        Dependence structure & Fixed at equation~\eqref{eq:bwors_construals} & Random                            \\
        Number of options    & Fixed $H_q=5$                                           & \textcolor{black}{Random in \{3, 5, 7, 9, 11\}}             \\
        Question skewness    & Random in $[-0.2, 0.2]$                                            & Random in $[-0.2, 0.2]$      \\
        \addlinespace[0.8em]
        \hline      
        \hline
        \end{tabular}

    \end{table}

        \subsection{Performance measures}
        
        We compare methods using four performance measures. The first is \emph{Construal Partition Accuracy} (CPA), which measures the rate at which a method correctly estimates the number of construals in the data. 
        The second measure is \emph{Mean Absolute Deviation} (MAD), which quantifies the average absolute deviation between the estimated and true number of construals, providing a different perspective on estimation performance. 

        The third measure is the \emph{Scaled Normalized Mutual Information} (SNMI). This is a scaled variant of Normalized Mutual Information (NMI), which evaluates classification accuracy on a scale from $0$ to $1$, with higher values indicating a greater accuracy in assigning respondents to their true construals.\footnote{\textcolor{black}{We also computed the Rand index, the Fowlkes--Mallows index, the adjusted versions of both indexes, and the Normalized Partition Accuracy (see \cite{gagolewski2025normalised} for a review of clustering performance metrics). Results are comparable to those obtained using SNMI and are available upon request.}} 
        
        \textcolor{black}{Although NMI is frequently used to evaluate construal clusters (\cite{boutyline_2017, sotoudeh_dimaggio_2023}), recent literature on clustering evaluation has found certain limitations in it. Vinh et al. (\citeyear{vinh2009information, vinh2010information}) proposed the \textit{Adjusted Mutual Information} (AMI) to overcome some of those limitations. Both NMI and AMI, however have been shown to be positively biased towards methods estimating larger number of construals (\cite{romano2014standardized}; \cite{amelio_2015}; \cite[p.~388]{boutyline_2017}; \cite{mahmoudi_2024}). Romano et al. (\citeyear{romano2014standardized}) presented an standardization of NMI, called \textit{Standard MI} (SMI). SMI does not present the aforementioned bias but its computation is complex. For this reason, we implemented SNMI introduced by Amelio and Pizzuti (\citeyear{amelio_2015}). SNMI scales NMI by a factor inversely proportional to the difference between the true and estimated number of construals. The scaling factor is more efficient and addresses NMI's bias. For comparability with previous CCM evaluation studies, Appendix~\ref{app:extra_results} also reports (unscaled) NMI figures.}
        
        The final performance measure we report is \emph{Correlational Dissimilarity} (CDIS). This measure, introduced in this study, quantifies the distance between the estimated and true dependency structures of construals within each dataset. \textcolor{black}{The CCM literature is generally interested in recovering the within-construal dependency structures. For instance, \citeauthor{baldasarri_goldberg_2014} (\citeyear{baldasarri_goldberg_2014}) estimated construal-specific correlations from the numerical representation of survey responses. In our simulation, construal dependency is characterized by the copula correlation matrices $\Sigma_k$. Therefore, we can evaluate how different CCMs recover it.} CDIS measures the difference between the estimated and true sets of correlation matrices, \textcolor{black}{assessing how accurately the $\Sigma_k$'s are estimated from a given partition of the data. A CDIS of $1$ means that the correlation structure is estimated as accurately as if the true construal memberships were known. Higher values of CDIS indicate lower accuracy in estimating the true correlation matrices from a given partition: a larger difference between the true and estimated dependence structure. For instance, a CDIS of $2$ indicates that estimation of the $\Sigma_k$'s based on a given clustering is twice as bad as estimation based on the true clustering.}  We present in detail CDIS in Appendix~\ref{app:performance}.

        \textcolor{black}{One natural question regarding CDIS is its relation with other construal evaluation measures. Measures such as SNMI, which is popular in the literature, evaluate how well an algorithm recovers the true class membership of observations. In our view, CDIS complements this measures when evaluating construal clustering. Indeed, there is a stark relation between CDIS and other cluster evaluation measures, such as SNMI. When SNMI of a given method is close to one (its maximum), the method’s CDIS is similar to CDIS computed with the true membership assignment (see Figure \ref{fig:benchmarkCDIS} in Appendix \ref{app:performance}).}

\subsection{Results}\label{sec:results}

        Results for the simulation analysis are shown in Table~\ref{ref:table_results}. The rows for Experiment 2 report averages for cases where the true number of construals ranges from $2$ to $4$, which is the number of construals estimated by most empirical studies (\cite{sotoudeh_dimaggio_2023}, p.\ 1842). Appendix~\ref{app:reanalyses} presents results separately across the true number of construals.

        \begin{table}[tbp]
    \centering
    \begin{threeparttable}

    \caption{Performance measures for RCA, CCA, RRCA, and BCA across experiments.}
    \label{ref:table_results}

    {\color{black}\begin{tabular}{lcccc}
        \hline
        \addlinespace[0.4em]
        & (1) & (2) & (3) & (4) \\
        \addlinespace[0.2em]
        & RCA & CCA & RRCA & BCA \\
        \addlinespace[0.2em]
        \hline

        \addlinespace[1.0em]
        \multicolumn{5}{l}{\textit{Experiment 1: Exploring limitations}} \\
        \addlinespace[0.6em]

        CPA                    & 0\%     & 0\%     & 0\%     & 93.1\% \\
        MAD                    & 337.810 & 5.992   & 2.169   & 0.069  \\
        CDIS                   & 10.428  & 14.885  & 11.601  & 2.927  \\
        SNMI                   & 0       & 0.004   & 0.024   & 0.172  \\
        Unit-construals        & 100\%   & 11.1\%  & 0\%     & 0\%    \\
        Computation time (s)   & 4.168   & 0.027   & 0.268   & 0.003  \\

        \addlinespace[1.5em]

        \multicolumn{5}{l}{\textit{Experiment 2: General performance (2--4 construals)}} \\
        \addlinespace[0.6em]

        CPA                    & 4.28\%  & 23.19\% & 33.60\% & 40.16\% \\
        MAD                    & 3.021   & 1.436   & 0.877   & 0.723   \\
        CDIS                   & 8.839   & 7.681   & 7.293   & 6.555   \\
        SNMI                   & 0.275   & 0.393   & 0.363   & 0.393   \\
        Unit-construals        & 22.19\% & 21.68\% & 0.06\%  & 0\%     \\
        Computation time (s)   & 39.243  & 0.051   & 35.937  & 0.163   \\

        \addlinespace[0.8em]
        \hline
        \hline
    \end{tabular}}
    \vspace{1.2em}
    \begin{tablenotes}[flushleft]
        \footnotesize
        \item \textit{Notes:} \textcolor{black}{Results are averages across simulated datasets. In Experiment~1, $36.2\%$ of datasets are discarded for RCA due to computational errors. For CDIS computations, unit-construals were removed prior to calculating it.}
    \end{tablenotes}

    \end{threeparttable}
\end{table}

\subsubsection{Experiment 1} 

    Our simulation of the example in Section~\ref{sec:RCACCA} (Experiment 1) confirms the limitations of existing CCMs. RCA, CCA, and RRCA fail to detect the presence of two construals in the data---the three methods always overestimate the number of construals. In contrast, BCA correctly identifies two construals in $93.1\%$ of the simulated datasets. We also observe this pattern in MAD from the true number of construals, where BCA's MAD is two orders of magnitude smaller than that of existing CCMs. BCA's SNMI is an order of magnitude bigger than that of the other methods. However, NMI values in this study are substantially lower than those observed in previous simulation analyses (see \cite{boutyline_2017, sotoudeh_dimaggio_2023}). \textcolor{black}{We believe that this difference reflects the more complex simulation approach carried out in this paper.} Regarding CDIS, BCA performs almost $3$ times worse than the benchmark \textcolor{black}{(considering that the true construal membership is known)}. Existing CCMs perform roughly $10$ times worse. \textcolor{black}{Differences between BCA and the other CCMs are statistically significant ($p-\mathrm{value} < 0.001$).}

    Throughout our simulation analysis, we found that RCA and CCA produced clusters with a single respondent. We refer to them as \textit{unit-construals}. Since construal populations are at least $200$, unit-construals may be interpreted as spurious. In Experiment~1, RCA produced at least one unit-construal for every dataset. CCA did so for $11.1\%$ of the datasets. In contrast, RRCA and BCA do not produce unit-clusters.

\subsubsection{Experiment 2: General results}

    We now examine the general performance of CCMs under a broader set of conditions. Identifying construals in this experiment is more challenging than in Experiment~1, as it involves more construals and randomly generated dependence structures, which may increase similarity among them. On the other hand, the larger number of questions leads to more information in the datasets. The performance of BCA decreases relative to Experiment 1, while that of RCA, CCA, and RRCA improves. However, CCA and RCA performed poorly in Experiment 1, where their ability to detect the dependence structure was particularly weak. Overall, BCA outperforms existing CCMs in Experiment 2 as well.

    BCA correctly estimates the number of construals in about $40\%$ of the times. RRCA, the second-best method, does so around $30\%$ of the times \textcolor{black}{(statistically significant difference at $0.001$)}. The MAD from the true number of construals provides a more detailed picture: on average, BCA misses the true number of construals by less than one. RRCA is also the second-best for this measure, deviating on average by around $0.2$ more from the true number of construals \textcolor{black}{(difference significant at  $0.001$)}. Regarding CDIS, BCA's accuracy in estimating the underlying dependence structure decreases to $6.5$ times worse than the benchmark. This is expected, as the setup in  Experiment 2 leads to a more challenging problem. \textcolor{black}{Nonetheless, it is the best performing method, with statistically significant differences with respect to the others ($p-\mathrm{value} < 0.001$).}

    \subsubsection{Experiment 2: Sensitivity analysis}

\begin{figure}[htb]
    \centering
\begin{subfigure}{0.33\textwidth}
  \includegraphics[width=\linewidth]{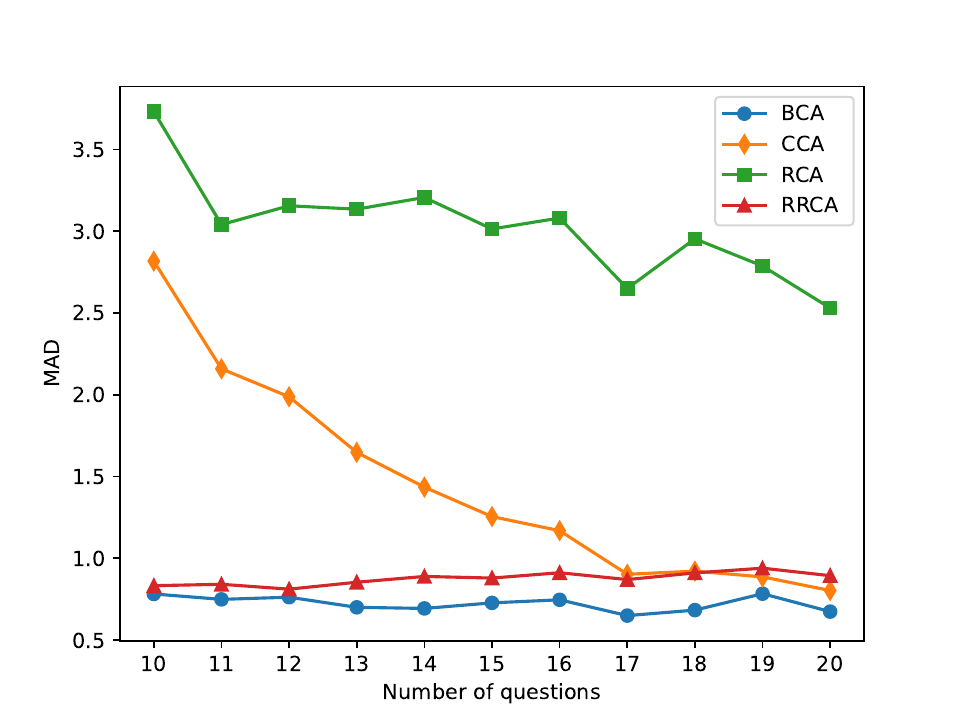}
  \caption{}
\end{subfigure}\hfil
\begin{subfigure}{0.33\textwidth}
  \includegraphics[width=\linewidth]{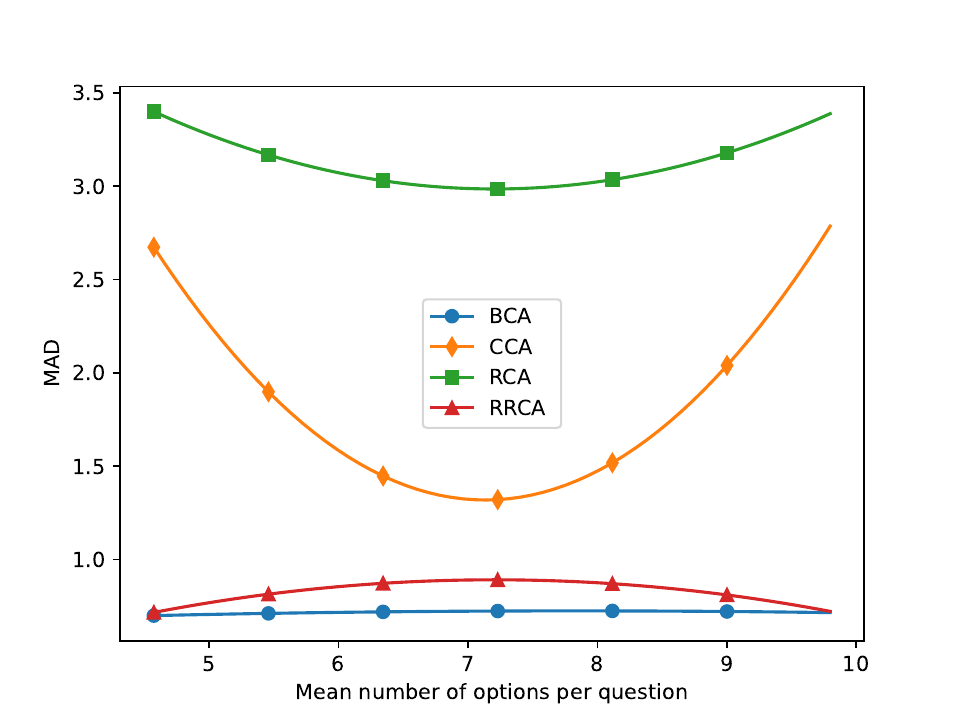}
  \caption{}
\end{subfigure}\hfil
\begin{subfigure}{0.33\textwidth}
  \includegraphics[width=\linewidth]{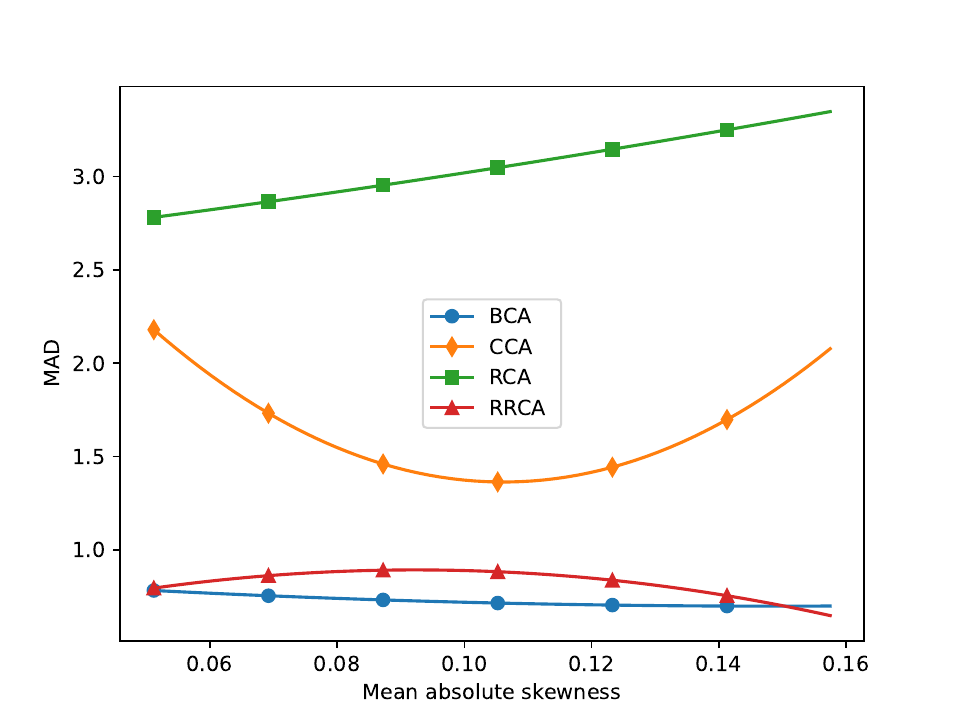}
  \caption{}
\end{subfigure}

\medskip
\begin{subfigure}{0.33\textwidth}
  \includegraphics[width=\linewidth]{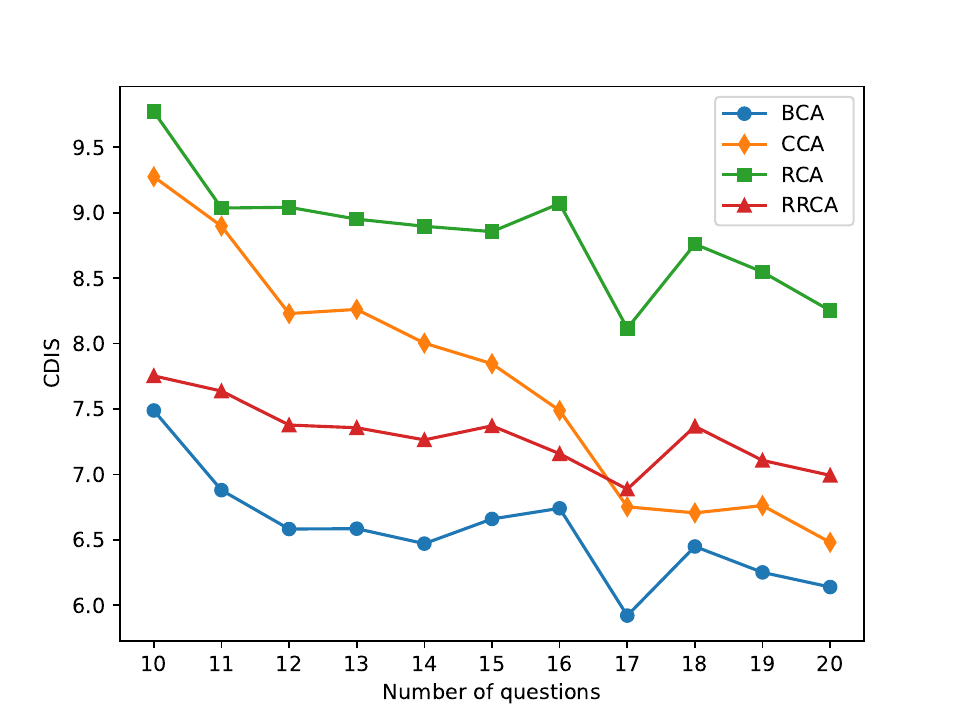}
  \caption{}
\end{subfigure}\hfil
\begin{subfigure}{0.33\textwidth}
  \includegraphics[width=\linewidth]{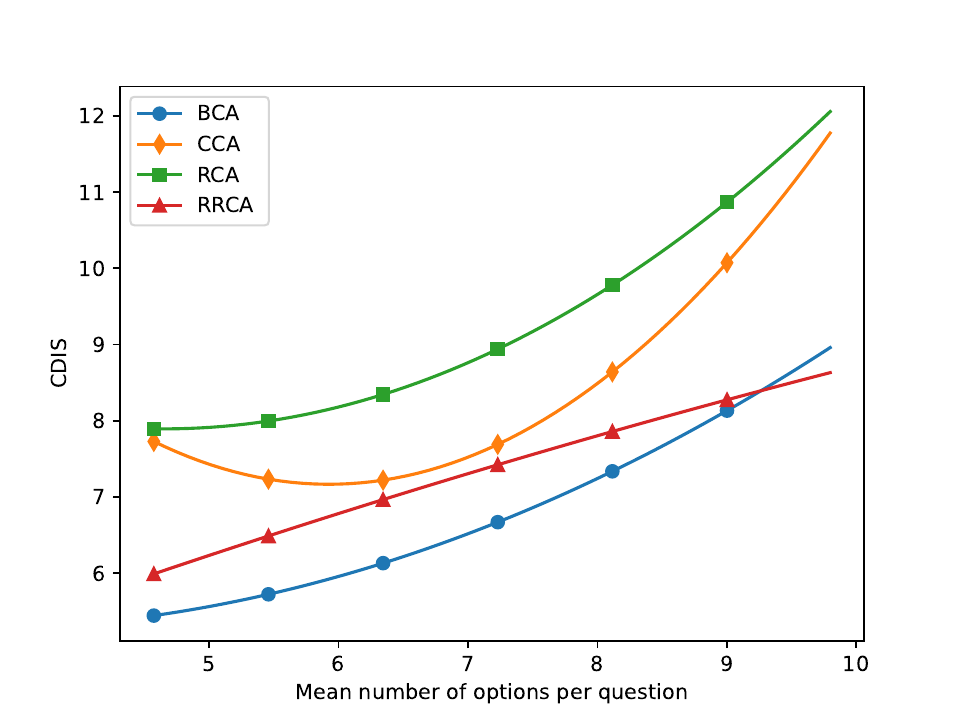}
  \caption{}
\end{subfigure}\hfil
\begin{subfigure}{0.33\textwidth}
  \includegraphics[width=\linewidth]{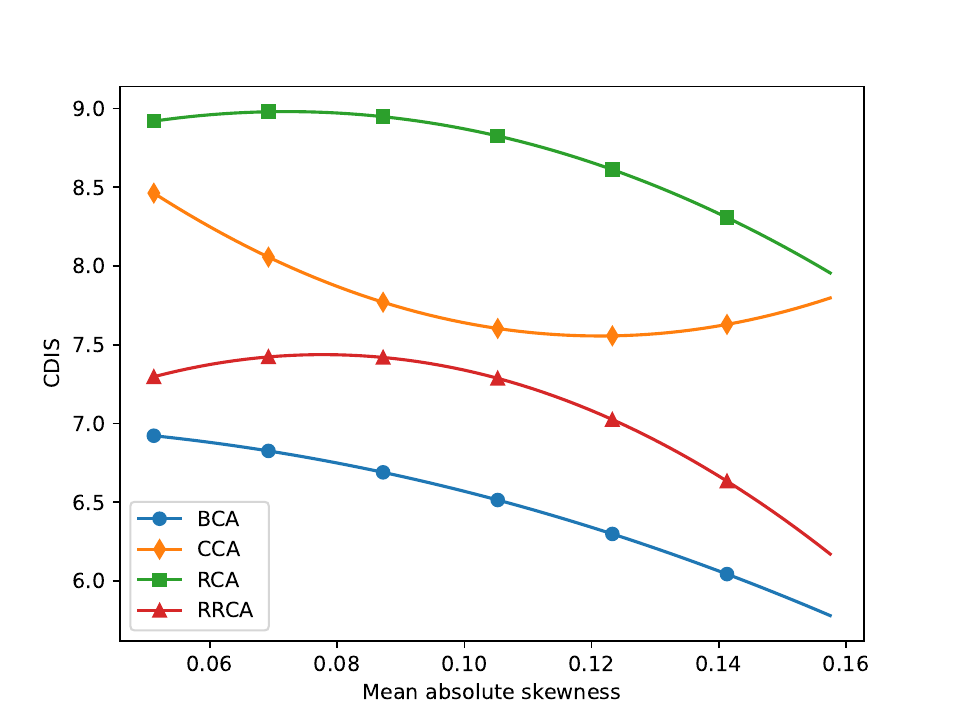}
  \caption{}
\end{subfigure}
\caption{\textcolor{black}{Sensitivity analysis (2--4 construals): MAD (Panels A-C) and CDIS (Panels D-F). Panels A and D: averages conditional on the number of questions. Panels B and E: quadratic fit conditional on the mean number of options per question. Panels C and F: quadratic fit conditional on the mean absolute value of the skewness parameter.}}
\label{fig:sensitivity_mad_cdis}
\end{figure}

    \textcolor{black}{ We analyze how the different CCMs perform conditional on three features of the underlying datasets. First, we compute the average performance of each method conditional on the number of questions present in the data. Second, we fit a quadratic regression of the methods' performance on the average number of options per question. Recall that options per question range from $3$ to $11$. Third, we also fit a quadratic regression of performance on the average absolute skewness parameter. The absolute value of skewness goes from $0$ (respondents divide equally between agreement and disagreement) and $0.2$ (the share of the population agreeing is $20$ percentage points greater than the one disagreeing, or vice versa). Figure~\ref{fig:sensitivity_mad_cdis} shows the results for  MAD and CDIS. Results for CPA generally parallel those for MAD. Results for SNMI are in line with those for CDIS, except in a couple of cases that we will highlight below. Figure~\ref{fig:sensitivity_cpa_snmi} in Appendix~\ref{app:extra_results} provides results for CPA and SNMI.
    }

\textcolor{black}{
Panel (A) in Figure~\ref{fig:sensitivity_mad_cdis} shows that MAD for BCA and RRCA is relatively constant across the number of questions. On the other hand, RCA's and CCA's MAD improves the more questions there are in the data. CCA's improvement is notorious, overtaking RRCA as the second best method for datasets with more than $17$ questions. Panel (D) shows an improvement in CDIS for all methods, with CCA again benefiting the most of the larger number of questions.
}

    The trends displayed in Panel~(B) indicate that BCA's, RCA's, and RRCA's MAD is relatively insensitive to the average number of options per question. This result is expected for BCA and RRCA, since both methods emphasize the agreement vs disagreement dimension, rather than the intensity of agreement or disagreement (see a more detailed discussion Section~\ref{sec:future_work}). CCA is an exception: its performance peaks around 7 options per question. SNMI across all methods also peaks around 7 options per question (see Panel~(E) of Figure~\ref{fig:sensitivity_cpa_snmi} in Appendix~\ref{app:extra_results}). In this case, CDIS shows a different trend: the performance of all methods  worsens as the number of options increases. This results is likely to stem from the combination of two facts: most method's performance is relatively flat in the number of options, but the benchmark (computed with the true membership vector) improves sharply with the number of options. \textcolor{black}{Hence, performance relative to the benchmark worsens.} 

\textcolor{black}{
Regarding the method's performance in terms of skewness, we again see that BCA and RRCA are relatively insensitive to this feature of the data (Panel~(C) for MAD). RCA worsens slightly as questions became, on average, more skewed; and CCA peaks when average skewness is around 0.1. The performance peak for somewhat skewed questions is observed for all method's SNMI. Results for CDIS are, in turn, opposite to what we found for the average number of questions. All method's CDIS decreases with the skewness parameter. We think that the same two driving forces are present in this case, but in opposite directions: a flat performance of the methods in the skewness parameter paired with the benchmark worsening as questions become more skewed.
}

\textcolor{black}{
To sum up, we find BCA's and RRCA's performance to be  consistent across different data scenarios. RCA and, specially, CCA are more reliant on the underlying data structure. CCA performs better when the number of questions is large, since this scenario allows for a finer adjacency measure (correlation between two respondent's answers). We find slight evidence, based just on SNMI, of all the methods performing better when the average number of options is around $7$ and questions are somewhat skewed. This evidence should be further investigated, particularly since the literature has recomemnded 5-point Likert scales (\cite{revilla2014choosing}).
}
    \subsubsection{Computational complexity}
    \textcolor{black}{
        The asymptotic computational complexity of all methods, with respect to the sample size, is $O(N^2)$. However, CCMs vary in terms of the complexity with respect to the number of questions. Both RCA and BCA are based on comparisons between question pairs and are, therefore, $O(Q^2)$. CCA computes correlations, an operation that is $O(Q)$. RRCA relies on ``second differences", hence it is $O(Q^4)$. This is reflected in the average wall-clock time that each CCM takes to compute their adjacency matrix (Table~\ref{ref:table_results}).\footnote{The timing analysis has been performed in a 13th Gen Intel(R) Core(TM) i7-13700, with 16GB of DDR5 RAM, and Microsoft Windows 10 Education as Operative System.} In Experiment~2, where $Q$ ranges from $10$ to $20$, CCA is the fastest method, while BCA is an order of magnitude slower. Nevertheless, BCA's computation time remains under a second. RCA and RRCA are two orders of magnitude slower than BCA. We note here that the performance of RCA is hampered by the edge removal process, which is a bootstrapping procedure. Results in Experiment~1 are slightly different, as $Q=3$. There, the bootstrapping procedure consumes a large share of computing time, so RCA is remarkably slower than the rest of the methods. BCA is even faster than CCA, since the later method implements an edge removal process (not based on bootstrapping).
    }

    \subsubsection{Robustness checks}
    
            We have also performed two additional experiments to asses the robustness of our results. A third experiment studies departures from Gaussianity, simulating data whose dependence is characterized by Student’s $t$ copulae with $3$ and $6$ degrees of freedom. The results are very similar to those obtained using the Gaussian copula (Tables~\ref{tab:studentt3} and \ref{tab:studentt6} in Appendix~\ref{app:extra_results}). This similarity is expected, since CCMs are not based on a Gaussianity assumption. A fourth experiment evaluates CCMs' performance when the bipolar structure is absent, simulating data with no dependence structure (a single construal with identity correlation matrix). In this case, all CCMs perform poorly, with a small edge for BCA and RRCA (Table~\ref{tab:structureless} in Appendix~\ref{app:extra_results}).

\section{Empirical Analysis}
{\label{sec:reanalyses}}

        The results of our simulations indicate that BCA can identify construals more accurately than existing CCMs. An equally important question, however, is whether the construal clusters estimated by BCA substantively diverge from those generated by other methods. We explore this question in detail by applying BCA, RCA, CCA, and RRCA to reanalyze two datasets previously studied in construal analyses: the 2004 American National Election Studies (ANES) survey on political attitudes and the 1993 General Social Survey (GSS) module on musical preferences (\cite{baldasarri_goldberg_2014,boutyline_2017, goldberg_2011}).\footnote{In our analyses, we apply the same data processing and recoding procedures as in these studies; consult Appendix~\ref{app:reanalyses} for details.} We also conduct a more succinct analysis of how the population of construals across CCMs differs when they are applied onto two other datasets. Our results indicate that BCA yields substantively  different results than other CCMs, beginning with the number of reported construals and extending to their underlying dependence structures, their sociodemographic composition and behavioral outcomes. Details on the analyses and full results are provided in Appendix~\ref{app:reanalyses}.


        \subsection{Political opinions: ANES 2004} We analyze responses to $29$ bipolar public opinion questions from $611$ respondents. These data were analyzed in detail by Baldassarri and Goldberg (henceforth, BG) in their study of political belief systems in the US (\citeyear{baldasarri_goldberg_2014}, pp.~60--61). These questions have $2$, $3$, $4$, or $7$ response points. BG classified these items into four domains: economic issues; civil rights (dealing mainly with affirmative action and positions on Black disadvantage); morality (including abortion and opinions on traditionalist worldviews); and foreign policy.

        BG identify three construals organizing how Americans bundle together their political opinions. One is the standard liberal-conservative (lib-con) ideological axis. When opinion variables are recoded so higher values correspond to increasingly right-wing positions, respondents on this axis exhibit widespread positive correlations in their opinions across items and domains. Overall, this construal places individuals with progressive views in all domains at one end of the axis and those with consistently conservative stances across these domains, on the other. A second construal is an ``alternative" political axis in which support for economically progressive issues is negatively correlated with support for liberal moral issues. This axis opposes fiscally conservative progressives to pro-redistribution conservatives, who combine progressive economic views with traditionalist cultural views. The third and final axis identified by BG is an ``agnostic" one, populated by individuals who hold issue positions largely dissociated from one another. Our RCA reanalysis successfully replicates this three-construal structure; graphical representations of these results are shown in Appendix~\ref{app:reanalyses} (Figure~\ref{fig:construal_politics_results}).
         
\begin{figure}[htbp]
    \centering


    \includegraphics[width=470pt]{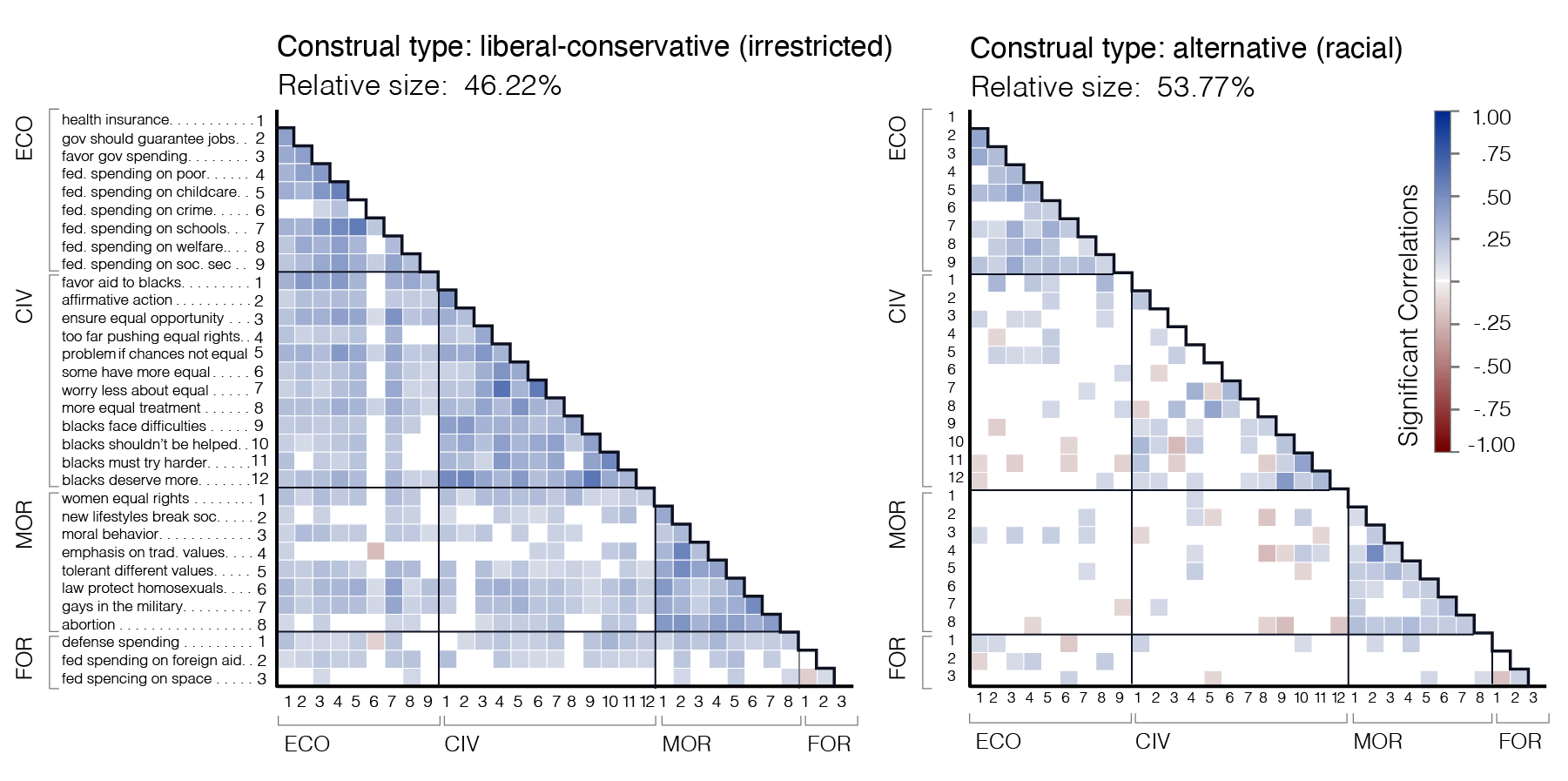}

    \captionsetup{width=470pt}
    \caption{BCA's dependence structures of political construals.}
    \caption*{\footnotesize\textit{Notes:} Non-significant correlations are shown in white.}

    \label{fig:politics_bca_estimates}
\end{figure}

         Analyzing the 2004 ANES data with BCA produces different results than RCA. Figure~\ref{fig:politics_bca_estimates} plots the correlation matrices of the two construal clusters produced by BCA. In the correlation matrix of Cluster I, shown in the left, three-quarters of survey items (77\%) are positively and significantly correlated with one another. This dependence structure aligns well with BG’s characterization of a liberal–conservative axis. But its size, on the other hand, is more than $50\%$ larger than the one estimated by BCA (see Table~\ref{tab:politics_construal_populations}). 

         The dependence structure of Construal II, by contrast, does not fit the characteristics of either the agnostic or the alternative axis described by BG. This construal is not agnostic, as it exhibits significant positive correlations between items pertaining to economic and moral domains. But it does not fit well with the characteristics that BG ascribed to their alternative axis. The dependence structure of construal II exhibits negative correlations between economic and civil rights opinions rather than between economic and moral ones. Therefore, construal II is best characterized as a novel ``alternative racial" axis marked by a trade-off between liberal positions on racial relations and affirmative action, on the one hand, and economic positions, on the other. As Table~\ref{tab:politics_construal_populations} shows, this type of political axis is only identified  by BCA. The construals identified by CCA or RRCA can be best characterized either as standard, unrestricted liberal–conservative axis, with widespread positive correlations between items across domains, or as more restricted liberal–conservative variants in which positive correlations are concentrated within a subset of domains (see Appendix~\ref{app:reanalyses}, Figure~\ref{fig:construal_politics_results}).

\begin{table}[htbp]
    \centering

     %
     \caption{Construal relative sizes for political preferences, ANES 2004.}
     \vspace{0.4em}
     \hrule
     \vspace{0.6em}

    \begin{tabular*}{470pt}{@{}p{3.1cm}@{\extracolsep{\fill}}ccccc@{}}
        \addlinespace[0.8em]
        & \multicolumn{2}{c}{Liberal-conservative} 
        & \multicolumn{2}{c}{Alternatives} 
        & Agnostic \\
        \cmidrule(lr){2-3} \cmidrule(lr){4-5}

        & Restricted       & Unrestricted  
        & Moral            & Racial        
        &                  \\
        \addlinespace[0.4em]

        & (1)              & (2)            
        & (3)              & (4)            
        & (5)              \\
        \addlinespace[1.0em]

        \midrule
        \addlinespace[0.8em]

        BCA\dotfill   & 46.22\% (1)   & ---           & ---           & 53.77\% (1) & ---        \\
        RCA\dotfill   & 30.32\% (1)   & ---           & 38.19\% (1)   & ---         & 31.47\% (1) \\
        CCA\dotfill   & 51.30\% (2)   & 48.68\% (1)   & ---           & ---         & ---        \\
        RRCA\dotfill  & 50.16\% (2)   & 49.17\% (2)   & ---           & ---         & ---        \\

        \addlinespace[0.6em]
        \hline
        \hline
    \end{tabular*}
    \vspace{1.2em}
    \caption*{\footnotesize\textit{Notes:} Relative size refers to the share of respondents classified in each construal. The numbers in parentheses indicate the number of estimated construals within categories. Dependence structures of construal estimates for RCA, CCA, and RRCA are shown in Appendix~\ref{app:reanalyses}, Figure~\ref{fig:construal_politics_results}.}

    \label{tab:politics_construal_populations}
\end{table}
        
        Furthermore, BCA leads to substantively different conclusions on how social characteristics shape respondents’ assignment to political opinion construals. We estimate logistic regression models predicting membership in the (restricted) liberal–conservative construal generated by each method. Most sociodemographic predictors exhibit similar effects across methods. The exception is racial identification. In models regressing BCA classifications, African American respondents are substantially more likely to be classified into the liberal–conservative construal ($\text{odds ratio} = 15.737$, $p-\mathrm{value} < 0.001$). This result underscores the central role that BCA assigns to race in ideological sorting.

        Finally, we find that the construal categories generated by BCA perform better as predictors of partisanship than those generated by alternative methods. We estimate OLS models regressing a $7$-point partisanship scale (from Strong Democrat to Strong Republican), on construal membership and sociodemographic characteristics. Ranking as a member of the BCA alternative racial axis (Construal II) predicts a $1.513$-point shift toward Republican identification. The magnitude of this estimated effect is more than 50\% larger than that associated with being part of any RCA, CCA or RRCA cluster. Of equal importance, models that use membership in BCA categories as predictors explain significantly more variation in partisanship than those including categories of ascription to construals from other methods. Whereas the adjusted $R^2$ of models using BCA categories is $0.240$, the adjusted $R^2$ of models employing construal memberships from other CCMs ranges between $0.151$ and $0.162$ (see Table~\ref{ref:apx_determinants_partisanship} in Appendix~\ref{app:reanalyses}).
  
    
        \subsection{Musical preferences: GSS 1993} This dataset consists of $1{,}532$ observations on seventeen $5$-point Likert-scale items measuring respondents' musical tastes. These data have been previously analyzed by Goldberg (\citeyear{goldberg_2011}) and Boutyline (\citeyear{boutyline_2017}). They generated two competing accounts of how Americans arranged their musical preferences. Goldberg (\citeyear{goldberg_2011}), using RCA, identified three musical preference construals. One is an “omni/univore” construal characterized by positive correlations in musical preferences across genres. Another is a “high/low-brow” construal in which preferences for genres high in cultural capital (such as opera and classical music) are negatively correlated with those for low-capital genres (such as rap and heavy metal). A third is a “contemporary/traditional” construal that opposes preferences for established American musical genres to preferences for more recent ones. By contrast, using CCA, Boutyline (\citeyear{boutyline_2017}) found that musical tastes were organized around one contemporary/traditional construal and three variants of an ``omni/univore" construal.

        
    

        Table \ref{table:music_bca_estimates} shows the number, type, and relative sizes of the construals estimated by CCMs. BCA reports only two construals, both of roughly equal size. They exhibit clearly distinctive dependence structures. One aligns with a contemporary/traditional form and the other with a high/low-brow type (correlation matrices are available in Figure \ref{fig:construal_music_results}, Appendix \ref{app:reanalyses}). Importantly, BCA does not retrieve an omni/univore construal, which is the dominant typology reported by all other methods. 
        \begin{table} [htbp]
        \centering
        \centering
         
                \caption {\textcolor{black}{Construal relative sizes for musical preferences, 1993 GSS.}}
                \vspace{0.4em}
        

                \begin{tabular*} {427pt}{@{\extracolsep{\fill}}lcccc}
                    \hline \\
                    \vspace{0.1cm}\
                                        & (1)           & (2)               & (3)                    \\
                                        & Omni/Univore  & High/Low-brow     & Contempo/Traditional   \\
                    \addlinespace[0.5em]
                    \hline \\		
                    BCA. . . . . .     &  ---          & 53.77\% (1)       & 46.22\% (1)            \\
                    RCA. . . . . .     &  44.45\% (1)  & 30.74\% (1)       & 24.80\% (1)            \\
                    CCA. . . .  . .    &  75.92\% (3)  & ---               & 24.08\% (1)            \\
                    RRCA. . . . .      &  77.03\% (3)  & ---               & 22.97\% (1)            \\
                    \addlinespace[0.5em]
                    \hline 
                    \hline
                \end{tabular*}
            
            \vspace{1.2em}
            \caption*{\footnotesize\textit{Notes:} Relative size refers to the share of respondents classified in each construal. Dependence structures of construal estimates for RCA, CCA, and RRCA are shown in Appendix~\ref{app:reanalyses}, Figure~\ref{fig:construal_music_results}.} 
            \label{table:music_bca_estimates}                         
        \end{table}
    
        Sociodemographic determinants also behave differently as predictors of BCA construal memberships. Education, gender, and residence in a small town are significant negative predictors of membership in the contemporary–traditional construal under BCA. By contrast, these variables are not statistically significant predictors of assignment to the contemporary/traditional construals identified by other methods. Income exhibits the opposite pattern: it is a significant negative predictor under other methods but does not reach statistical significance when predicting membership in BCA’s contemporary–traditional construal (see Table~\ref{ref:apx_determinants_conttrad} in Appendix~\ref{app:reanalyses}).

        Extending the analysis to cultural consumption practices, we find again that using BCA construals as predictors yields distinctive results. We run OLS models predicting attendance at three types of high-brow events (dance, art, and classical music) using construal categories as predictors, controlling for sociodemographic characteristics. Consistent with the results from the partisanship analyses, models using BCA categorizations as regressors exhibit improved predictive performance. Membership in BCA’s high–low-brow construal is a positive and statistically significant predictor of attendance at high-brow events, and it is the strongest such predictor among all construal membership categories across CCMs. The adjusted $R^2$ of the model that uses BCA categories as predictors of event attendance is also the largest ($0.171$; full results shown in Table~\ref{ref:apx_music_go2events_determinants} in Appendix~\ref{app:reanalyses}). 

        \subsection{Additional Analyses} We conclude this section by examining CCM clustering results derived from two more empirical databases to assess whether BCA continues to diverge from other methods in core aggregate descriptive properties—namely, the number of construals identified and their relative population sizes. We analyze the 1988 GSS module on attitudes toward science and religion. This dataset was previously analyzed using construal methods by DiMaggio et al. (\citeyear{dimaggio_etal_2018}). It comprises responses from $1{,}481$ individuals to $15$ questions gauging opinions on science and religion. The second is the “Towards Gender Harmony” dataset (TGHD; \cite{kosakowska2024towards}), a survey designed to measure perceptions of gender roles. We analyze data from $1{,}214$ respondents in the United Kingdom on items assessing how desirable it is for a woman to be associated with $12$ different traits. In Appendix~\ref{app:reanalyses}, Tables~\ref{ref:apx_s&r_constvars} and~\ref{ref:apx_women_constvars} list the set of variables analyzed for each dataset.

         RCA generates eleven construals when applied to the 1988 GSS data. The three largest construals, which are comparable in size, together account for roughly 60\% of the sample. The remaining respondents are divided between one construal containing 16.38\% of the sample and seven additional construals of marginal and roughly equal size (0.12\%) each. Other methods identify fewer construals and distribute respondents more evenly across them. Results for the TGHD differ more sharply between BCA and other methods. For this dataset, BCA identifies four construals with a highly unbalanced population distribution: 68.62\% of respondents belong to the largest construal. Each of the remaining three construals contains fewer than 15\% of the analyzed sample. This distribution differs from the four roughly equal-sized construals identified by RCA, which gather 88.8\% of the sample. The rest of the sample is assigned by RCA to single-member construals. The distribution identified by BCA also differs from the structures produced by RRCA and CCA, which identify three roughly equally sized construals, followed by a fourth small construal in the case of CCA. See Table~\ref{ref:apx_extra_analyses} for complete results.

         \color{black}\subsection{\textcolor{black}{Computational cost}} We recorded the wall-clock time required for each CCM to cluster the datasets analyzed; the results are reported in Table~\ref{tab:time_empirical}. BCA and CCA were consistently the most time-efficient methods. Their running times across datasets were below one second. RCA, by contrast, exhibited substantially longer running times, reaching up to two and a half minutes for the GSS~1993 musical preference dataset. RRCA was the least time-efficient method. Its running time was highly sensitive to the number of questions under analysis. RRCA’s computing times ranged from 43 seconds (GSS~1988) to more than four minutes for ANES~2004. These findings align with our simulation experiments and underscore BCA’s favorable scalability.

    
    
\color{black}

\section{Limitations and Further Work}
\label{sec:limitations}

\subsection{Performance with Larger Numbers of Construals}

    Most empirical studies report at most four construal clusters (see \cite[p.~1842]{sotoudeh_dimaggio_2023} for an overview). Our empirical analysis supports these results: except for what appear to be spurious unit-construals, all methods found four or less construals in the studied datasets. Moreover, in some cases as the 1993 General Social Survey for musical tastes, RRCA reports 4 construals with no significant differences between their dependence structures. Nevertheless, in our simulation analysis, we included cases with $5$ and $6$ construals in Experiment 2 to assess the robustness of CCMs. Table~\ref{ref:table_limitations} presents the results.

    \begin{table}[tbp]
    \centering
    \begin{threeparttable}

    \caption{\textcolor{black}{Performance measures for RCA, CCA, RRCA, and BCA (with Newman’s and Leiden’s partition algorithms) for Experiment~2 with $5$ and $6$ construals.}}
    \label{ref:table_limitations}

    {\color{black}
    \begin{tabular}{lcccc}
        \hline
        \addlinespace[0.4em]
        & (1) & (2) & (3) & (4) \\
        \addlinespace[0.2em]
        & RCA & CCA & RRCA & BCA \\
        \addlinespace[0.2em]
        \hline

        \addlinespace[1.0em]
        \multicolumn{5}{l}{\textit{Experiment~2: General performance ($5$--$6$ construals)}} \\
        \addlinespace[0.6em]

        \multicolumn{5}{l}{\textit{Newman's algorithm}} \\
        \addlinespace[0.6em]

        CPA                & 30.22\% & 31.41\% & 19.23\% & 13.82\% \\
        MAD                & 1.160   & 1.048   & 1.244   & 1.404   \\
        CDIS               & 6.284   & 6.763   & 6.437   & 6.208   \\
        SNMI               & 0.500   & 0.449   & 0.320   & 0.308   \\
        Unit-construals    & 4.17\%  & 21.32\% & 0.05\%  & 0\%     \\

        \addlinespace[0.8em]
        \hline
        \addlinespace[0.6em]

        \multicolumn{5}{l}{\textit{Leiden's algorithm}} \\
        \addlinespace[0.6em]

        CPA                & 19.48\% & 51.09\% & 40.61\% & 41.65\% \\
        MAD                & 1.060   & 0.632   & 0.715   & 0.721   \\
        CDIS               & 5.199   & 5.734   & 5.867   & 6.144   \\
        SNMI               & 0.613   & 0.574   & 0.480   & 0.465   \\
        Unit-construals    & 0.20\%  & 13.68\% & 0\%     & 0\%     \\

        \addlinespace[0.8em]
        \hline
        \hline
    \end{tabular}
    }

    \begin{tablenotes}[flushleft]
        \footnotesize
        \item \textit{Notes:} \textcolor{black}{For CDIS computations, unit-construals were removed prior to calculating it.}
    \end{tablenotes}

    \end{threeparttable}
\end{table}

    Under Newman’s partition algorithm, CPA of RRCA and BCA drops when the data contain $5$ or $6$ construals, whereas the performance of RCA and CCA improves. MAD scores exhibit a similar pattern: performance drops of BCA and RRCA and improvements for RCA and CCA. This pattern is expected, as RCA and CCA tend to overestimate the number of classes (Table~\ref{ref:table_results} shows that RCA exhibits, by far, the highest MAD in both experiments, followed by CCA). By contrast, BCA and RRCA tend to reduce the frequency with which high numbers of construals are detected, which naturally leads to lower accuracy in the $5$--$6$ construals case. As shown in Appendix~\ref{app:extra_results}, the largest decline in precision occurs at $6$ construals rather than at $5$. Indeed, BCA and RRCA rarely partition the data into $6$ construals, doing so in only $2.82\%$ and $5.32\%$ of the $5{,}000$ simulations, respectively. 

Regarding CDIS, the results are somewhat different from those obtained for CPA and MAD. Under Newman’s partition algorithm, CDIS decreases across methods \textcolor{black}{when 5-6 construals are present}. Nonetheless, BCA still achieves the best CDIS.

    Overall, the results for $5$ and $6$ construals depart from those observed in our previous analysis. RCA and CCA, which perform poorly in Experiment~1 and in Experiment~2 with $2-4$ construals, show a substantial improvement in performance. Both methods tend to overestimate the number of classes; consequently, their performance improves as the number of classes present in the simulated data increases. However, no method performs consistently across performance measures: although CCA stands out in CPA and MAD, RCA is best regarding SNMI and CDIS. By contrast, RRCA and BCA display similar behavior as the number of construals grows: their performance deteriorates under Newman’s partition algorithm.

    From a broader perspective, however, these results highlight that data environments with large number of construals are particularly challenging to analyze. Importantly, CDIS values are relatively similar across methods in the 5--6 construal case: the largest CDIS is just 18.2\% larger than the smallest value. For comparison, this difference is 34.9\% in the case of 3--4 construals (Table~\ref{ref:table_results}). According to our DGP, latent positions are mapped to a finite number of observable responses. Therefore, construals whose latent dependence structures are distinct may appear similar in observed responses. This becomes more likely as the number of construals increases. We believe that convergence of CDIS in the 5--6 construal case comes from the benchmark case (that estimated with the true membership assignment) performing worse. This highlights the difficulties arising when a large number of construals is present.

\subsection{\textcolor{black}{Sensitivity to the partitioning algorithm}} 

\textcolor{black}{
Our simulations show that all CCMs are sensitive to partitioning algorithm choices. Moreover, which partitioning algorithm leads to better performance depends on the true number of construals. When there are 2--3 construals in the data, Newman's partitioning algorithm leads to the best performance across all CCMs (see Tables \ref{tab:2construals} and \ref{tab:3construals}). When compared to other CCMs using Leiden's algorithm, BCA is still the best performing method. Nevertheless, it performs notably worse than BCA with Newman's algorithm.
}

\textcolor{black}{
Our results indicate the presence of a four-construal threshold in CCMs’ performance under the Leiden algorithm. For datasets with more than four construals, all CCMs except RCA perform better under the Leiden algorithm than under Newman's one. The performance difference is most pronounced for RRCA: when four construals are present, it is the best-performing CCM when paired with the Leiden algorithm (see Table~\ref{tab:4construals}). However, when $5$--$6$ construals are present in the data, Leiden's partitioning algorithm leads to substantial performance gains across methods, as shown in Table~\ref{ref:table_limitations}. These gains are most pronounced for BCA and CCA. When paired with the Leiden algorithm, both methods catch up with RRCA, and CCA emerges as the best-performing method. RCA also improves, but its performance remains well below that of the other methods (except on the CDIS metric, where it ranks as the best-performing).
}  



    
\textcolor{black}{
Overall, our results indicate that BCA paired with the Newman algorithm is the best-performing method when the data are organized around two and three construals. BCA-Newman's performance remains competitive under $4$ construals, but its relative performance drops for more fragmented construal structures. Our findings further indicate that RRCA–Leiden and CCA–Leiden are the best-performing methods when four and five to six construals characterize the data, respectively; in both cases, BCA–Leiden ranks as the second-best-performing method.} 

\textcolor{black}{In light of these results, we recommend BCA paired with the Newman algorithm in cases where theory or prior empirical evidence suggests that fewer than five construals are likely present. Both the theoretical frameworks and empirical findings of prior construal analyses suggest that this is very often the case (\cite{sotoudeh_dimaggio_2023}, p.~1842). Finally, when the possibility of a larger number of construals is of theoretical or empirical interest, BCA–Leiden is a robust alternative: it ranks as the second-best-performing method under more fragmented structures, it performs well when the number of construals is small, and it is specifically tailored to handle bipolar data usual in construal analysis.
}


\subsection{Directions for Future Work}
\label{sec:future_work}

    The first avenue for future work delineated by our findings is to further examine the performance of CCMs with different partitioning algorithms. Sotoudeh and DiMaggio (\citeyear{sotoudeh_dimaggio_2023}) reported performance growths for RCA when the Louvain's algorithm was employed, and, consequently, adopted it for RRCA. Our results present mixed evidence. When the data contains $2$ to $4$ construals, BCA with Newman's algorithm outperforms BCA with Leiden's one. However, for all CCMs, the use of Leiden's algorithm surpasses Newman's in most performance measures when $5$ or $6$ construals are present in the data. This behavior warrants further investigation.

    More importantly, we believe that BCA's performance with a larger number of construals could be improved by refining how the adjacency matrix is computed. Note that BCA maps all possible responses to a given question onto three categories: \texttt{positive}, \texttt{neutral}, and \texttt{negative}. As the number of construals increases, distinctions may originate more from differences in opinion intensity than from shifts in opinion semispace. However, BCA’s response categorization ignores these intensity differences, making them irrelevant.

    This issue may be addressed by developing BCA as a familiy of methods, rather than as a single CCM applied uniformly across all cases, as existing CCMs are. Incorporating an intensity measure could allow BCA to capture finer differences in cases with many construals. One possible generalization of the polarity function is:
    \begin{equation*}
        \tilde{P}(u,v) = \frac{2}{Q(Q-1)}\sum_{k=1}^{Q-1}\sum_{l=k+1}^Q\pi(u_{lk},v_{lk})\mu(u_{lk},v_{lk}),
    \end{equation*}
    where $\mu(u_{lk},v_{lk})$ represents the intensity term. Since there is no canonical way to measure differences in opinion intensities, a family of methods, each corresponding to a different choice of intensity $\mu$, could be preferable. Selecting different intensity functions $\mu$ based on specific characteristics of the dataset would provide greater flexibility, allowing the method to adapt to different applications.

    Additionally, refining BCA could allow it to handle continuous cases, broadening the scope of CCM research beyond survey data. Once the domain of an item admits a conventional partition into opposite subsets, BCA may be applied. An example of a continuous data structure that admit such oriented decomposition include greyscales and color representations such as the HSL (Hue, Saturation, Lightness) color model (\cite{W3CColor3,JobloveGreenberg1978,Bergstedt1987}).
    Moreover, an extension of BCA could allow items to have multidimensional response spaces, provided that each dimension preserves a bipolar structure. In this setting, polarity would be aggregated across dimensions by averaging coordinate-wise comparisons, yielding a multidimensional measure that retains BCA’s core invariance properties (see Appendix \ref{app:polarity}).
 
    Finally, it remains unclear whether BCA can be applied to data that do not exhibit a bipolar structure. BCA is designed for bipolar response formats, where each item’s answer space admits a decomposition into two semispaces labeled “positive” and “negative”, and where adjacency is naturally tied to how respondents’ answers move between these semispaces. For non-bipolar formats such as nominal or unordered multiple-choice items, there is typically no canonical notion of opposition or orientation. Thus, extending polarity would require additional modeling assumptions and should be addressed in each particular case.


\section{Conclusions}
\label{sec:conclusion}
    
    The development of CCMs has significantly advanced research on affinity structures. The introduction of RCA in 2011 gave rise to research agendas focused on identifying and analyzing the sociological correlates of different ways of interpreting the world, from religion and science to politics. Subsequent methodological innovations, including CCA and RRCA, have expanded the applicability of CCMs and broadened their contributions to sociological research. Continuing this trajectory, this study introduces \textit{Bipolar Class Analysis} (BCA), a novel method for detecting shared attitudinal patterns among survey respondents. BCA is designed to address the limitations of existing CCMs in opinion data analysis, clustering respondents based on how their opinions on social issues shift between rejection and support zones.

    By analyzing simulated survey responses where individuals map latent opinions onto observable answers, we found that BCA outperforms RCA, CCA, and RRCA in accurately estimating construals in empirically relevant scenarios. BCA more reliably identifies the number of construals and their dependence structures. Moreover,  we show that applying BCA to real-world datasets produces empirical results that differ from those of existing CCMs. Overall, our results indicate that grouping respondents based on how their responses shift across positive and negative opinion semispaces may provide a more effective adjacency measure than other approaches that rely on variable distance terms. Nevertheless, we do identify limitations of BCA, which open avenues for further research into refining this method.

    \textcolor{black}{As a practical guideline, and in light of our simulation results, we contend that CCMs that are not specifically tailored to handle bipolar data may lead to misleading results, as they perform poorly in Experiment~1, the simplest analytical setting we investigate. In this simple setting, BCA ranks as the best-performing CCM. In more complex settings, BCA paired with Newman’s partition algorithm remains the best-performing method when the number of construals ranges from two to four, the range most frequently estimated in empirical studies. Against this backdrop, BCA-Newman appears to be the most appropriate CCM to consider as a first approach when analyzing a real dataset.}
    
    \textcolor{black}{When the number of construals is higher, Newman’s partition algorithm does not appear to be the most appropriate choice for any CCM. In this setting, CCA-Leiden performs best, followed closely by BCA-Leiden. Although CCA achieves a roughly 10\% advantage in Construal Partition Accuracy, BCA produces no unit-construals, which may be substantively desirable.  Opinion intensity becomes more salient in clustering as construal fragmentation increases. We believe that when a large number of construals is present, this is the primary factor underlying performance differentials between CCA, which, unlike BCA, incorporates intensity similarity into its similarity scores. Prior to conducting analyses we therefore recommend carefully assessing whether a high number of construals is theoretically or empirically plausible given extant findings in the relevant literature.}

   Summing up, we believe that Bipolar Class Analysis (BCA) represents a significant step forward in fully leveraging the analytical potential of construal clustering techniques for empirical research. Its computational procedures are specifically designed to align with the cognitive and structural properties of bipolar opinion survey items. The clusters it produces are more accurate than those generated by existing CCMs. However, an important challenge remains: developing appropriate adjacency measures that account for opinion intensity. Further investigation into the performance of these measures, as well as the development of alternative approaches, presents a promising direction for future research.

\section{Ethical approval}

This article does not contain any studies with human participants performed by any of the authors. 

\section{Informed consent}

This article does not contain any studies with human participants performed by any of the authors.

\printbibliography

@inproceedings{amelio_2015,
  title={Is normalized mutual information a fair measure for comparing community detection methods?},
  author={Amelio, Alessia and Pizzuti, Clara},
  booktitle={Proceedings of the 2015 IEEE/ACM international conference on advances in social networks analysis and mining 2015},
  pages={1584--1585},
  year={2015},
  doi = {10.1145/2808797.2809344}
}

@article{baldasarri_goldberg_2014,
 ISSN = {00029602, 15375390},
 author = {Delia Baldassarri and Amir Goldberg},
 journal = {American Journal of Sociology},
 number = {1},
 pages = {45--95},
 publisher = {The University of Chicago Press},
 title = {Neither Ideologues nor Agnostics: Alternative Voters’ Belief System in an Age of Partisan Politics},
 volume = {120},
 year = {2014},
 doi = {10.1086/676042}
}

@article{batyrshin_etal_2017,
 ISSN = {00029602, 15375390},
 author = {Ildar Batyrshin and Fernando Monroy-Tenorio and Alexander Gelbukh and Luis Alfonso Villa-Vargas and Valery Solovyev and Nailya Kubysheva},
 journal = {Acta Politechnica Hungarica},
 volume = {14},
 number = {3},
 pages =  {33-57},
 title = {Bipolar Rating Scales},
 year = {2017},
 doi = {10.12700/APH.14.3.2017.3.3}
}

@article{blondel_2011,
doi = {10.1088/1742-5468/2008/10/P10008},
year = {2011},
volume = {2008},
number = {10},
pages = {P10008},
author = {Vincent D Blondel and Jean-Loup Guillaume and Renaud Lambiotte and Etienne Lefebvre},
title = {Fast unfolding of communities in large networks},
journal = {Journal of Statistical Mechanics: Theory and Experiment},
}

@article {boutyline_2017,
author = {Andrei Boutyline },
title = {Improving the Measurement of Shared Cultural Schemas with Correlational Class Analysis: Theory and Method},
journal = {Sociological Science},
volume = {4},
number = {15},
issn = {2330-6696},
url = {http://dx.doi.org/10.15195/v4.a15},
doi = {10.15195/v4.a15},
pages = {353--393},
year = {2017},
}

@article{brensinger_sotoudeh_2022,
  title={Party, race, and neutrality: investigating the interdependence of attitudes toward social groups},
  author={Brensinger, Jordan and Sotoudeh, Ramina},
  journal={American Sociological Review},
  volume={87},
  number={6},
  pages={1049--1093},
  year={2022},
  publisher={SAGE Publications Sage CA: Los Angeles, CA},
  doi = {10.1177/00031224221135797}
}

@book{cohen_2022,
 title     = {Practical Linear Algebra for Data Science: From Core Concepts to Applications Using Python},
  author    = {Cohen, Mike X.},
  year      = {2022},
  publisher = {O'Reilly Media},
  address   = {Sebastopol, CA},
  isbn      = {978-1-098-12061-0},
  url       = {https://www.oreilly.com/library/view/practical-linear-algebra/9781098120603/}
}

@article{daenekindt_2017, 
    title={On the structure of dispositions. Transponsability of and oppositions between aesthetic dispositions}, 
    journal={Poetics},
    volume={62}, 
    author={Daenekindt, Stijn}, 
    year={2017}, 
    pages={43–52},
    doi = {10.1016/j.poetic.2017.01.004}
}

@article{dimaggio_goldberg_2018, 
title={Searching for Homo Economicus: Variation in Americans’ Construals of and Attitudes toward Markets}, 
volume={59},
DOI={10.1017/S0003975617000558}, 
number={2}, journal={European Journal of Sociology}, 
author={DiMaggio, Paul and Goldberg, Amir}, 
year={2018}, 
pages={151–189}}

@article{dimaggio_etal_2018,
title = {Culture out of attitudes: Relationality, population heterogeneity and attitudes toward science and religion in the U.S.},
journal = {Poetics},
volume = {68},
pages = {31-51},
year = {2018},
issn = {0304-422X},
doi = {https://doi.org/10.1016/j.poetic.2017.11.001},
author = {Paul DiMaggio and Ramina Sotoudeh and Amir Goldberg and Hana Shepherd},
}

@incollection{fisher_1997,
  title={Copulas},
  author={Fisher, N.I.},
  booktitle = {Encyclopedia of Statistical Sciences, Volume 1},
  editor={Kotz, Samuel and Balakrishnan, Narayanaswamy and Read, Campbell B and Vidakovic, Brani},
  year={1997},
  publisher={John Wiley \& Sons}
}

@article{goldberg_2011,
  title={Mapping shared understandings using relational class analysis: The case of the cultural omnivore reexamined},
  author={Goldberg, Amir},
  journal={American Journal of Sociology},
  volume={116},
  number={5},
  pages={1397--1436},
  year={2011},
  publisher={University of Chicago Press Chicago, IL},
  doi = {10.1086/657976}
}

@book{greene_2010,
    place={Cambridge}, 
    title={Modeling Ordered Choices: A Primer}, 
    publisher={Cambridge University Press}, 
    author={Greene, William H. and Hensher, David A.}, 
    year={2010},
    isbn      = {978-0521766555},
    doi       = {10.1017/CBO9780511845062}
}

@article{krosnick_1999,
  author  = {Jon A. Krosnick},
  title   = {Survey Research},
  journal = {Annual Review of Psychology},
  volume  = {50},
  number  = {1},
  pages   = {537--567},
  year    = {1999},
  publisher = {Annual Reviews},
  doi     = {10.1146/annurev.psych.50.1.537},
}

@article{malhotra_krosnick_2007,
    title={The Effect of Survey Mode and Sampling on Inferences about Political Attitudes and Behavior: Comparing the 2000 and 2004 ANES to Internet Surveys with Nonprobability Samples}, 
    volume={15}, 
    DOI={10.1093/pan/mpm003}, 
    number={3}, 
    journal={Political Analysis}, 
    author={Malhotra, Neil and Krosnick, Jon A.},
    year={2007},
    pages={286--323}}

@article{lewandowski_2009,
title = {Generating random correlation matrices based on vines and extended onion method},
journal = {Journal of Multivariate Analysis},
volume = {100},
number = {9},
pages = {1989--2001},
year = {2009},
issn = {0047-259X},
doi = {https://doi.org/10.1016/j.jmva.2009.04.008},
author = {Daniel Lewandowski and Dorota Kurowicka and Harry Joe}
}

@article{mahmoudi_2024,
author={Mahmoudi, Amin
and Jemielniak, Dariusz},
title={Proof of biased behavior of Normalized Mutual Information},
journal={Scientific Reports},
year={2024},
month={04},
day={19},
volume={14},
number={1},
pages={9021},
issn={2045-2322},
doi={10.1038/s41598-024-59073-9},
url={https://doi.org/10.1038/s41598-024-59073-9}
}

@article{malhotra_etal_2009,
 author = {Malhotra, Neil and Krosnick, Jon A. and Thomas, Randall K.},
    title = {Optimal Design of Branching Questions to Measure Bipolar Constructs},
    journal = {Public Opinion Quarterly},
    volume = {73},
    number = {2},
    pages = {304--324},
    year = {2009},
    month = {05},
    issn = {0033-362X},
    doi = {10.1093/poq/nfp023},
   }

@book{nelsen_2006,
  author    = {Nelsen, Roger B.},
  title     = {An Introduction to Copulas},
  publisher = {Springer},
  year      = {2006},
  series    = {Springer Series in Statistics},
  edition   = {2nd},
  address   = {New York},
  isbn      = {978-0387286594},
  doi       = {10.1007/0-387-28678-0}
}

@article{newman_2004,
author       = {Newman, Mark E.\ J.},
  title        = {Analysis of weighted networks},
  journal      = {Physical Review E},
  volume       = {70},
  number       = {5},
  pages        = {056131},
  year         = {2004},
  publisher    = {American Physical Society},
  doi          = {10.1103/PhysRevE.70.056131}
}

@article{newman_2006,
  author       = {Newman, Mark E.\ J. },
  title        = {Finding community structure in networks using the eigenvectors of matrices},
  journal      = {Physical Review E},
  volume       = {74},
  number       = {3},
  pages        = {036104},
  year         = {2006},
  publisher    = {American Physical Society},
  doi          = {10.1103/PhysRevE.74.036104}
}

@Inbook{ostrom_1987,
author="Ostrom, Thomas M.",
editor="Hippler, Hans-J.
and Schwarz, Norbert
and Sudman, Seymour",
title="Bipolar Survey Items: An Information Processing Perspective",
bookTitle="Social Information Processing and Survey Methodology",
year="1987",
publisher="Springer New York",
address="New York, NY",
pages="71--85",
isbn="978-1-4612-4798-2",
doi="10.1007/978-1-4612-4798-2_4",
}

@incollection{ostrom_etal_1992,
author="Ostrom, Thomas M.
and Betz, Andrew L.
and Skowronski, John J.",
editor="Schwarz, Norbert
and Sudman, Seymour",
title="Cognitive Representation of Bipolar Survey Items",
bookTitle="Context Effects in Social and Psychological Research",
year="1992",
publisher="Springer New York",
address="New York, NY",
pages="297--311",
isbn="978-1-4612-2848-6",
doi="10.1007/978-1-4612-2848-6_20",
}

@book{schwarz_sudman_1996,
author    = {Norbert Schwarz and Seymour Sudman},
  title     = {Answering Questions: Methodology for Determining Cognitive and Communicative Processes in Survey Research},
  year      = {1996},
  publisher = {Jossey-Bass Publishers},
  address   = {San Francisco},
  isbn      = {978-0787901455},
}

@article{sotoudeh_dimaggio_2023,
author = {Ramina Sotoudeh and Paul DiMaggio},
title ={Coping With Plenitude: A Computational Approach to Selecting the Right Algorithm},
journal = {Sociological Methods \& Research},
volume = {52},
number = {4},
pages = {1838--1882},
year = {2023},
doi = {10.1177/00491241211031273},
}

@article{robert1995simulation,
author       = {Robert, Christian P.},
  title        = {Simulation of truncated normal variables},
  journal      = {Statistics and Computing},
  volume       = {5},
  number       = {2},
  pages        = {121--125},
  year         = {1995},
  publisher    = {Springer},
  doi          = {10.1007/BF00143942}
}

@article{NYtimes_republicans,
 author  = {Cohn, Nate},
 date    = {2023-08-17},
 title   = {The 6 Kinds of Republican Voters},
 journal = {New York Times},
 url     = {hhttps://www.nytimes.com/interactive/2023/08/17/upshot/six-kinds-of-republican-voters.html},
 urldate = {2023-08-17}
}

@inproceedings{vinh2009information,
  title={Information theoretic measures for clusterings comparison: is a correction for chance necessary?},
  author={Vinh, Nguyen Xuan and Epps, Julien and Bailey, James},
  booktitle={Proceedings of the 26th annual international conference on machine learning},
  pages={1073--1080},
  year={2009}
}

@misc{vinh2010information,
  title={Information theoretic measures for clusterings comparison: Variants, Properties, Normalization and Correction for Chance. 2Journal of Machine Learning Research 11 (2010), 2837--2854},
  author={Vinh, Nguyen Xuan and Epps, Julien and Bailey, James},
  year={2010}
}

@inproceedings{romano2014standardized,
  title={Standardized mutual information for clustering comparisons: one step further in adjustment for chance},
  author={Romano, Simone and Bailey, James and Nguyen, Vinh and Verspoor, Karin},
  booktitle={International conference on machine learning},
  pages={1143--1151},
  year={2014},
  organization={PMLR}
}

@book{mccutcheon1987latent,
  title={Latent class analysis},
  author={McCutcheon, Allan L},
  volume={64},
  year={1987},
  publisher={Sage}
}

@article{hunzaker_2019,
author = {M.B. Fallin Hunzaker and Lauren Valentino},
title ={Mapping Cultural Schemas: From Theory to Method},
journal = {American Sociological Review},
volume = {84},
number = {5},
pages = {950-981},
year = {2019},
doi = {10.1177/0003122419875638},
URL = {https://doi.org/10.1177/0003122419875638},
eprint = {https://doi.org/10.1177/0003122419875638}
}

@article{bertero2024inequality,
  title={Inequality Belief Systems: What They Look Like, How to Study Them, and Why They Matter},
  author={Bertero, Arturo and Franetovic, Gonzalo and Mijs, Jonathan JB},
  journal={Social Indicators Research},
  volume={174},
  number={2},
  pages={445--472},
  year={2024},
  publisher={Springer}
}

@article{lindner2024stances,
  title={What do stances on immigrants' welfare entitlement mean? Evidence from a correlational class analysis},
  author={Lindner, Thijs and Daenekindt, Stijn and de Koster, Willem and van der Waal, Jeroen},
  journal={The British Journal of Sociology},
  volume={75},
  number={3},
  pages={271--289},
  year={2024},
  publisher={Wiley Online Library}
}

@article{batzke2025cognitive,
  title={From cognitive coherence to political polarization: A data-driven agent-based model of belief change},
  author={Batzke, Marlene CL and Steiglechner, Peter and Lorenz, Jan and Edmonds, Bruce and Kalvas, Franti{\v{s}}ek},
  journal={Political Psychology},
  year={2025},
  publisher={Wiley Online Library}
}

@article{van2022support,
  title={Support for European Union membership comes in various guises: Evidence from a correlational class analysis of novel dutch survey data},
  author={Van den Hoogen, Elske and Daenekindt, Stijn and de Koster, Willem and van der Waal, Jeroen},
  journal={European Union Politics},
  volume={23},
  number={3},
  pages={489--508},
  year={2022},
  publisher={SAGE Publications Sage UK: London, England}
}

@article{ganz2025subcoalition,
  title={Subcoalition Cluster Analysis: A New Method for Modeling Conflict in Organizations},
  author={Ganz, Scott C and Schiff, Daniel S},
  journal={Management Science},
  year={2025},
  publisher={INFORMS}
}

@article{bayan2024modularity_review,
  title = {Bayan algorithm: Detecting communities in networks through exact and approximate optimization of modularity},
  author = {Aref, Samin and Mostajabdaveh, Mahdi and Chheda, Hriday},
  journal = {Phys. Rev. E},
  volume = {110},
  issue = {4},
  pages = {044315},
  numpages = {23},
  year = {2024},
  month = {Oct},
  publisher = {American Physical Society},
  doi = {10.1103/PhysRevE.110.044315},
  url = {https://link.aps.org/doi/10.1103/PhysRevE.110.044315}
}

@article{traag2019louvain,
  title={From Louvain to Leiden: guaranteeing well-connected communities},
  author={Traag, Vincent A and Waltman, Ludo and Van Eck, Nees Jan},
  journal={Scientific reports},
  volume={9},
  number={1},
  pages={1--12},
  year={2019},
  publisher={Nature Publishing Group}
}

@article{gagolewski2025normalised,
  title={Normalised clustering accuracy: An asymmetric external cluster validity measure},
  author={Gagolewski, Marek},
  journal={Journal of Classification},
  volume={42},
  number={1},
  pages={2--30},
  year={2025},
  publisher={Springer}
}

@article{kosakowska2024towards,
  title={Towards Gender Harmony Dataset: Gender Beliefs and Gender Stereotypes in 62 Countries},
  author={Kosakowska-Berezecka, Natasza and Besta, Tomasz and Jurek, Pawe{\l} and Olech, Micha{\l} and Sobiecki, Jurand and Bosson, Jennifer and Vandello, Joseph A and Best, Deborah and Zawisza, Magdalena and Safdar, Saba and others},
  journal={Scientific Data},
  volume={11},
  number={1},
  pages={392},
  year={2024},
  publisher={Nature Publishing Group UK London}
}

@article{revilla2014choosing,
  title={Choosing the number of categories in agree--disagree scales},
  author={Revilla, Melanie A and Saris, Willem E and Krosnick, Jon A},
  journal={Sociological methods \& research},
  volume={43},
  number={1},
  pages={73--97},
  year={2014},
  publisher={Sage Publications Sage CA: Los Angeles, CA}
}

@patent{Bergstedt1987,
  author  = {Bergstedt, Gar A.},
  title   = {Apparatus and method for modifying displayed color images},
  number  = {US4694286A},
  year    = {1987},
  month   = sep,
  day     = {15},
  note    = {Filed 1983-04-08; assignee: Tektronix, Inc.}
}

@misc{W3CColor3,
  author       = {{W3C}},
  title        = {{CSS Color Module Level 3}},
  howpublished = {W3C Recommendation},
  year         = {2011},
  month        = jun,
  day          = {7},
  url          = {https://www.w3.org/TR/2011/REC-css3-color-20110607/},
  note         = {Editors: Tantek \c{C}elik, Chris Lilley, L. David Baron. Additional Authors: Steven Pemberton, Brad Pettit.}
}

@inproceedings{JobloveGreenberg1978,
  author    = {Joblove, George H. and Greenberg, Donald},
  title     = {Color spaces for computer graphics},
  booktitle = {Proceedings of the 5th Annual Conference on Computer Graphics and Interactive Techniques (SIGGRAPH '78)},
  series    = {SIGGRAPH '78},
  pages     = {20--25},
  year      = {1978},
  publisher = {ACM},
  address   = {New York, NY, USA},
  doi       = {10.1145/965139.807362}
}

@article{taylor2020concept,
  title={Concept class analysis: A method for identifying cultural schemas in texts},
  author={Taylor, Marshall A and Stoltz, Dustin S},
  journal={Sociological Science},
  volume={7},
  pages={544--569},
  year={2020}
}

@article{bonald2018hierarchical,
  title={Hierarchical graph clustering using node pair sampling},
  author={Bonald, Thomas and Charpentier, Bertrand and Galland, Alexis and Hollocou, Alexandre},
  journal={arXiv preprint arXiv:1806.01664},
  year={2018}
}

\cleardoublepage

\appendix

\thispagestyle{empty}

\begin{center}
    \huge Supplementary material 
\end{center}

\cleardoublepage

\pagestyle{plain}

\setcounter{page}{1}

\renewcommand{\thefigure}{\thesection.\arabic{figure}}
\renewcommand{\thetable}{\thesection.\arabic{table}}

\counterwithin{figure}{section}
\counterwithin{table}{section}

\setcounter{footnote}{0}

\numberwithin{equation}{section}

\section{Formal Definition of Polarity}
\label{app:polarity}

    In this paper, we introduce Bipolar Class Analysis (BCA), a new Construal Clustering Method (CCM) specifically designed for bipolar survey items. Like other CCMs, BCA constructs an adjacency matrix out of the answers of pairs of respondents, but it differs in how this matrix is computed.

    Let $u =(u_1,\dots,u_Q)$ and $v=(v_1,\dots,v_Q)$ denote the responses of two survey respondents $u$ and $v$, respectively. Define $u_{kl} = (u_k,u_l)$ and $v_{kl}=(v_k,v_l)$ as the paired responses to questions $k$ and $l$ for each respondent. The \textit{polarity function} $P$ is given by 
    \[
    P(u,v) = \frac{2}{Q(Q-1)}\sum_{k=1}^{Q-1}\sum_{l=k+1}^Q\pi(u_{kl}, v_{kl}),
    \]
    where $\pi$ is the \textit{pairwise polarity function}. In this appendix, we formally define $\pi$.

    The pairwise polarity function measures whether respondents' observed answers converge or diverge in terms of how they switch between opinion semispaces---that is, how they their responses move relative to the neutrality element. There are four possible types of movement between opinion semispaces, which we formalize through the following \emph{movement functions}:

            \begin{alignat*}{2}
            M_{++}(u_{kl}) &= 
            \begin{cases} 
            1, & \text{if } u_k \in \mathcal{P}_k, \ u_l \in \mathcal{P}_l, \\ 
            0, & \text{otherwise};
            \end{cases} & \quad
            M_{--}(u_{kl}) &= 
            \begin{cases} 
            1, & \text{if } u_k \in \mathcal{N}_k, \ u_l \in \mathcal{N}_l, \\ 
            0, & \text{otherwise};
            \end{cases} \\
            M_{-+}(u_{kl}) &= 
            \begin{cases} 
            1, & \text{if } u_k \in \mathcal{N}_k^\circ, \ u_l \in \mathcal{P}_l^\circ, \\ 
            0, & \text{otherwise};
            \end{cases} & \quad
            M_{+-}(u_{kl}) &= 
            \begin{cases} 
            1, & \text{if } u_k \in \mathcal{P}_k^\circ, \ u_l \in \mathcal{N}_l^\circ, \\ 
            0, & \text{otherwise.}
            \end{cases}
            \end{alignat*}

        Here, $\mathcal{N}^\circ$ consists of all elements in $\mathcal{N}$ except the neutrality element. More formally, $\mathcal{N}^\circ=\mathcal{N}\cap\mathcal{P}^c$, where $\mathcal{P}^c$ denotes the complement of $\mathcal{P}$. Similarly,  $\mathcal{P}^\circ=\mathcal{P}\cap \mathcal{N}^c$. For any answer pair  $u_{kl}$, at least one of the functions listed above must take the value $1$.

    Next, we define the \emph{joint movement function} $f_{A,B}$ between pairs of observed responses.  This function assigns a value of $0$ or $1$ to each pair $(u_{kl},v_{kl})$  and is given by
        \begin{equation}{\label{productmovementfunction}}
        \begin{split}
                f_{A,B}\colon(\mathcal{A}_{k}\times\mathcal{A}_{l})\times(\mathcal{A}_{k}\times\mathcal{A}_{l})&\to\{0,1\}\\
            (u_{kl},v_{kl})&\mapsto M_A(u_{kl})M_{B}(v_{kl}),
        \end{split}
        \end{equation}where $A,B\in\{++,- -,-+,+-\}$.

        Having defined the movement and joint movement functions, we  now define $\pi$, distinguishing between two cases. For ease of notation, we let $n_{kl} = (n_k,n_l)$, where $n_i$ denotes the neutrality element in $\mathcal{A}_i$, the space of possible responses to question $i$.

        If $u_{kl},v_{kl}\neq n_{kl}$, we define
            \begin{equation}
            {\label{orientation}}
                \pi(u_{kl},v_{kl})=
               \begin{cases}
                     \ \ 1, & \text{ if }f_{A,A}(u_{kl},v_{kl})=1 \text{ for at least one }A;\\
                     -1, & \text{ if }f_{A,A^*}(u_{kl},v_{kl})=1\text{ for at least one }A;\\
                    \ \ 0, & \text{otherwise.}
                \end{cases}
            \end{equation}Here, the map $*\colon A\mapsto A^*$ is defined as 
            \begin{align}
            (++)^*=- -, \ (- -)^*=++, \ (-+)^*=+-, \text{ and }(+-)^*=-+;
            \end{align} 
            If either $u_{kl}= n_{kl}$ or $v_{kl} = n_{kl}$ (for $u_{kl}=v_{kl} = n_{kl}$, we set $\pi(n_{kl},n_{kl})=1$), we define
            \begin{equation}
            {\label{orientationnuet1}}
                \pi(n_{kl}, v_{kl})= \left\{ \begin{array}{l}
                1, \ \text{if } M_{++}(v_{kl})=1 \text{ or } M_{- -}(v_{kl})=1;  \\
                0, \ \text{otherwise.}
                \end{array} \right.
            \end{equation}
        and
            \begin{equation}
            {\label{orientationneut2}}
                \pi(u_{kl}, n_{kl})= \left\{ \begin{array}{l}
                     1, \ \text{if } M_{++}(u_{kl})=1 \text{ or } M_{- -}(u_{kl})=1;  \\
                     0, \ \text{otherwise.}
                \end{array} \right.
            \end{equation}

    \subsection{Multidimensional approach}\label{sec:multidimensionality}

            A natural extension of BCA covers items whose response options are multi-dimensional but still bipolar in each dimension. Concretely, fix $d\geq 1$ and assume that the answer space for each question $q$ can be written as a Cartesian product
\[
A_q=\prod_{j=1}^d A^{j}_q,
\]
where each factor $A^{j}_q$ carries the same type of bipolar structure as before (a decomposition into positive and negative opinion semispaces, together with a neutrality option). 

Following our conventions, an answer is then a $Q$-tuple $(a_1,\ldots,a_Q) \in A_1\times\cdots \times A_Q$ with $a_q = (a^1_q, \ldots,a_q^d)\in A_q$ and $a_q^j\in A^j_q$ for $q=1,\ldots,Q$ and $j=1,\ldots,d$.  
For a pair of items $(k,l)$ and respondents $u,v$, let $u_{kl}= (u_k,u_l)$ and $v_{kl} = (v_k,v_l)$, and write $u_{kl}^{j}=(u_k^j,u_l^j)$ and $v_{kl}^{j}=(v_k^j,v_l^j)$. Define an aggregated polarity function  $\hat\pi$ by averaging coordinate-wise,
\[
\hat{\pi}(u_{kl},v_{kl})=\frac{1}{d}\sum_{j=1}^d \pi\!\left(u_{kl}^{j},v_{kl}^{j}\right),
\]
where $\pi$ is the pairwise polarity function. The resulting \textit{(multidimensional) polarity function} $\hat P(u,v)$ is obtained from~(3.1) by replacing $\pi$ with $\hat\pi$, yielding
\[
\hat{P}(u,v)=\frac{2}{Q(Q-1)}\sum_{k=1}^{Q-1}\sum_{l=k+1}^{Q}\hat{\pi}\left(u_{kl},v_{kl}\right).
\]
Note that $\hat{P}$ takes values in $[-1,1]$ and measures, across coordinates, how consistently two respondents relate the pairs of questions $(k,l)$ on average. The extension inherits invariance under a global reversal of orientation (swapping positive and negative opinion semispaces in every coordinate) and is invariant under permutation of the factors $A^j_q$ (the same relabelling of coordinate indices $j$ for every item and every respondent). However, for interpretation, it requires that coordinates are meaningfully aligned across items.

\begin{landscape}

\mbox{}%

\vfill

\begin{figure}[h]
    \resizebox{!}{0.16\textheight}{%
        \flowchart
    }
      \caption{BCA's algorithm flowchart.}
      \label{fig:flowchart}
      \end{figure}

      \vfill

      \mbox{}

\end{landscape}

\cleardoublepage

\section{Simulation Model}
\label{app:simulation}

\subsection{Data Generating Process (DGP)}

Our synthetic datasets consist of matrices $(x_{iq})_{i=1,q=1}^{N,Q}$, where $N$ is the number of respondents and $Q\geq 1$ is the number of questions in the survey. Each row $(x_{i1}, \dots, x_{iQ})$ is a realization of the random vector $X_i = (X_{i1}, \dots, X_{iQ})$. Here, we present the Data Generating Process (DGP) for $X_i$. 

We use the following notation throughout the appendix: $H_q$ is the number of options for question $q$ and $\mathcal{A}_q=\{a_{q1},\dots, a_{qH_q}\}$ is the answer space for question $q$. We assume that the elements in the answer space are totally ordered: $a_{q1} \leq_q a_{q2} \leq_q \cdots \leq_q a_{qH_q}$. We note that the support of $X_i$ is $\mathcal{A}_1 \times \cdots \times \mathcal{A}_Q$. Also, there are $K$ construals in the data. Their relative populations are denoted by $\pi_k = N_k/N$, where $N_k$ is the number of respondents adhering to construal $k$.

\subsubsection{Drawing latent positions}

First, we randomly choose the construal to which respondent $i$ belongs. The probability of belonging to construal $k$ is $\pi_k$. Then, if respondent $i$ belongs to construal $k$, the latent position $X_i^* = (X_{i1}^*, \dots, X_{iQ}^*)$ is drawn from the Gaussian copula $C_k(\cdot)$. The $Q$-dimensional Gaussian copula is supported on $[0, 1]^Q$ and its shape is determined by a correlation matrix $\Sigma_k$:
\begin{equation}
    C_k(x_1^*, \dots x_Q^*)= \Phi_{\Sigma_k}\left(\Phi^{-1}(x_1^*),\dots, \Phi^{-1}(x_Q^*)\right),
\end{equation}
where $\Phi$ is the cumulative distribution function (cdf) of a standard normal variable and $\Phi_\Sigma$ is the cdf of a multivariate normal vector with mean zero and covariance matrix $\Sigma$.

\subsubsection{Drawing the answers} Once the latent positions $X_i^*$ are drawn, the respondent bases her choices on where these lie relative to question-specific thresholds. For question $q$, the $H_q -1$ thresholds $(\tau_{q\eta})_{\eta=1}^{H_q-1}$ partition the $[0, 1]$ interval into $H_q$ subintervals. We have that $X_{iq}=a_{q\eta}$ if and only if $X_{iq}^*$ falls in the $\eta$-th subinterval. Formally,
\begin{equation} \label{eq:latent_choice}
    X_{iq}=a_{q\eta} \text{ if and only if } X_{iq}^* \in (\tau_{q,\eta-1}, \tau_{q\eta}) \text{ for any } \eta = 1, \dots, H_q.
\end{equation}
In the preceding equation, for convenience, we set $\tau_{q0} = 0$ and $\tau_{qH_q}=1$. Also, since $X_{iq}^*$ is continuous uniform, $\operatorname{P}(X_{iq}^* = \tau_{q\eta}) = 0$ for every $\eta$. Hence, it is irrelevant to include or exclude the boundaries of each interval.

\subsection{Parameter generation} The DGP takes as inputs, among others, a collection of correlation matrices $\{\Sigma_1, \dots, \Sigma_K\}$ and a collection of question-specific thresholds $\{(\tau_{1\eta})_{\eta=1}^{H_1-1}, \dots, (\tau_{Q\eta})_{\eta=1}^{H_Q-1} \}$. Here we describe how these parameters are randomly generated. 

\subsubsection{Correlation matrix generation} For each construal $k$, the correlation matrix $\Sigma_k$ must be positive definite. To ensure this, we first generate a $Q(Q-1)/2$ vector of partial correlations. Each element of the vector is drawn from a mixture of beta distributions and then translated to $[-1, 1]$. Precisely, we follow these steps to generate a partial correlation:
\begin{enumerate}
    \item Draw a random integer $c$ from $\{-1, 0, 1 \}$, with the following probabilities: $\operatorname{P}(c = 0) = 0.4$ and $\operatorname{P}(c = -1)  = \operatorname{P}(c = 1) = 0.3$.
    \item Draw from a random beta distribution $B \sim  \operatorname{Beta}(\alpha_c, \beta_c)$, where $(\alpha_{-1}, \beta_{-1}) = (1, 40)$, $(\alpha_{0}, \beta_{0}) = (40, 40)$, and $(\alpha_{1}, \beta_{1}) = (40, 1)$.
    \item Scale and shift the result: the partial correlation is $2B - 1$.
\end{enumerate}
This process ensures that partial correlations tend to bunch around $-1$, $0$, or $1$ ---modeling negative, none, or positive dependence between questions. We then follow the algorithm in \cite{lewandowski_2009} to build $\Sigma_k$ from the vector of partial correlations.

\subsubsection{Threshold generation} Thresholds model question-specific features, for example, whether most people agree with a statement. Section~\ref{sec:bipolar_data} has argued about the importance of the ``first sign, then magnitude" paradigm in survey responses. We base our algorithm on drawing a ``neutral position" $n_q^*$ that splits the $[0, 1]$ interval into two: one region favoring agreement and other favoring disagreement. Here, we provide an algorithm to randomly generate thresholds $(\tau_{q\eta})_{\eta=1}^{H_q-1}$ for a question $q$ with an odd number of answer choices $H_q$. Also, we consider that subinterval lengths are symmetric. The algorithm can be readily extended to other cases.

The first step of the algorithm is to draw the ``neutral position" $n_q^*$. The location of $n_q^*$ in $[0, 1]$ is determined by the skewness parameter $b_q$. This parameter measures the difference between positive and negative answers: $b_q = \operatorname{P}(X_{iq} \in \mathcal{P}_q) - \operatorname{P}(X_{iq} \in \mathcal{N}_q)$. The skewness parameter is drawn from a continuous uniform in $[-0.2, 0.2]$. Then, we set $n_q^* = (1 - b_q)/2$. 

To build the thresholds, we think of them as departing away from the ``neutral position" $n_q^*$. Thus, we draw  $\bar{H}_q =(H_q-1)/2$ departure distances: $(\pi_{q\eta})_{\eta=1}^{\bar{H}_q}$. The first distance $\pi_{q1}$ gives half the length of the subinterval for which the neutral answer is chosen ---the neutral option is the $(\bar{H}_q +1)$-th option. That is, we set
\begin{equation}
    \tau_{q\bar{H}_q} = n_q^*-\pi_{q1} \text{ and } \tau_{q,\bar{H}_q+1}=n_q^*+\pi_{q1}.
\end{equation}
The half-length parameter is drawn as $\pi_{1q} = \ell_{q1} \cdot Z$, where $\ell_{q1} = \min\{n_q^*, 1 - n_q^*\}$ and $Z$ is a truncated normal distribution with parameters $\mu = 1/H_q$, $\mu^-=0$, $\mu^+=1$, and $\sigma = 0.025$ (see \cite{robert1995simulation}). This ensures that the thresholds are in $[0, 1]$.

The next distances $(\pi_{iq\eta})_{\eta=2}^{\bar{H}_q}$ measure the length of the remaining subintervals. We build them recursively for $\eta = 2, \dots, \bar{H}_q$. First, we compute $\ell_{q\eta} = \ell_{q, \eta -1} - \pi_{q, \eta - 1}$. Then, we draw $\pi_{q\eta}$ from a truncated normal distribution with parameters $\mu = \ell_{q1}/\bar{H}_q$, $\mu^-=0$, $\mu^+=\ell_{q\eta}$, and $\sigma = 0.025$. We then set the next two thresholds as
\begin{equation}
    \tau_{q, \bar{H}_q+\eta}= \tau_{q, \bar{H}_q+\eta-1} +\pi_{q\eta} \text{ and }  \tau_{q, \bar{H}_q+1-\eta} = \tau_{q, \bar{H}_q+2-\eta} - \pi_{q\eta}.
\end{equation}

\subsection{Simulation algorithm}

We present the steps to generate a synthetic dataset with the characteristics of Experiment 2 (see Section~\ref{sec:simulations}). This algorithm can be adapted to other simulation procedures. 
\begin{enumerate}
    \item Draw the number of construals $K$ from $\{2, 3, 4, 5, 6\}$, all options being equiprobable.
    \item Draw the number of questions $Q$ from $\{10, 11, \dots, 19, 20\}$, all options being equiprobable.
    \item Generate the $Q \times Q$ dimensional correlation matrices $\{\Sigma_1, \dots, \Sigma_K\}$, as described above.
    \item For each construal $k$:
    \begin{enumerate}
        \item Draw the population $N_k$ from $\{200, 201, \dots, 499, 500\}$, all options being equiprobable.
        \item Draw $N_k$ latent position vectors $(X_i^*)_{i=1}^{N_k}$ from a Gaussian copula with correlation $\Sigma_k$.
    \end{enumerate}
    \item For each question $q$:
    \begin{enumerate}
        \item Draw the number of choices $H_q$ from $\{3, 5, 7\}$, all options being equiprobable.
        \item Draw the skewness parameter $b_q$ from a continuous uniform in $[-0.2, 0.2]$.
        \item Generate $H_q - 1$ thresholds $(\tau_{q\eta})_{\eta=1}^{H_q - 1}$, as described above.
    \end{enumerate}
    \item For each respondent $i$ and question $q$, generate $X_{iq}$ following equation~\eqref{eq:latent_choice}.
\end{enumerate}

\subsection{A note about uniform positions} We have decided to model the latent positions for each question ($X_{iq}^*$) as uniform random variables for three reasons: (i) it is easy to interpret them, (ii) it is easy to simulate them, and (iii) it is observationally equivalent to a DGP with question-specific absolutely continuous marginal distributions $F_q(\cdot)$. Here, we develop the last point.

Consider a DGP where, when respondent $i$ belongs to construal $k$, her latent position vector $X_i^*$ follows an absolutely continuous multivariate distribution $F_k(\cdot)$ with support on $\mathbb{R}^Q$. Denote the corresponding marginal distributions by $\{F_{k1}(\cdot), \dots, F_{kQ}(\cdot)\}$. We assume that the marginal distributions are question-specific and do not depend on the construal: $F_{kq}(\cdot) = F_q(\cdot)$ for every $k=1, \dots, K$ and $q=1,\dots, Q$. Also, consider that there are question-specific thresholds $(\tau_{q\eta})_{\eta=1}^{H_q - 1}$ that partition $\mathbb{R}$ into $H_q$ subintervals. Our point is that we can find an alternative DGP, whose latent positions are uniformly distributed, that will lead to exactly the same probability distribution for the observed answers $X_i$.

By Sklar's Theorem (Theorem~2.10.9 in \cite{nelsen_2006}), there exists a unique copula $C_k(\cdot)$ such that the absolutely continuous cdf can be decomposed as
\begin{equation}
    F_k(x_1^*, \dots, x_Q^*) = C_k(F_{k1}(x_1^*), \dots, F_{kQ}(x_Q^*)) = C_k(F_{1}(x_1^*), \dots, F_{Q}(x_Q^*))
\end{equation}
Therefore, we can consider the uniformly distributed latent positions $\tilde{X}^*_{iq} = F_q(X^*_{iq})$, whose dependence is given by the copula $C_k(\cdot)$ when $i$ belongs to construal $k$. Also, by defining the alternative thresholds $\tilde{\tau}_{q\eta} = F_q(\tau_{q\eta})$, we get the equivalence:
\begin{equation}
    X_{iq}^* \in (\tau_{q,\eta-1}, \tau_{q\eta}) \text{ if and only if } \tilde{X}_{iq}^* \in (\tilde{\tau}_{q,\eta-1}, \tilde{\tau}_{q\eta}) \text{ for any } \eta = 1, \dots, H_q.
\end{equation}
Hence, by equation~\eqref{eq:latent_choice}, both DGPs have the same distribution for the observed answers $X_i$ (they are observationally equivalent).

\cleardoublepage

\section{Correlation Dissimilarity}{\label{app:performance}}


    Our simulations consider that the dependence structure for each construal $k$ is given by copulae with correlation matrix $\Sigma_k$. It is then natural to ask whether each method is able to recover the correlation structure $(\Sigma_k)_{k=1}^K$ from the data. There are two problems to address when answering this question: (i) that the number of construals estimated by the method may differ from the true number of construals and (ii) that the labelling of each construal is arbitrary (i.e., what a certain method labels as ``the first construal'' may differ from the group that was labelled as ``the first construal'' when generating the data).

    We propose a method to measure how well each method estimates the underlying correlation matrices, which overcomes these issues. The inputs to the method are a collection of true and estimated correlation matrices, $(\Sigma_k)_{k=1}^K$ and $(\hat{\Sigma}_k)_{k=1}^{\hat{K}}$, respectively. The number of true ($K$) and estimated ($\hat{K}$) construals may differ. Additionally, the researcher must provide a measure of dissimilarity between two matrices, which we denote by $d$. We base our assessment of the methods on the Frobenius distance
    
        \begin{equation*}
            d(A, B) = \sqrt{\trace{(A-B)(A-B)'}},
        \end{equation*}
    where $A$ and $B$ are matrices and $A'$ denotes the transpose of $A$.\footnote{The choice of the Frobenius distance is not particularly relevant. Indeed, in case the dissimilarity measure comes from a norm, the ranking of methods will not depend on the chosen norm. This comes from all norms being equivalent.} For more information about the Frobenius distance, see (\cite{cohen_2022}).

    As we have discussed, our measure of performance must deal with the fact that there is no automatic way to pair an estimated construal with a true one. That is, for a given estimated construal $\hat{k} \in \{1,\dots, \hat{K}\}$ we do not know the corresponding true construal $k\in \{1,\dots, K\}$. Thus, we cannot directly compute $d(\Sigma_k, \hat{\Sigma}_{\hat{k}})$. The problem is exacerbated when the number of estimated construals differs from the number of true construals. To deal with these inconveniences, we will work with all possible pairings between $\{1,\dots,\hat{K}\}$ and $\{1,\dots,K\}$. Our measure of performance will then be the minimum of matrix dissimilarity across all possible pairings.

    We define a \textit{paring} between $\{1,\dots,\hat{K}\}$ and $\{1,\dots,K\}$ as an injective function 
    $p\colon \{1,\dots,\hat{K}\} \to \{1,\dots,K\}$. This function assigns a unique true construal $k=p(\hat{k})$ to each estimated construal $\hat{k}$. Note, however, that no injective function exists when $\hat{K}>K$, that is, when the method overestimates the number of construals. To deal with this, we propose to augment the true number of construals by introducing $\hat{K}-K$ artificial ones. The correlation structure of these artificial construals is $I_Q$, the $Q$-dimensional identity matrix. That is, issues for these artificial construals are not related, indicating that the construals are not really present. So, when $\hat{K}>K$, we propose to compute the dissimilarity between the collection of estimated correlation matrices $(\hat{\Sigma}_k)_{k=1}^{\hat{K}}$ and the augmented collection of $\hat{K}$ true correlation matrices $(\Sigma^1,\dots, \Sigma^K, I_Q,\dots, I_Q)$.\footnote{An alternative to this approach would be to consider injective functions $p\colon\{1,\dots,K\}\to\{1,\dots,\hat{K}\}$ when $\hat{K}>K$. We discuss why we believe that this other approach could be misleading in Remark~\ref{rema:other_approach}.}

    To introduce the measure of dissimilarity between $(\hat{\Sigma}_{k})_{k=1}^{\hat{K}}$ and $(\Sigma_{k})_{k=1}^K$, consider first that $\hat{K} \leq K$. Denote by $P$ the set of all possible pairings between $\{1,\dots,\hat{K}\}$ and $\{1,\dots, K\}$. Given a dissimilarity measure between pairs of matrices $d$, the dissimilarity between collections, denoted $\mathcal{D}$, is
    
        \begin{equation}{\label{distancematrix}}
            \mathcal{D}\left((\hat{\Sigma}_{k})_{k=1}^{\hat{K}}, (\Sigma_{k})_{k=1}^K\right)=\min_{p\in P}\left\{\max_{\hat{k}=1,\dots,\hat{K}} d\left( \hat{\Sigma}_{\hat{k}}, \Sigma_{p(\hat{k})}\right)\right\}, \text{ when } \hat{K}\leq K.
        \end{equation}

    As we have discussed, when $\hat{K}>K$, we augment the collection of true construals by adding construals whose correlation structure is the identity. That is, we define the collection $(\tilde{\Sigma}_k)_{k=1}^{\hat{K}}$, with $\tilde{\Sigma}_k=\Sigma_k$, for $k\leq K$, and $\tilde{\Sigma}_k=I_Q$, for $k=K+1,\dots, \hat{K}$. In this case, the dissimilarity between the collections is given by
    
        \begin{equation*}
            \mathcal{D}\left((\hat{\Sigma}_{k})_{k=1}^{\hat{K}}, (\Sigma_{k})_{k=1}^K\right)= \mathcal{D}\left((\hat{\Sigma}_{k})_{k=1}^{\hat{K}}, (\tilde{\Sigma}_{k})_{k=1}^{\hat{K}}\right), \text{ when } \hat{K} > K.
        \end{equation*}

    \begin{rema}\label{rema:other_approach}
        An alternative to augmenting the number of true construals when $\hat{K}>K$ is to consider, in that case, pairings as injective functions $p\colon\{1,\dots,K\}\to\{1,\dots,\hat{K}\}$. This leaves, however, some estimated construals unpaired with true construals. Thus, we believe that this other approach unfairly benefits methods that overestimate the number of construals. Indeed, a method can trick this alternative measure by estimating a lot of construals, even if some of those are very dissimilar to the true construals. This is so since the minimum in equation~\eqref{distancematrix} will be achieved when the dissimilar estimated construals are left unpaired. 
    \end{rema}


    \subsection{A Benchmark to asses method performance} One of the problems of the above measure of dissimilarity is that there is no clear notion of whether a method estimated properly the underlying correlation structure. This is exacerbated since the procedure to estimate the correlation structure using the correlation of the numerical values where responses are mapped may be biased. To address this we propose to compare the correlation dissimilarity of each method to a benchmark.

    Let  $(X_{i})_{i=1}^{N}$ be a synthetic dataset generated as described in Appendix~\ref{app:simulation}. For each synthetic observation $i$, we know the construal it belongs to. That is, if there are $K$ true construals in the synthetic data, we observe the true construal membership function $c\colon\{1,\dots, N\} \to \{1,\dots, K\}$, where $c(i)=k$ means that observation $i$ belongs to construal $k$. With this information, we can estimate $\bar{\Sigma}_k$ using only the observations $i$ such that $c(i)=k$. This correlation matrix differs from the true correlation matrix for construal $k$ ($\Sigma_k$) for three reasons: (i) there is some sample variation, (ii) we only observe answers $X_i$ and not the latent positions $X_i^*$, and (iii) we use the numerical value where $X_i$ is mapped to.

    When we estimate the correlation matrices $(\hat{\Sigma}_k)_{k=1}^{\hat{K}}$ using a given method, we also face the above three sources of noise. However, on top of them, each method must estimate the construal membership function. Therefore, we believe that $\mathcal{D}( (\bar{\Sigma}_{k})_{k=1}^K,  (\Sigma_{k})_{k=1}^K ) = \mathcal{D}^b$ provides a good benchmark to measure the performance of the method when estimating the underlying correlation structure.

\begin{figure}[h]
        \centering
        \includegraphics[width=0.65\linewidth]{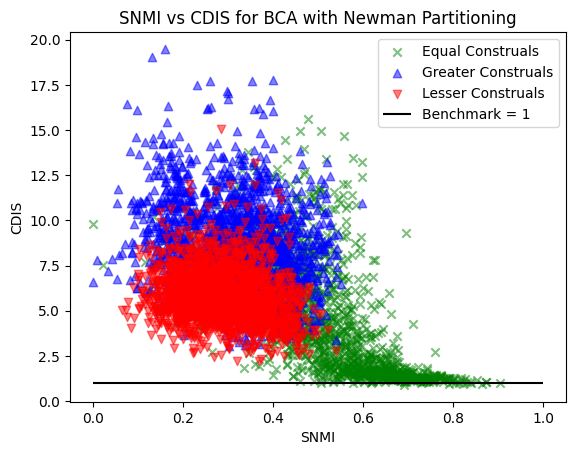}
        \caption{Point cloud showing the relationship between SNMI values and CDIS for BCA using Newman’s partition algorithm in Experiment~2, across all numbers of construals.}
        \label{fig:benchmarkCDIS}
        \caption*{\footnotesize\textit{Notes:} The figure distinguishes between points where the estimated number of construals matches the true number (x-shaped markers in green), exceeds it (upper triangles in blue), or falls below it (lower triangles in red).}
    \end{figure}

    Once we compute $\mathcal{D}^b$, we define the \textit{Correlational Dissimilarity (CDIS)} of a method by
        \begin{equation*}
            \operatorname{CDIS} = \frac{\mathcal{D}\left((\hat{\Sigma}_{k})_{k=1}^{\hat{K}}, (\Sigma_{k})_{k=1}^K\right)}{\mathcal{D}^b},
        \end{equation*}
    where $(\hat{\Sigma}_k)_{k=1}^{\hat{K}}$ are derived using the method's clusters. The magnitude of the above ratio is meaningful. For instance, if the ratio is 2 for a given method, one can say that the ``estimation of the underlying correlation structure with the method is twice as bad as if we knew the true membership of each observation''. Furthermore, normalizing by the benchmark provides the clearest link between CDIS and classical accuracy measures such as NMI, SNMI, or the Rand index: as the validity measure approaches its maximum, CDIS tends to $1$ (see Figure~\ref{fig:benchmarkCDIS}).

\cleardoublepage

\section{Additional Simulation Results}
\label{app:extra_results}
\enlargethispage{5\baselineskip}
\begin{table}[h!]
\centering
\resizebox{0.85\columnwidth}{!}{%
    \begin{tabular}{rcccc}
    & RCA & CCA & RRCA & BCA\\
     \hline \\
     \multicolumn{5}{l}{\textit{Experiment $1$: Exploring limitations}}\\
     \hdashline\\
     \multicolumn{5}{l}{\textit{Newman}}\\
     \addlinespace
      CPA&  0\% & 0\%  & 0\%  & 93.1\% \\
      MAD&  337.810 & 5.992 & 2.169 & 0.069 \\
      CDIS& 10.427 & 14.855 & 11.601 & 2.927  \\
      SNMI&  0 & 0.004  &  0.024 & 0.172  \\
      NMI&0.172&0.083&0.067&0.176\\
      Unit-construals & 100\% & 11.1\% & 0\%   & 0\%  \\
      Compu. Error&36.2\%&0\%&0\%&0\%\\\hline\\

      \multicolumn{5}{l}{\textit{Leiden}}\\
      \addlinespace
      CPA&  0\% & 0\%  & 0\%  & 89.4\% \\
      MAD&  337.475 & 5.992 & 1.789 & 0.106 \\
      CDIS& 11.091 & 14.855 & 10.755 & 3.116  \\
      SNMI&  0 & 0.004  &  0.029 & 0.169  \\
      NMI&0.172&0.083&0.068&0.175\\
      Unit-construals & 100\% & 11.1\% & 0\%   & 0\%  \\
      Compu. Error&36.2\%&0\%&0\%&0\%\\\hline\hline
      \end{tabular}}
\caption{Complete performance table for Experiment $1$.}
\end{table}

\newpage

\begin{table}[h!]
\centering
\resizebox{0.85\columnwidth}{!}{%
    \begin{tabular}{rcccc}
    & RCA & CCA & RRCA & BCA\\
     \hline \\
      \multicolumn{5}{l}{\textit{Experiment $2$: General performance (2 construals)}}\\
      \hdashline\\
      \multicolumn{5}{l}{\textit{Newman}}\\
      \addlinespace
      CPA&  0\% & 8.99\%  & 13.02\%  & 26.76\%\\
      MAD&  4.784 & 1.792 & 1.279 & 0.977\\
      CDIS& 10.951 & 9.441 & 8.992 & 7.598\\
      SNMI&  0.070 & 0.278  &  0.290 & 0.354\\
      NMI&0.492&0.551&0.490&0.519\\
      Unit-construals & 45.52\% & 19.73\% & 0.1\%   & 0\%\\
      NOC\footnote{Number of Occurences of each Construal}&0&90&155&314\\\hline
      \addlinespace
      \multicolumn{5}{l}{\textit{Leiden}}\\
      \addlinespace
      CPA& 0\% & 0.92\%  &  2.69\% & 5.37\%\\
      MAD& 4.847 & 1.974 & 1.495 & 1.339\\                    
      CDIS& 10.871 & 10.137 &  9.863 & 8.778\\
      SNMI& 0.058 & 0.254  &  0.283 & 0.297\\
      NMI&0.498&0.580&0.549&0.544\\
      Unit-construals   & 37.09\% & 16.53\% & 0\%   & 0\%\\
      NOC&0&9&26&54\\\hline\hline
\end{tabular}}
\caption{Complete performance table for Experiment $2$ - Construals $2$.}
\label{tab:2construals}
\end{table}

\newpage

\begin{table}[h!]
\centering
\resizebox{0.85\columnwidth}{!}{%
    \begin{tabular}{rcccc}
    & RCA & CCA & RRCA & BCA\\
     \hline \\
      \multicolumn{5}{l}{\textit{Experiment $2$: General performance (3 construals)}}\\
      \hdashline\\
      \multicolumn{5}{l}{\textit{Newman}}\\
      \addlinespace
      CPA&  0.37\% & 26.23\%  & 38.96\%  & 46.42\%\\
      MAD&  2.608 & 1.375 & 0.775 & 0.624\\
      CDIS& 8.758 & 9.441 & 6.621 & 6.057\\
      SNMI&  0.161 & 0.288  &  0.284 & 0.324\\
      NMI&0.484&0.566&0.549&0.495\\
      Unit-construals & 12.83\% & 21.32\% & 0.09\%   & 0\%\\
      NOC&9&706&1424&1671\\\hline
      \addlinespace
      \multicolumn{5}{l}{\textit{Leiden}}\\
      \addlinespace
      CPA& 0.18\% & 17.83\%  &  26.22\% & 30.85\%\\
      MAD& 2.787 & 1.345 & 0.887 & 0.821\\                    
      CDIS& 9.133 & 7.491 &  7.139 & 6.212\\
      SNMI& 0.391 & 0.469  &  0.471 & 0.399\\
      NMI&0.604&0.638&0.606&0.480\\
      Unit-construals   & 5.37\% & 15.75\% & 0\%   & 0\%\\
      NOC&2&524&837&963\\\hline\hline
\end{tabular}}
\caption{Complete performance table for Experiment $2$ - Construals $3$.}
\label{tab:3construals}
\end{table}

\newpage

\begin{table}[h!]
\centering
\resizebox{0.85\columnwidth}{!}{%
    \begin{tabular}{rcccc}
    & RCA & CCA & RRCA & BCA\\
     \hline \\
      \multicolumn{5}{l}{\textit{Experiment $2$: General performance (4 construals)}}\\
      \hdashline\\
      \multicolumn{5}{l}{\textit{Newman}}\\
      \addlinespace
      CPA&  12.92\% & 34.16\%  & 48.44\%  & 46.77\%\\
      MAD&  1.699 & 1.144 & 0.585 & 0.575\\
      CDIS& 6.798 & 5.543 & 6.323 & 6.054\\
      SNMI&  0.458 & 0.471  &  0.400 & 0.400\\
      NMI&0.660&0.592&0.458&0.457\\
      Unit-construals & 6.98\% & 22.60\% & 0\%   & 0\%\\
      NOC&434&1371&2147&2017\\\hline
      \addlinespace
      \multicolumn{5}{l}{\textit{Leiden}}\\
      \addlinespace
      CPA& 3.43\% & 40.31\%  &  60.10\% & 53.33\%\\
      MAD& 1.770 & 0.908 & 0.435 & 0.509\\                    
      CDIS& 7.342 & 5.866 &  5.640 & 5.815\\
      SNMI& 0.491 & 0.566  &  0.540 & 0.513\\
      NMI&0.732&0.675&0.593&0.577\\
      Unit-construals   & 1.04\% & 14.79\% & 0\%   & 0\%\\
      NOC&154&1389&2136&2010\\\hline\hline
\end{tabular}}
\caption{Complete performance table for Experiment $2$ - Construals $4$.}
\label{tab:4construals}
\end{table}

\newpage

\begin{table}[h!]
\centering
\resizebox{0.85\columnwidth}{!}{%
    \begin{tabular}{rcccc}
    & RCA & CCA & RRCA & BCA\\
     \hline \\
      \multicolumn{5}{l}{\textit{Experiment $2$: General performance (5 construals)}}\\
      \hdashline\\
      \multicolumn{5}{l}{\textit{Newman}}\\
      \addlinespace
      CPA&  23.52\% & 33.53\%  & 27.79\%  & 21.67\%\\
      MAD&  1.324 & 1.053 & 0.908 & 1.020\\
      CDIS& 6.200 & 6.714 & 6.323 & 6.234\\
      SNMI&  0.494 & 0.457  &  0.346 & 0.335\\
      NMI&0.625&0.549&0.406&0.457\\
      Unit-construals & 5.15\% & 23.03\% & 0\%   & 0\%\\
      NOC&1318&1324&954&749\\\hline
      \addlinespace
      \multicolumn{5}{l}{\textit{Leiden}}\\
      \addlinespace
      CPA& 12.44\% & 52.87\%  &  55.98\% & 53.94\%\\
      MAD& 1.247 & 0.656 & 0.466 & 0.489\\                    
      CDIS& 5.557 & 5.723 &  5.640 & 6.108\\
      SNMI& 0.589 & 0.582  &  0.509 & 0.487\\
      NMI&0.737&0.647&0.550&0.577\\
      Unit-construals   & 0.39\% & 15.16\% & 0\%   & 0\%\\
      NOC&970&1601&1547&1517\\\hline\hline
\end{tabular}}
\caption{Complete performance table for Experiment $2$ - Construals $5$.}
\end{table}

\newpage

\begin{table}[h!]
\centering
\resizebox{0.85\columnwidth}{!}{%
    \begin{tabular}{rcccc}
    & RCA & CCA & RRCA & BCA\\
     \hline \\
      \multicolumn{5}{l}{\textit{Experiment $2$: General performance (6 construals)}}\\
      \hdashline\\
      \multicolumn{5}{l}{\textit{Newman}}\\
      \addlinespace
      CPA&  37.23\% & 29.20\%  & 10.27\%  & 5.60\%\\
      MAD&  0.988 & 1.042 & 1.595 & 1.806\\
      CDIS& 6.372 & 6.816 & 6.466 & 6.180\\
      SNMI&  0.494 & 0.441  &  0.505 & 0.280\\
      NMI&0.588&0.516&0.369&0.370\\
      Unit-construals & 3.15\% & 19.53\% & 0.10\%   & 0\%\\
      NOC&1568&815&266&141\\\hline
      \addlinespace
      \multicolumn{5}{l}{\textit{Leiden}}\\
      \addlinespace
      CPA& 26.86\% & 43.24\%  &  26.65\% & 26.65\%\\
      MAD& 0.866 & 0.606 & 0.976 & 0.964\\                    
      CDIS& 4.823 & 5.745 &  5.907 & 6.181\\
      SNMI& 0.589 & 0.565  &  0.639 & 0.441\\
      NMI&0.730&0.616&0.516&0.509\\
      Unit-construals   & 0\% & 12.11\% & 0\%   & 0\%\\
      NOC&1940&1038&412&431\\\hline\hline
\end{tabular}}
\caption{Complete performance table for Experiment $2$ - Construals $6$.}
\end{table}

\newpage

\begin{table}[h!]
\centering
\resizebox{0.85\columnwidth}{!}{%
    \begin{tabular}{rcccc}
    & RCA & CCA & RRCA & BCA\\
     \hline \\
      \multicolumn{5}{l}{\textit{Experiment $3$: Non-gaussian copulae (Student's $t$ with 3}}\\
      \multicolumn{5}{l}{\textit{\phantom{Experiment $3$:} degrees of freedom's copula) (2-4 construals)}}\\
      \hdashline\\
      \multicolumn{5}{l}{\textit{Newman}}\\
      \addlinespace
      CPA&  2.30\% & 19.84\%  & 30.82\%N  & 36.72\%\\
      MAD&  4.046 & 1.703 & 0.954 & 0.780\\
      CDIS& 8.601 & 7.577 & 7.050 & 6.309\\
      SNMI&  0.221& 0.361  &  0.352 & 0.389\\
      NMI&0.579&0.573&0.477&0.500\\
      Unit-construals & 36.39\% & 24.26\% & 0.16\%   & 0.16\%\\\hline\hline
\end{tabular}}
\caption{Complete performance table for Experiment $3$ - Construals 2-4 for a Student's $t$ with $3$ degrees of freedom's copula.}
\label{tab:studentt3}
\end{table}

\begin{table}[h!]
\centering
\resizebox{0.85\columnwidth}{!}{%
    \begin{tabular}{rcccc}
    & RCA & CCA & RRCA & BCA\\
     \hline \\
      \multicolumn{5}{l}{\textit{Experiment $3$: Non-gaussian copulae (Student's $t$ with 6}}\\
      \multicolumn{5}{l}{\textit{\phantom{Experiment $3$:} degrees of freedom's copula) (2-4 construals)}}\\
      \hdashline\\
      \multicolumn{5}{l}{\textit{Newman}}\\
      \addlinespace
      CPA&  2.50\% & 22.87\%  & 43.41\%  & 36.73\%\\
      MAD&  3.693 & 1.544 & 0.865 & 0.691\\
      CDIS& 8.850 & 7.550 & 7.181 & 6.276\\
      SNMI&  0.238 & 0.384  &  0.372 & 0.405\\
      NMI&0.584&0.586&0.493&0.506\\
      Unit-construals & 30.88\% & 25.88\% & 0.50\%   & 0.16\%\\\hline\hline
  \end{tabular}}
\caption{Complete performance table for Experiment $3$ - Construals 2-4 for a Student's $t$ with $6$ degrees of freedom's copula.}
\label{tab:studentt6}
\end{table}

\begin{table}[h!]
\centering
\resizebox{0.85\columnwidth}{!}{%
    \begin{tabular}{rcccc}
    & RCA & CCA & RRCA & BCA\\
     \hline \\
      \multicolumn{5}{l}{\textit{Experiment $4$: Data with no structure}}\\
      \hdashline\\
      \multicolumn{5}{l}{\textit{Newman}}\\
      \addlinespace
      CPA&  0\% & 0\%  & 0\%  & 0\%\\
      MAD&  4.404 & 4.966 & 3.563 & 3.967\\
      CDIS& 7.209 & 7.452 & 6.216 & 6.358\\
      SNMI&  0 & 0  &  0 & 0\\
      NMI&0&0&0&0\\
      Unit-construals & 0\% & 11.2\% & 0\%   & 0\%\\\hline

      \addlinespace
      \multicolumn{5}{l}{\textit{Leiden}}\\
      \addlinespace
      CPA& 0\% & 0\%  &  0\% & 0\%\\
      MAD& 8.211 & 7.008 & 4.924 & 7.245\\                    
      CDIS& 10.640 & 9.683 &  7.857 & 9.127\\
      SNMI&  0 & 0  &  0 & 0\\
      NMI&0&0&0&0\\
      Unit-construals   & 0\% & 10.4\% & 0\%   & 0\%\\\hline\hline
  \end{tabular}}
\caption{Complete performance table for the experiment using data with no underlying structure. Note that the NMI (and SNMI) are $0$ by definition, since all individuals belong to the same construal.}
\label{tab:structureless}
\end{table}

\newpage

\begin{figure}[htb]
    \centering
\begin{subfigure}{0.32\textwidth}
  \includegraphics[width=\linewidth]{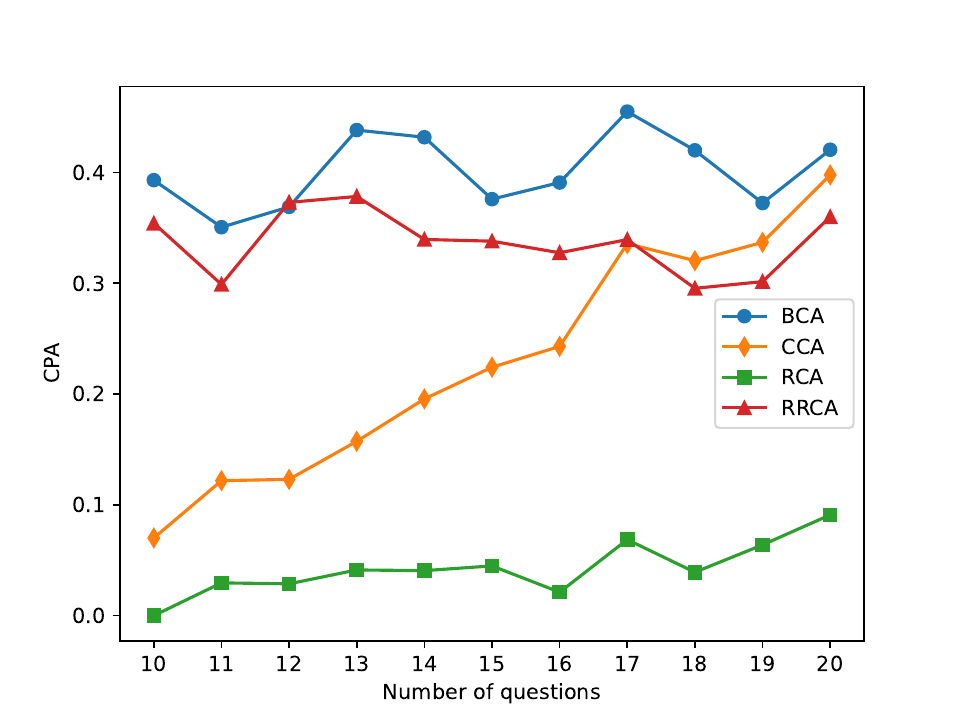}
  \caption{}
\end{subfigure}\hfil
\begin{subfigure}{0.32\textwidth}
  \includegraphics[width=\linewidth]{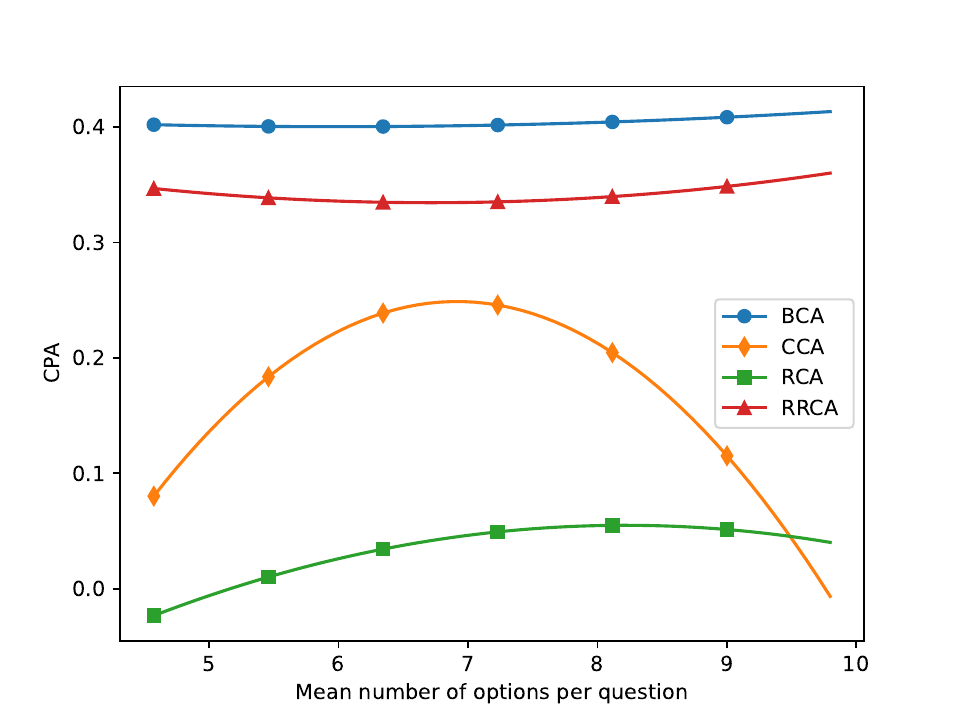}
  \caption{}
\end{subfigure}\hfil
\begin{subfigure}{0.32\textwidth}
  \includegraphics[width=\linewidth]{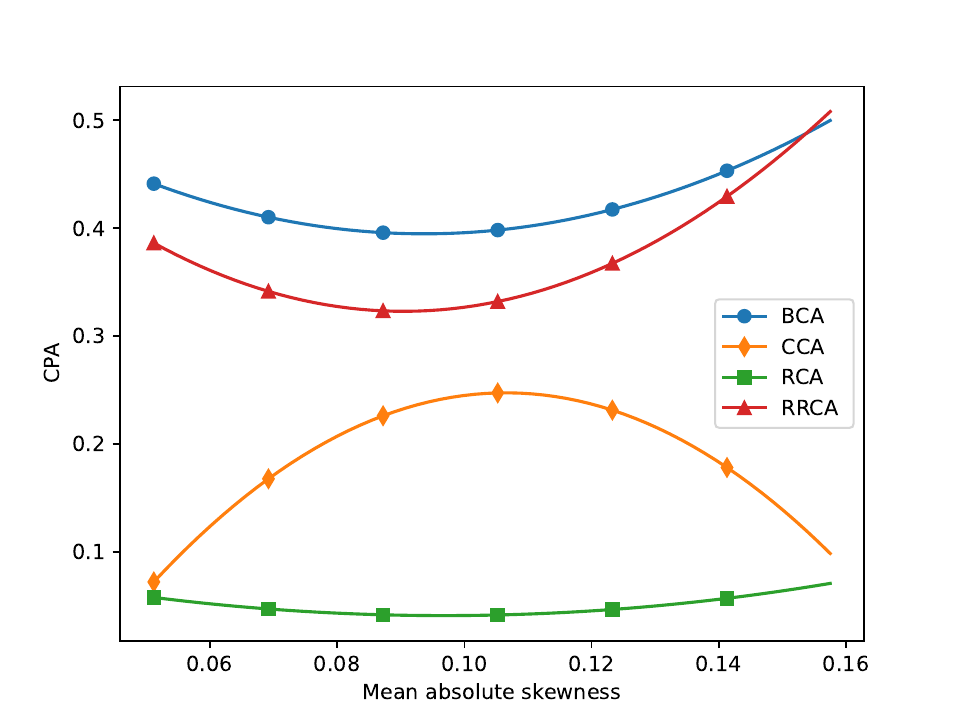}
  \caption{}
\end{subfigure}

\medskip
\begin{subfigure}{0.32\textwidth}
  \includegraphics[width=\linewidth]{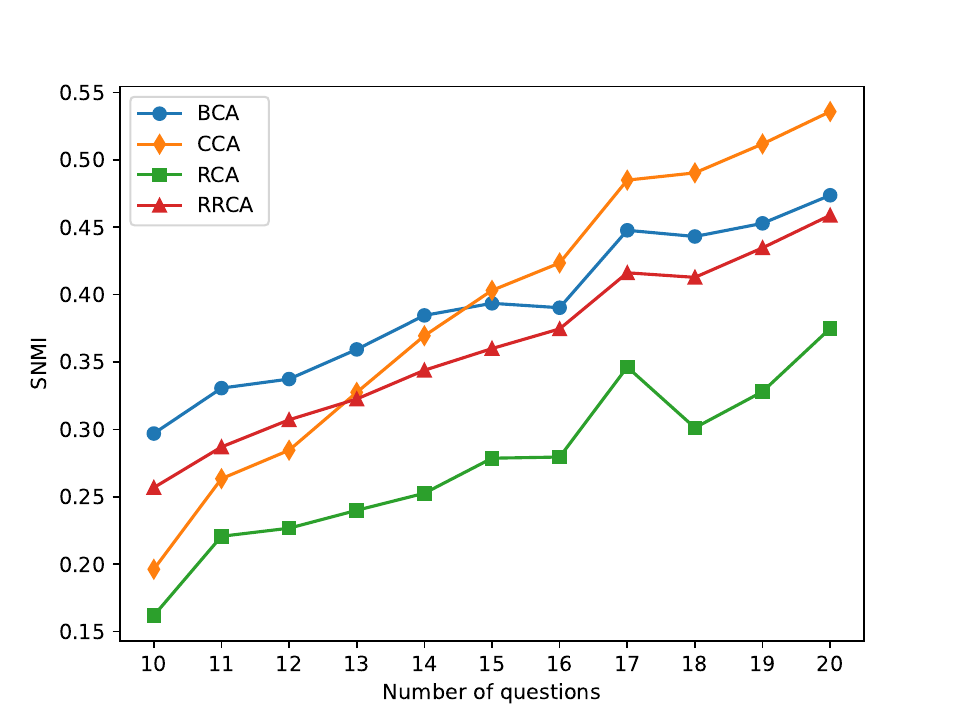}
  \caption{}
\end{subfigure}\hfil
\begin{subfigure}{0.32\textwidth}
  \includegraphics[width=\linewidth]{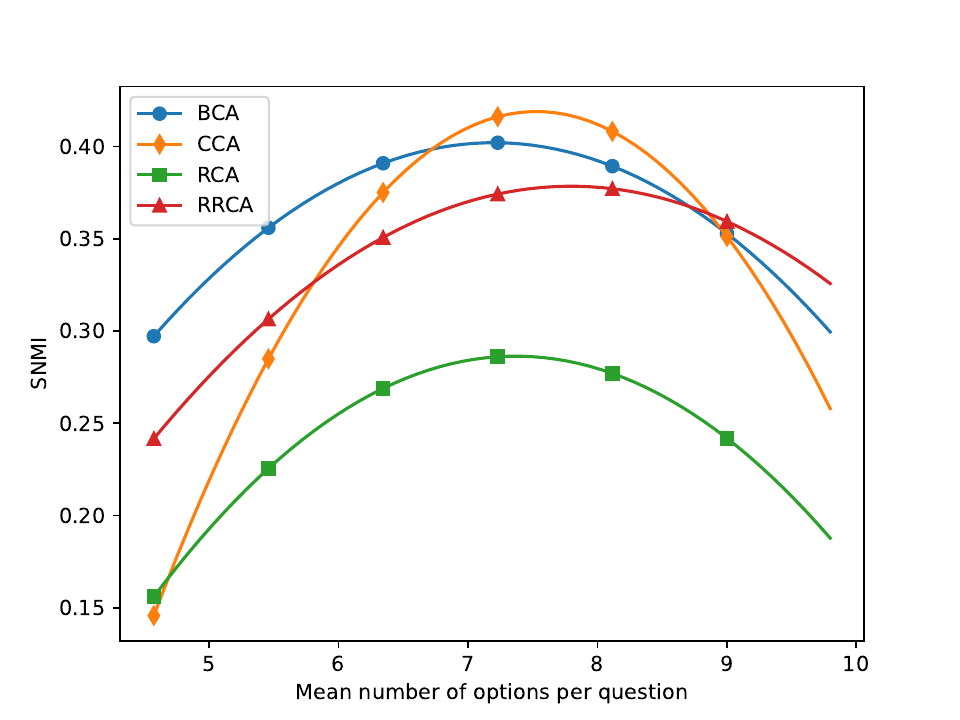}
  \caption{}
\end{subfigure}\hfil
\begin{subfigure}{0.32\textwidth}
  \includegraphics[width=\linewidth]{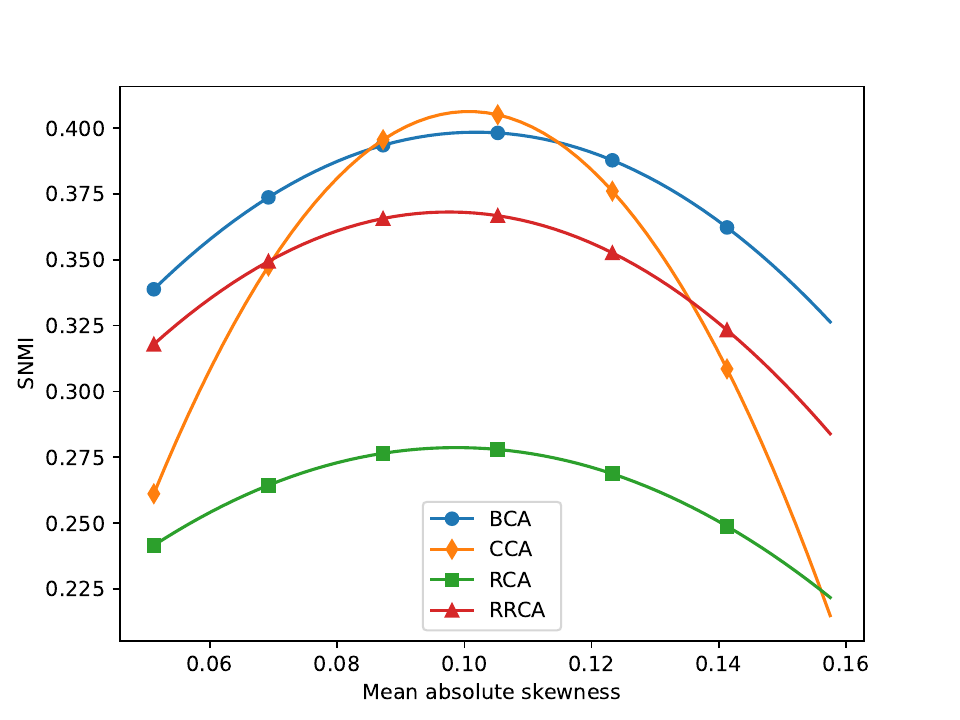}
  \caption{}
\end{subfigure}
\caption{Sensitivity analysis (2--4 construals): CPA (Panels A-C) and SNMI (Panels D-F). Panels A and D: averages conditional on the number of questions. Panels B and E: quadratic fit conditional on the mean number of options per question. Panels C and F: quadratic fit conditional on the mean absolute value of the skewness parameter.}
\label{fig:sensitivity_cpa_snmi}
\end{figure}

\cleardoublepage

\section{Additional Empirical Analyses Information}
\label{app:reanalyses}

    \vspace{1.2em}
    \noindent\text{Analytical Strategy}
    \vspace{0.8em}

        Section~\ref{sec:reanalyses} assesses how substantively similar the results produced by BCA are to those generated by alternative methods. To this end, it proceeds through two complementary sets of analyses.

         The first consists of a reanalysis of the 2004 ANES data on political opinions and the 1993 GSS data on musical preferences. These datasets were selected because they have been used in prior construal analyses, allowing us to assess whether BCA results converge with established baseline findings. These reanalyses assess whether BCA yields findings that differ from those reported in prior work relative to three outcomes of key substantive relevance: first, the number, type, and dependence structure of the estimated construals; second, the association between sociodemographic variables and construal membership; and third, the performance of construal membership as a predictor of non-attitudinal outcomes. To ensure comparability with prior findings, unless otherwise stated, we follow exactly the same data selection, data processing, and analytical procedures conducted by earlier research (\cite{goldberg_2011, baldasarri_goldberg_2014, boutyline_2017}).

        The second set of analyses is an additional check on whether further evidence of heterogeneity in results between BCA and other CCMs can be found beyond the 2004 ANES and the 1993 GSS data. This set reanalyzes the 1988 GSS data on science and religion, previously examined in a construal study by DiMaggio, Sotoudeh and Goldberg (\citeyear{dimaggio_etal_2018}), as well as the Towards Gender Harmony dataset on perceptions of masculinity and femininity (\cite{kosakowska2024towards}). Here we focus merely on assessing whether CCMs converge in estimating the same number and relative population sizes of construals.

    \vspace{1.2em}
    \noindent\text{Construal Estimations}
    \vspace{0.8em}

        \subsubsection{Data selection} 

            
            Table~\ref{ref:apx_politics_constvars} lists the twenty-nine bipolar survey items used to estimate public opinion construals in the US during 2004. These items, which have 2, 3, 4, or 7 response points, are identical to those used by Baldassarri and Goldberg (\citeyear{baldasarri_goldberg_2014}, Supplementary Information, pp.~7–8).

            \begin{table}[htbp]
            \centering
            \footnotesize
            \begin{threeparttable}
            \caption{Public opinion items used for construal estimation}
            \label{ref:apx_politics_constvars}
        
            \begin{tabular*}{\textwidth}{@{}R{0.7cm}@{\hspace{1.2em}}L{13.5cm}@{\extracolsep{\fill}}c@{}}

                \toprule
                \addlinespace[0.8em]
                \multicolumn{2}{@{}l@{}}{\textbf{Items and measurement target}} & \textbf{Response } \\
                \multicolumn{2}{@{}l@{}}{\textbf{}} & \textbf{Points} \\
                \midrule
                \addlinespace[1.0em]
        
                \textit{Economy} \\
                \addlinespace[0.6em]
        
                 1 & \textsc{vcf\oldstylenums{0806}}. Preference for public or privately funded health insurance. & 7 \\
                 2 & \textsc{vcf\oldstylenums{0809}}. Assessing that gov. should or shouldn't guarantee jobs and living standards. & 7 \\
                 3 & \textsc{vcf\oldstylenums{0839}}. Favor more or less gov services and spending. & 7 \\
                 4 & \textsc{vcf\oldstylenums{0886}}. Favor increases or decreases to fed. spending on aid to the poor. & 3 \\
                 5 & \textsc{vcf\oldstylenums{0887}}. Favor increases or decreases to fed. spending on child care. & 3 \\
                 6 & \textsc{vcf\oldstylenums{0888}}. Favor decreases or increases to fed. spending on dealing with crime. & 3 \\
                 7 & \textsc{vcf\oldstylenums{0890}}. Favor increases or decreases to fed. spending on public schools. & 3 \\
                 8 & \textsc{vcf\oldstylenums{0894}}. Favor increases or decreases to fed. spending on welfare programs. & 3 \\
                 9 & \textsc{vcf\oldstylenums{9049}}. Favor increases or decreases to fed. spending on social security. & 3 \\
        
                \addlinespace[1.0em]
                \multicolumn{2}{@{}l@{}}{\textit{Civil Rights}} \\
                \addlinespace[0.6em]
        
                 1 & \textsc{vcf\oldstylenums{0830}}. Support for gov efforts to improve social conditions of blacks and minorities. & 7 \\
                 2 & \textsc{vcf\oldstylenums{0867}a}. Support for giving preferences to blacks in hirings and promotion. & 4 \\
                 3 & \textsc{vcf\oldstylenums{9013}}. Agreement that society should do everything necessary to ensure equal opportunity. & 5 \\
                 4 & \textsc{vcf\oldstylenums{9014}}. Agreement that pushing equal rights has gone too far. & 5 \\
                 5 & \textsc{vcf\oldstylenums{9015}}. Agreement that lack of equal chances is a big problem in the US. & 5 \\
                 6 & \textsc{vcf\oldstylenums{9016}}. Agreement that it is not a big problem if some people have more of a chance in life. & 5 \\
                 7 & \textsc{vcf\oldstylenums{9017}}. Agreement that the US would better if there were less worries about how equal people are. & 5 \\
                 8 & \textsc{vcf\oldstylenums{9018}}. Agreement that there would be fewer problems in the US if people were treated more equally. & 5 \\
                 9 & \textsc{vcf\oldstylenums{9039}}. Agreement that slavery and historical discrimination obstruct upward mobility for blacks. & 5 \\
                10 & \textsc{vcf\oldstylenums{9040}}. Agreement that blacks shuold advance without special favors. & 5 \\
                11 & \textsc{vcf\oldstylenums{9041}}. Agreement that if blacks tried harder they would be just as well off as whites. & 5 \\
                12 & \textsc{vcf\oldstylenums{9042}}. Agreement that blacks have gotten less than they deserve in recent times. & 5 \\
        
                \addlinespace[1.0em]
                \multicolumn{2}{@{}l@{}}{\textit{Morality}} \\
                \addlinespace[0.6em]
        
                1 & \textsc{vcf\oldstylenums{0834}}. Assessing that women should have an equal role than men in business and government or women's place should be at home. & 7 \\
                2 & \textsc{vcf\oldstylenums{0851}}. Agreement that newer lifestyles are contributing to the breakdown of society. & 5 \\
                3 & \textsc{vcf\oldstylenums{0852}}. Agreement that views on moral behavior should be adjusted to changes in the world. & 5 \\
                4 & \textsc{vcf\oldstylenums{0853}}. Agreement that the US would have less problems if there was less emphasis on traditional family ties. & 5 \\
                5 & \textsc{vcf\oldstylenums{0854}}. Agreement that there should be more tolerance for people who live according to different moral standards. & 5 \\
                6 & \textsc{vcf\oldstylenums{0876}}. Favor or oppose laws to protect gays and lesbians against job discrimination. & 2 \\
                7 & \textsc{vcf\oldstylenums{0877}}. Favor or oppose allowing homosexuals to serve in the army. & 2 \\
                8 & \textsc{vcf\oldstylenums{0838}}. Assessing that abortion should never be permitted by law or should always be legal as a matter of personal choice. & 4 \\
        
                \addlinespace[1.0em]
                \multicolumn{2}{@{}l@{}}{\textit{Foreign Affairs}} \\
                \addlinespace[0.6em]
        
                1 & \textsc{vcf\oldstylenums{0843}}. Favor decreases or increases in defense spending. & 7 \\
                2 & \textsc{vcf\oldstylenums{0892}}. Favor increases or decrease in fed. spending on foreign aid. & 3 \\
                3 & \textsc{vcf\oldstylenums{9048}}. Favor increases or decrease in fed. spending on space, science, and technology. & 3 \\
        
                \bottomrule
            \end{tabular*}
        
            \begin{tablenotes}[flushleft]
                \footnotesize
                \item 
                \textit{Notes:} Data are drawn from the 2004 ANES cumulative dataset. Labels in small caps refer to item variable names. The Response Points column reports the number of response categories available for each item. Variable set and domain classifications follow Baldassarri and Goldberg (\citeyear{baldasarri_goldberg_2014}). All items are coded so that higher values indicate more conservative positions; item wording follows this orientation.
            \end{tablenotes}
            
        \end{threeparttable}
        \end{table}

            Table~\ref{ref:apx_music_constvars} reports the set of the bipolar survey items from the 1993 GSS used to analyze musical tastes. This battery of questions corresponds to the one employed by Goldberg (\citeyear{goldberg_2011}) and Boutyline (\citeyear{boutyline_2017}). All items are measured using five-point Likert scales.

            \begin{table}[htbp]
            \centering
            \footnotesize
            \begin{threeparttable}
            \caption{Musical preference items used for construal estimation}
            \label{ref:apx_music_constvars}
                    
            \begin{tabular}
            {@{}rL{8.5cm}@{}}

            \multicolumn{2}{@{}l@{}}{\text{Measurement target:}} \\
            \multicolumn{2}{@{}l@{}}{\text{Liking or disliking musical genres}\textit{(five-point response scale)}} \\
            \addlinespace[0.2em]  
                    
                    \toprule
                        \addlinespace[0.6em]  
                        \textbf{Items} &  \\
                        \addlinespace[0.2em]  
                    \midrule
                    \addlinespace[0.6em]                
                
                    1 & \textsc{country}. Country/Western music. \\
                    2 & \textsc{conrock}$^1$. Contemporary pop and rock music. \\
                    3 & \textsc{oldies}. Oldies rock music. \\
                    4 & \textsc{classicl}. Classical music; symphony and chamber music. \\
                    5 & \textsc{opera}. Opera music. \\
                    6 & \textsc{musicals}. Broadway musicals and show tunes. \\
                    7 & \textsc{bigband}. Big Band music. \\
                    8 & \textsc{jazz}. Jazz music. \\
                    9 & \textsc{reggae}. Reggae music. \\
                    10 & \textsc{rap}. Rap music. \\
                    11 & \textsc{hvymetal}. Heavy metal music. \\
                    12 & \textsc{latin}. Latin, mariachi, and salsa music. \\
                    13 & \textsc{folk}. Folk music. \\
                    14 & \textsc{moodeasy}. Mood and easy listening music. \\
                    15 & \textsc{blugrass}. Bluegrass music. \\
                    16 & \textsc{gospel}. Gospel music. \\
                    17 & \textsc{blues}. Blues and R\&B music. \\
                    
                    \bottomrule
                    \end{tabular}
                    
                \begin{tablenotes}[flushleft]
                \footnotesize
                    \item  \textit{Notes:} Data drawn from the 1993 GSS. Labels correspond to names used in correlation matrices: musical genres refers to the musical genres contained in each musical preference variable.
                    \vspace{0.3em}
                    \item [$^1$] Variable shown in correlation matrixes as \textsc{pop}.
                \end{tablenotes}
                
                \end{threeparttable}
                \end{table}

            Table~\ref{ref:apx_s&r_constvars} displays the fifteen bipolar items from the 1988 GSS used to estimate construals of attitudes toward science and religion. Variable selection follows DiMaggio, Sotoudeh, and Goldberg (\citeyear{dimaggio_etal_2018}, p.~36). These items have between two and six response points.

            \begin{table}[htbp]
                \centering
                \footnotesize
                \begin{threeparttable}
                \caption{Items on attitudes towards science and religion used for construal estimation}
                \label{ref:apx_s&r_constvars}
            
                \begin{tabular*}{\textwidth}{@{}R{0.7cm}@{\hspace{1.2em}}L{13.5cm}@{\extracolsep{\fill}}c@{}}
    
                    \toprule
                    \addlinespace[0.8em]
                    \multicolumn{2}{@{}l@{}}{\textbf{Item and measurement target}} & \textbf{Response } \\
                    \multicolumn{2}{@{}l@{}}{\textbf{}} & \textbf{Points} \\
                    \midrule
                    \addlinespace[1.0em]
            
                    \addlinespace[0.6em]
            
                     1 & \textsc{conclerg}. Confidence in organized religion. & 3 \\
                     2 & \textsc{neargod}. Assessment of feeling close to good. & 5 \\
                     3 & \textsc{god}. Belief in God. & 6 \\
                     4 & \textsc{decbible}. Importance of Bible in making life decisions. & 5 \\
                     5 & \textsc{decchurch}. Importance of teachings of church or synagogue to make life decisions. & 5 \\
                     6 & \textsc{consci}. Confidence on the Scientific Community. & 3 \\
                     7 & \textsc{scichng}. Agree or disagree that one trouble with science is that it makes ways of life too fast. & 2 \\
                     8 & \textsc{scimoral}. Agree or disagree that one bad effect of science is that it breaks down people's ideas about right and wrong. & 2 \\
                     9 & \textsc{scipry}. Agree or disagree that scientist always seem to be prying into things that they really ought to stay out of. & 2 \\
                     10 & \textsc{scisolvew}. Disagree or agree that science will solve social problems like crime and mental illness. & 2 \\
                     11 & \textsc{dejavu}. Frequency of respondent having thought to be somewhere she had been before, but knew was impossible. & 4 \\
                     12 & \textsc{esp}. Frequency of respondent having thought to be in touch with someone far away from hers. & 4 \\
                     13 & \textsc{spirits.} Frequency of respondent having thought to have been in touch with someone who had died. & 4 \\
                     14 & \textsc{visions}. Frequency of respondent having seen events that happened at a great distance as they were happening. & 4 \\
                     15 & \textsc{grace}. Frequency of respondent having thought to be close to a powerful, spiritual force that seemed to lift her out of herself. & 4 \\
            
                    \bottomrule
                \end{tabular*}
            
                \begin{tablenotes}[flushleft]
                    \footnotesize
                    \item 
                    \textit{Notes:} Data drawn from the 1988 GSS dataset. Labels in small caps refer to item variable names. The Response Points column reports the number of response categories available for each item. 
                \end{tablenotes}
                
            \end{threeparttable}
        \end{table}

            Finally, Table~\ref{ref:apx_women_constvars} presents the twelve seven-point bipolar survey items from the TGHD used to analyze perceptions of desirable attributes for women. 

            \begin{table}[htbp]
                \centering
                \footnotesize
                \begin{threeparttable}
                \caption{Attitudes on desirable women traits used for construal estimation}
                \label{ref:apx_women_constvars}
                        
                \begin{tabular}
                {@{}rL{8.5cm}@{}}
    
                \multicolumn{2}{@{}l@{}}{\text{Measurement target:}} \\
                \multicolumn{2}{@{}l@{}}{\text{Assessing social desirability of personal attributes for women}\textit{ (seven-point response scale)}} \\
                
                \addlinespace[0.2em]  
                        
                        \toprule
                            \addlinespace[0.6em]  
                            \textbf{Items} &  \\
                            \addlinespace[0.2em]  
                        \midrule
                        \addlinespace[0.6em]                
                    
                        1 & \textsc{v\oldstylenums{77}}. Compassionate. \\
                        2 & \textsc{v\oldstylenums{78}}. Helpful to others. \\
                        3 & \textsc{v\oldstylenums{79}}. Sympathetic. \\
                        4 & \textsc{v\oldstylenums{80}}. Understanding of others. \\
                        5 & \textsc{v\oldstylenums{81}}. Sensitive. \\
                        6 & \textsc{v\oldstylenums{83}}. Aware of others' feelings. \\
                        7 & \textsc{v\oldstylenums{84}}. Cooperative. \\
                        8 & \textsc{v\oldstylenums{85}}. Devoted to others. \\
                        9 & \textsc{v\oldstylenums{87}}. Warm. \\
                        10 & \textsc{v\oldstylenums{88}}. Supportive. \\
                        11 & \textsc{v\oldstylenums{116}}. Competent. \\
                        12 & \textsc{v\oldstylenums{123}}. Capable. \\
                       
                        \bottomrule
                        \end{tabular}
                        
                    \begin{tablenotes}[flushleft]
                    \footnotesize
                        \item  \textit{Notes:} Data drawn from the TGHD dataset (\citeyear{kosakowska2024towards}). Labels in small caps correspond to variable names. Item codes reflect original survey numbering.
                        \vspace{0.3em}
                    \end{tablenotes}
                    
                    \end{threeparttable}
            \end{table}

        \subsubsection{Data processing and methods implementation} 
    
             As a first step, we harmonized the interpretive orientation of all variables. ANES variables were recoded so that lower values correspond to more left-wing positions and higher values to more right-wing positions. For GSS variables measuring musical preferences, lower values indicate stronger dislike and higher values indicate stronger liking of a given genre. GSS variables measuring attitudes toward science and religion were coded so that lower values indicate less support for science and higher values greater support. Finally, for TGHD items, lower values indicate lower perceived desirability and higher values greater perceived desirability. We used list-wise deletion to adress missing data.
    
            A second data-processing step involved transforming variable scales to meet the scale requirements of each CCM. For BCA, the neutral point of bipolar items was set at zero, with values on the negative and positive sides spaced at equal one-unit intervals. For RCA, CCA, and RRCA, variables were rescaled to a 0–1 interval with equally spaced increments between values.
        
            We implemented RCA, CCA, and RRCA using the same software versions employed in the simulation analyses: RCA version~2.0, CCA version~0.2.1, and the RRCA variant implemented in the repository available at \url{https://github.com/raminasotoudeh/coping-with-plenitude}). This RRCA version uses the Louvain instead of the Newman partition algorithm.

        \subsubsection{Results} 

        Construal estimates for political opinions and musical preferences are shown in Figures~\ref{fig:construal_politics_results} and~\ref{fig:construal_music_results}, respectively; construal estimates for attitudes toward science and religion and toward perceptions of attribute desirability for women are available upon request. For the ANES 2004 analysis, we typified a construal as a unrestricted lib con if positive correlations predominated across all domain intersections, and  as a restricted lib con if it exhibited generally positive correlations but were not a majority in at least of one domain intersection. For the 1993 GSS data, we characterized the defining trait of an omni-univore cluster type the predominance of positive correlations and the absence of negative correlations; and, as contempo-trad the presence of negative correlations between  rap, reggae, and pop, on one hand, and gospel, blues, bluegrass, or big band, on the other. 

            
            \begin{figure}[htbp]
                \centering
                \caption{Political opinions: Dependence structures of estimated construals}
                \includegraphics[width=450pt]{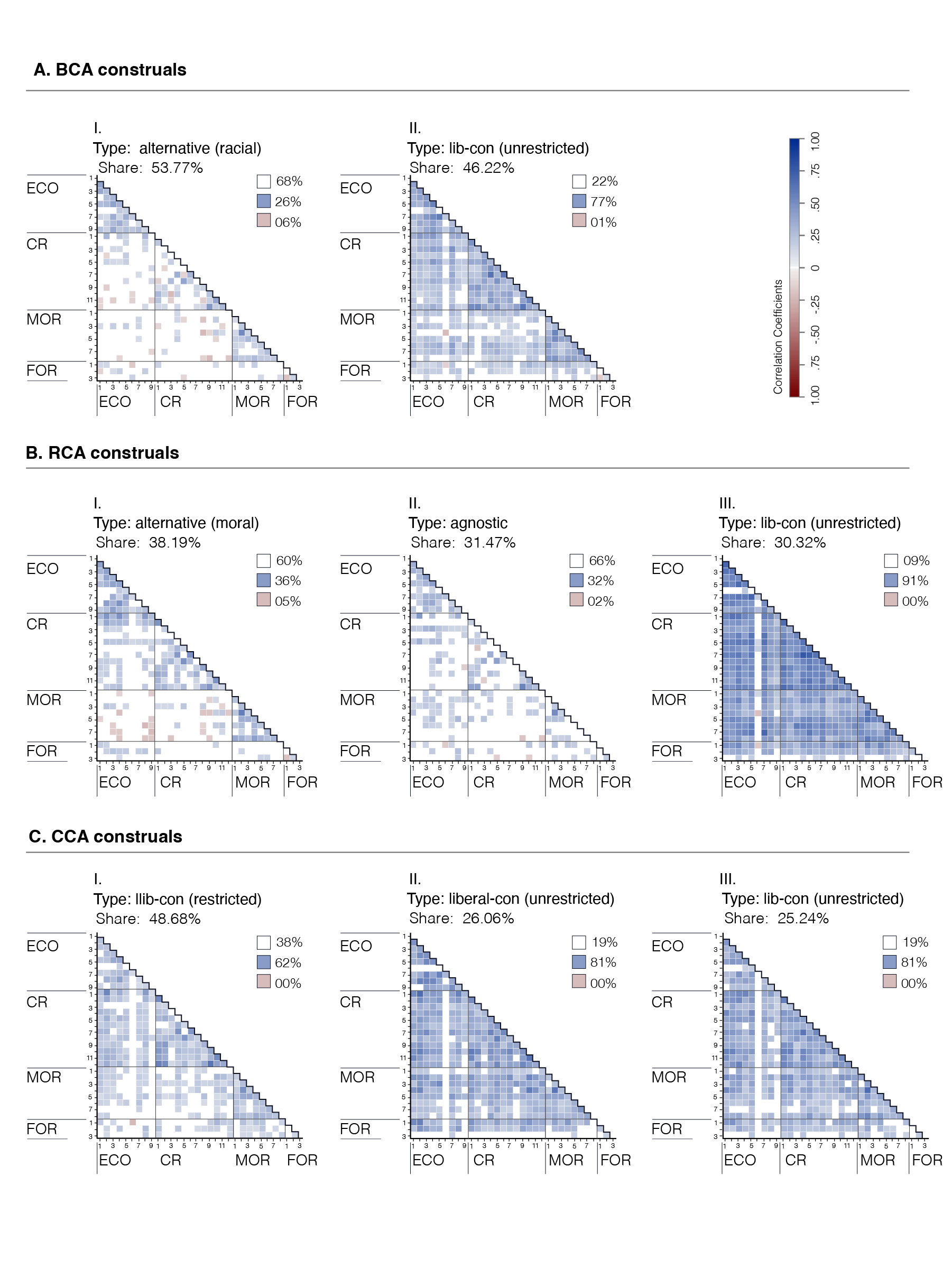}
                \label{fig:construal_politics_results}
            \end{figure}
    
            \begin{figure}[htbp]
                \ContinuedFloat
                \centering      
                \includegraphics[width=450pt]{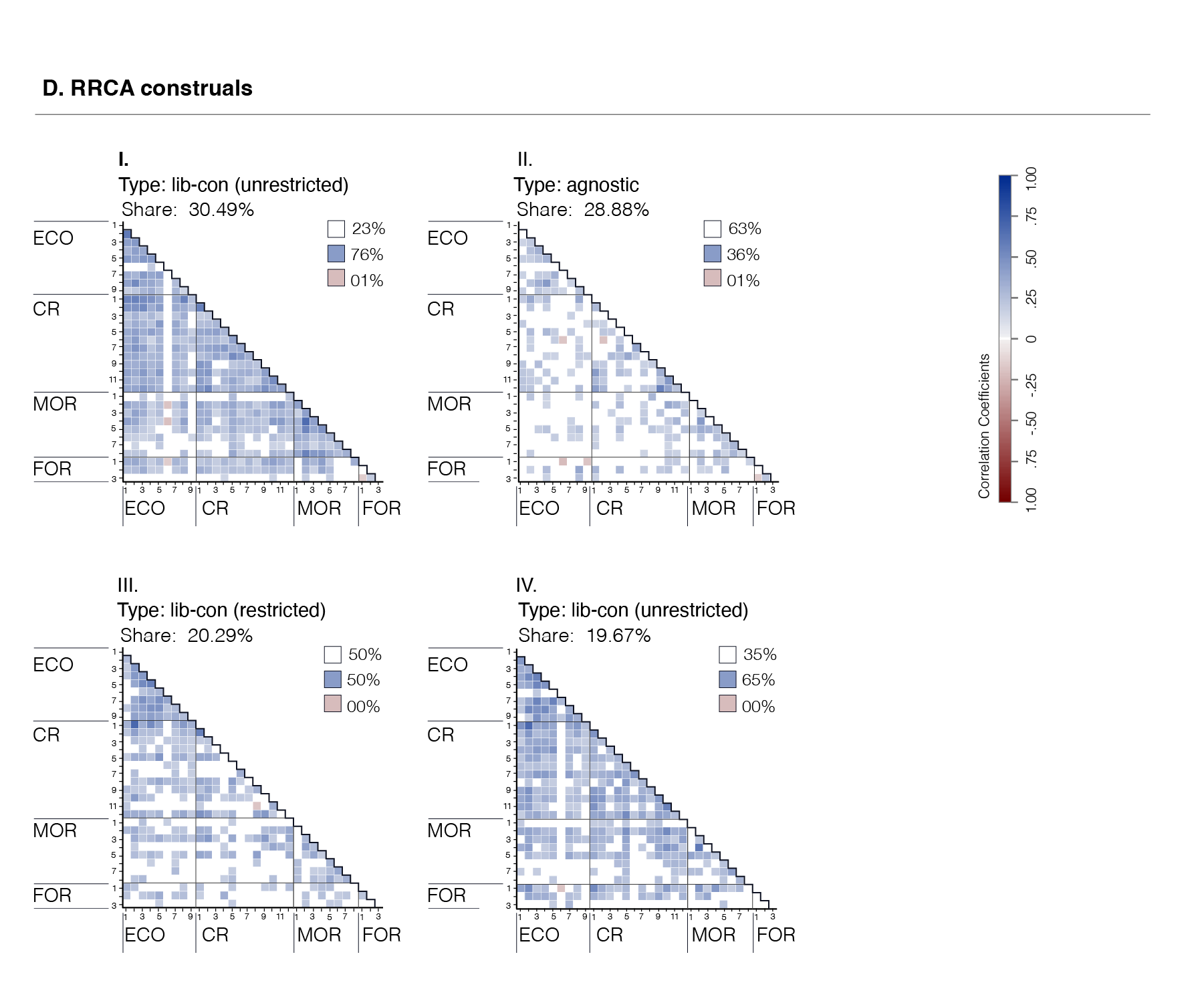}                
                \caption{(cont.)}
                \caption*{\footnotesize\textit{Notes:} White cells indicate non-significant correlations. Share figures report the relative proportion of respondents in each construal. \textsc{eco}: Economy; \textsc{cr}: Civil Rights; \textsc{mor}: Morality; \textsc{for}: Foreign Affairs. Numbers along the axes refer to item indices listed in Table~\ref{ref:apx_politics_constvars}.}
                \rule{\linewidth}{0.2pt}
                
            \end{figure}

                \begin{figure}[htbp]
                    \centering
                    \caption{Musical preferences: Dependence structures of estimated construals}
                    \includegraphics[width=450pt]{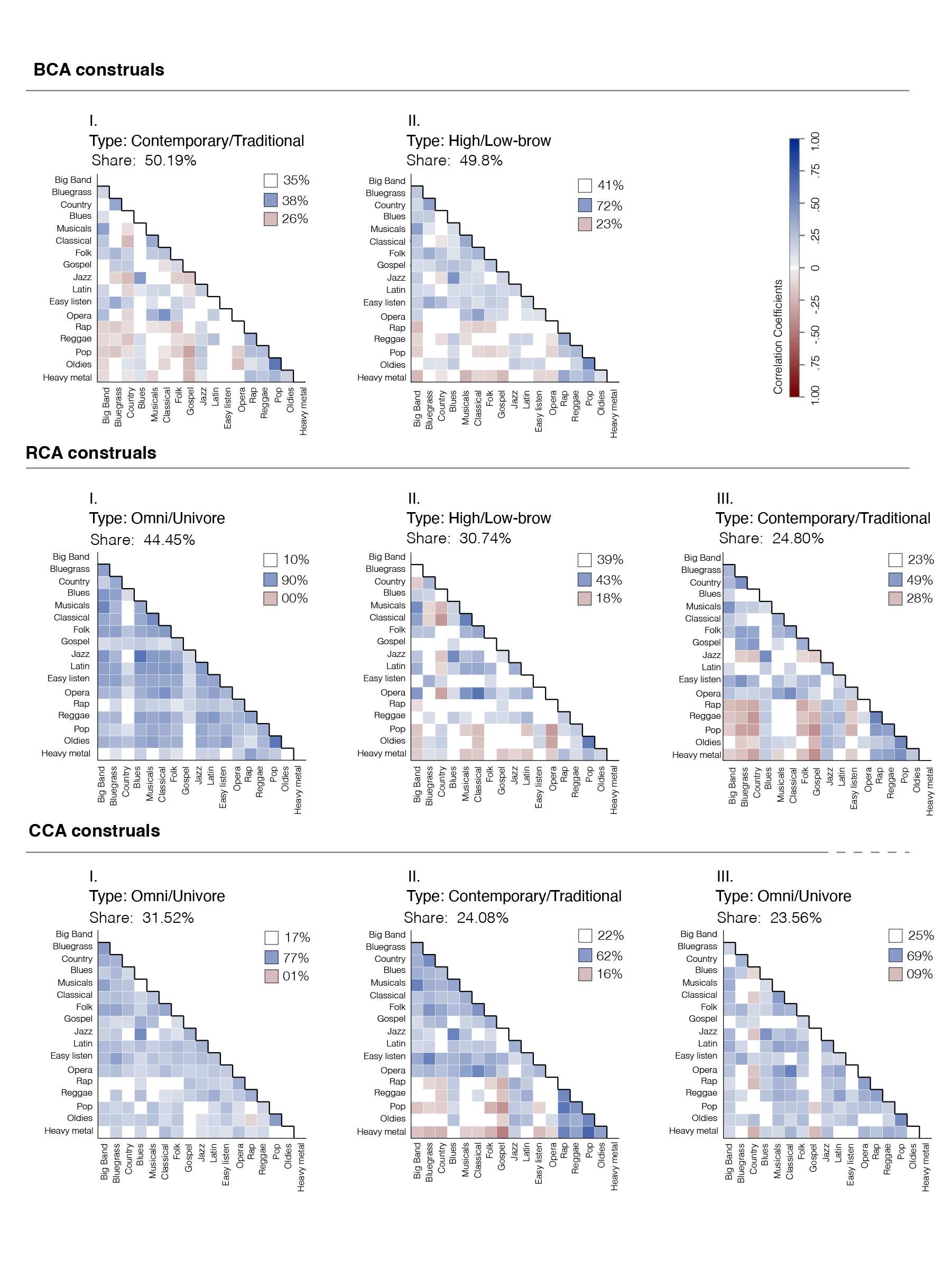}
                    \label{fig:construal_music_results}
                \end{figure}
    
                \begin{figure}[htbp]
                    \ContinuedFloat
                    \centering
                    \includegraphics[width=450pt]{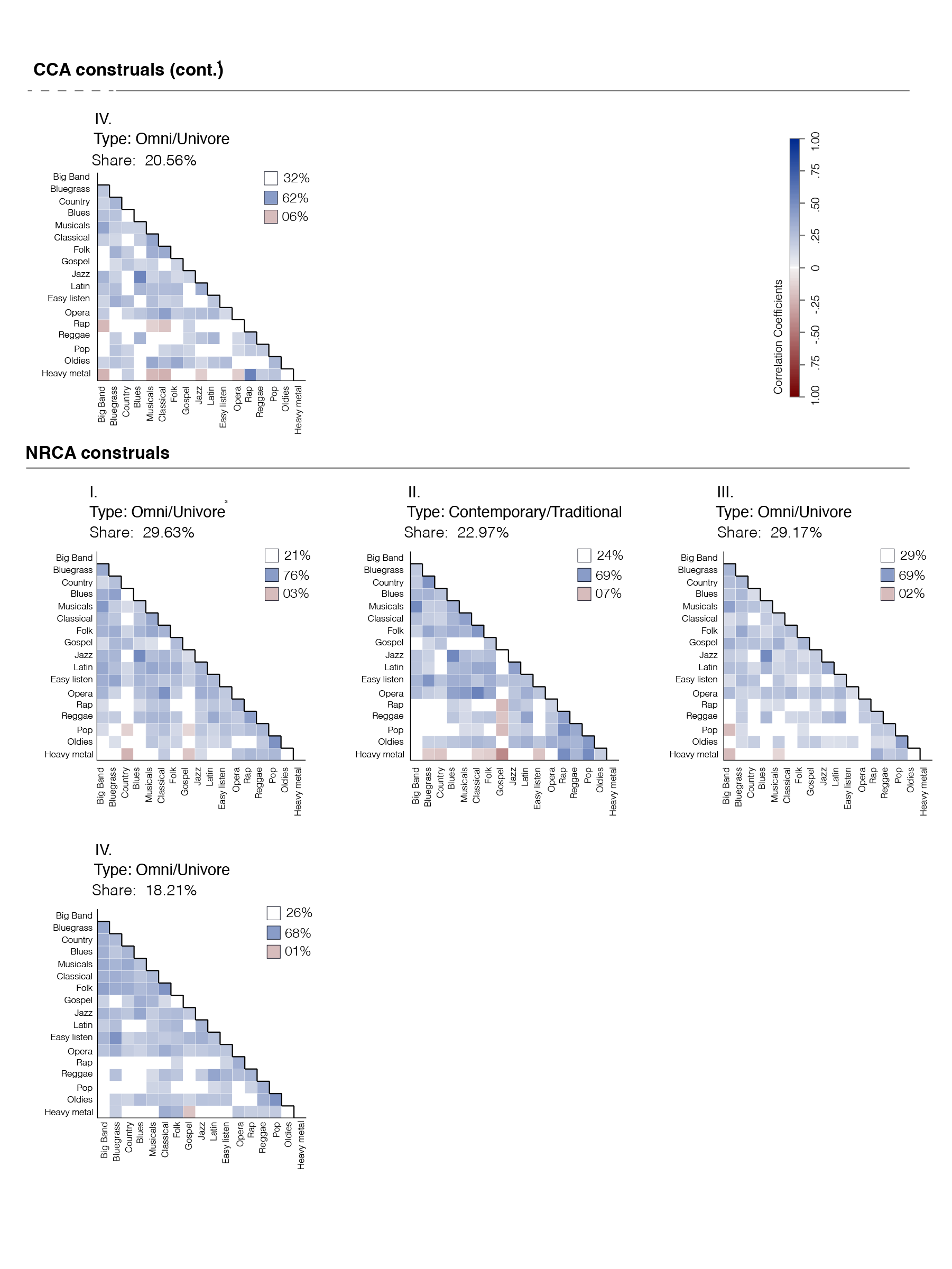}
                    \caption{(cont.)}
                \caption*{\footnotesize
                \textit{Notes:} White cells indicate non-significant correlations. Share figures report the relative proportion of respondents assigned to each construal. Numbers along the axes correspond to item indices listed in Table~\ref{ref:apx_politics_constvars}. \\ 
                $^1$ Cumulative population of CCA construals in the plot do not add to 100\%; the method identifies an additional construal with a marginal population of 4 whose items are perfectly correlated with one another.
                }
                \rule{\linewidth}{0.2pt}
                \end{figure}

    \vspace{1.2em}
    \noindent\text{Regression analyses: variables and model specification}
    \vspace{0.8em} 


            At the end of Section~6, we summarize results from four regression analyses examining how construal membership relates to sociodemographic characteristics and politically and culturally relevant outcomes across methods. The following paragraphs describe these analyses in detail and present the corresponding results.

        \vspace{1.2em}
        \noindent\text{Political opinion analyses}
        \vspace{0.8em} 

            The first two analyses examine the association between respondents’ membership in political opinion construals and their sociodemographic characteristics and political self-identifications. In Analysis 1, the dependent variable is respondents’ construal membership type. In Analysis 2, the dependent variable is party identification, measured on a seven-point scale ranging from Strong Democrat (1) to Strong Republican (7).

            Both Analyses~1 and~2 include as regressors the same set of sociodemographic and political engagement variables employed by previous research (\cite{baldasarri_goldberg_2014}; Supporting Information Appendix, pp.~9). These variables are listed in Table~\ref{ref:apx_politics_regvars}. 

            \begin{table}[htbp]
            \centering
            \footnotesize
            \begin{threeparttable}
            \caption{Regression Models on Political Opinion: Sociodemographic and Political Engagement Variables}
            \label{ref:apx_politics_regvars}
            
            \begin{tabularx}{\textwidth}
                {@{}L{3.0cm}L{2.3cm}cX@{}}
                \toprule
                \textbf{Variable} 
                & \textbf{Type} 
                & \textbf{Range}
                & \textbf{Maximum and Minimum values} \\
                \midrule
                \addlinespace[0.6em]
                
                Age 
                & Continuous 
                & 17--96 
                & 17: age 17; 97: age 97 or more \\
                
                Church attendance 
                & Categorical 
                & 1--5 
                & 1: Every week; 5: Never \\
                
                Income 
                & Categorical 
                & 1--5 
                & 1: 0--16 percentile; 5: 96--100 percentile \\
                
                Gender
                & Binary 
                & 0, 1 
                & 0: Male; 1: Female \\
                
                Black 
                & Binary 
                & 0, 1 
                & 0: Other; 1: Black \\
                
                Southerner
                & Binary 
                & 0, 1 
                & 0: Other 1: Southern Census Region \\
                
                Education
                & Categorical 
                & 1--6 
                & 1: 8 grades or less; 6: bachelor's or more \\
                
                Professional status 
                & Binary 
                & 0, 1 
                & 0: Other; 1: Professional/ managerial occup. \\
                
                Interest in political campaigns$^1$ 
                & Categorical 
                & 1--4 
                & 1: Hardly at all; 4: Most of the time \\
                
                Political activism$^2$  
                & Additive Index 
                & 0--6 
                & 0: No participation; 6: Participation in 6 activities \\
    
                Political discussion$^3$  
                & Binary 
                & 0,1
                & 0: Does not discuss politics; 1: Discusses politics with family and friends\\
                
                \bottomrule
            \end{tabularx}
            
            \begin{tablenotes}[flushleft]
                \footnotesize
                \item \textit{Notes:} Data drawn from the 2004 ANES.
                \item[$^1$] Interest in political campaigns.
                \item[$^2$] Additive scale measuring respondents' engagement in the following activities:
                (1) trying to influence the vote of others;
                (2) attending political meetings or rallies;
                (3) working for a party or candidate;
                (4) wearing a campaign button or sticker;
                (5) donating money to a party or candidate;
                \item[$^3$] Respondent discusses politics with family or friends.
            \end{tablenotes}
            
            \end{threeparttable}
            \end{table}


            Table~\ref{ref:apx_determinants_libcon} reports results from Analysis~1, which estimates logistic regression models predicting ranking as a member of unrestricted liberal–conservative construals estimated by BCA (Model~1), RCA (Model~2), CCA (Model~3), and RRCA (Model~4). The key independent variables in these models are the socioeconomic and political engagement covariates listed in Table~\ref{ref:apx_politics_regvars}. An interaction term between religious attendance and income, following model specifications from previous analyses (\cite{baldasarri_goldberg_2014}, Supplementary Material, p.~18),  Odds ratios are reported for brevity and substantive clarity; estimated coefficients and standard errors are available upon request. 
        
            \begin{table}[htbp]
                \centering
                \caption{Logistic regressions of liberal/conservative construal memberships: odds ratios}
                \label{ref:apx_determinants_libcon}
                
                \begin{tabular*}{\textwidth}{@{\extracolsep{\fill}}L{3.6cm}cccc}
                \toprule
                \addlinespace[0.8em]
                         & (1)  & (2) & (3)  & (4)    \\
                        \addlinespace[0.6em]
                        & DV: BCA  & DV: RCA  & DV: CCA  & DV: RRCA    \\
                        & lib-con   & lib-con   & lib-con   & lib-con   \\
                        & membership  & membership  & membership  & membership   \\                        
                        \addlinespace[0.6em]
                
                \midrule
                \addlinespace[1.2em]
                \addlinespace[1.2em]            
                \text{\textit{Independent variables:}} & & & & \\
    
                \addlinespace[0.4em]
                
                 Age                    & 0.978***  & 0.988*   & 0.993    & 0.990       \\
                \addlinespace[0.4em]
                 Church attendance      & 0.962     & 0.825     & 1.098     & 1.010     \\
                \addlinespace[0.4em]
                 Income                 & 0.829     & 0.978     & 1.291     & 1.289     \\
                \addlinespace[0.4em]
                 Female                 & 1.189     & 1.203     & 0.774     & 0.967     \\
                \addlinespace[0.4em]
                 Black                  & 15.737***  & 1.999**  & 0.976     & 2.582***  \\
                \addlinespace[0.4em]
                 South                  & 0.859     & 0.824     & 0.970     & 0.934     \\
                \addlinespace[0.4em]
                 Education              & 1.319***  & 1.301***  & 1.343**** & 1.294***  \\
                \addlinespace[0.4em]
                 Professional status
                                        & 1.684**     & 2.000**   & 1.293     & 1.565**  \\
                \addlinespace[0.4em]
                 Income $\times$ church attendance
                                        & 0.962     & 1.042     & 1.010     & 0.973     \\
                \addlinespace[1.2em]
                Political interest
                                        & 1.342*     & 1.345     & 1.358*   & 1.659***  \\
                \addlinespace[0.4em]
                Political activity
                                        & 1.077     & 1.164*   & 0.891     & 0.906      \\        
                \addlinespace[0.4em]
                Political discussion
                                        & 0.963     & 2.522**   & 1.006     & 1.143     \\
            
                \addlinespace[1.2em]
                
                \midrule
                AIC                      & 666.895  & 625.157   & 720.084 & 719.236 \\
                $N$                      & 550      & 550       & 550     & 550         \\
                \bottomrule
                \end{tabular*}
                
                \caption*{\footnotesize
                \textit{Notes:} Odds ratios reported; estimated coefficients and standard errors available upon request.  * $p<0.10$, ** $p<0.05$, *** $p<0.01$, **** $p<0.0001$.
                }
            \end{table}

            Table~\ref{ref:apx_determinants_partisanship} reports the results of Analysis~2, which estimates OLS regression models of partisan orientation. The key regressors are dummy variables indicating respondents' membership in one of the five substantive types of construals identified across CCMs: unrestricted liberal-conservative (reference category), restricted liberal-conservative, alternative racial, alternative moral, and agnostic. In Model~1, uses construal assignations from BCA; in Model~2, from RCA; and in Models~3 and~4, from CCA and RRCA, respectively. models include as controls the sociodemograpic variables listed in Table~\ref{ref:apx_politics_regvars}.

            \begin{table}[htbp]
                \centering
                \footnotesize
                \begin{threeparttable}
                
                \caption{OLS regressions of party identification}
                \label{ref:apx_determinants_partisanship}
                
                \begin{tabular*}{\textwidth}{@{}L{3.6cm}@{\extracolsep{\fill}}cccc@{}}
                    
                    \toprule
                    \addlinespace[0.8em]
                            & (1) & (2) & (3) & (4)  \\
                            \addlinespace[0.6em]                       
                            & key regressors:     & key regressors:        &key regressors:       & key regressors:              \\
                            \addlinespace[0.4em]
                            & BCA     & RCA        & CCA       & RRCA              \\
                            & construal & construal  construal & construal       & construal \\
                            & membership & membership & membership & membership \\
                    \midrule
                    
                    \addlinespace[1.2em]
                    \text{Construal membership indicators \textit{(ref: unrestricted lib-con})}$^1$:  \\
                    \addlinespace[0.8em]

                    restricted lib-con                  &  ---       &    ---    & -0.587*** &  0.585**     \\
                                                        &  ---       &    ---    & (0.177)   &  (0.226)     \\

                    \addlinespace[0.8em]                                   
                    alternative, racial                 & 1.513****  &    ---    &    ---    &  ---         \\
                                                        & (0.175)    &    ---    &    ---    &  ---         \\
                    \addlinespace[0.8em]                                   
                    alternative, moral                  &  ---       & 0.858**** &  ---      &  ---         \\
                                                        &  ---       & (0.209)   &  ---      &  ---         \\
                    \addlinespace[0.8em]                                   
                    agnostic                            &  ---       & 0.734***  &  ---        & -0.219     \\
                                                        &  ---       &  (0.222)  &  ---        & (0.205)    \\

                    \addlinespace[1.2em]
                    \text{Sociodemographics} \\
                    \addlinespace[0.8em]
                    Education                           & 0.076     & 0.032      & -0.048     & -0.016     \\
                                                        & (0.069)   & (0.072)    & (0.073)    & (0.073)    \\
                    \addlinespace[0.8em]
                    Income                              & 0.092     & 0.194**    & 0.139*     & 0.175**    \\
                                                        & (0.073)   & (0.077)    & (0.078)    & (0.077)    \\
                    \addlinespace[0.8em]
                    Church attendance                   & 0.186***  & 0.222***   & 0.220***   & 0.208***   \\
                                                        & (0.054)   & (0.056)    & (0.056)    & (0.057)    \\
                    \addlinespace[0.8em]
                    Age                                 & -0.003    & 0.002      & 0.004      & 0.001      \\
                                                        & (0.005)   & (0.005)    & (0.005)    & (0.005)    \\
                    \addlinespace[0.8em]
                    Female                              & -0.227    & -0.254     & -0.253     & -0.297*    \\
                                                        & (0.161)   & (0.170)    & (0.170)    & (0.170)    \\
                    \addlinespace[0.8em]
                    Ethnicity: Black                    & -1.327****& -2.012**** & -2.107**** & -2.123**** \\
                                                        & (0.268)   & (0.267)    & (0.267)    & (0.271)    \\
                    \addlinespace[0.8em]
                    South                               & 0.165     & 0.172      & 0.210      & 0.203      \\
                                                        & (0.176)   & (0.186)    & (0.186)    & (0.187)    \\
                    \addlinespace[0.8em]
                    Professional status
                                                        & -0.438**  & -0.498**   & -0.660***  & -0.582***  \\
                                                        & (0.186)   & (0.197)    & (0.196)    & (0.197)    \\
                    
                    \addlinespace[1.2em]
                    Constant                            & 2.577**** & 2.539****  & 3.915****  & 3.471****  \\
                                                        & (0.513)   & (0.574)    & (0.565)    & (0.567)    \\
                    
                    \midrule
                    Adj. R2                             & 0.240     & 0.162      & 0.152      & 0.151      \\
                    $N$                                 & 545       & 545        & 545        & 545        \\
                    \bottomrule
                \end{tabular*}

                   \begin{tablenotes}[flushleft]
                    \footnotesize
                        \item  \textit{Notes:} Standard errors in parentheses.  * $p<0.10$, ** $p<0.05$, *** $p<0.01$, **** $p<0.0001$. 
                        \vspace{0.3em}
                        \item $^1$ Construal categories with the same label are not necessarily equivalent across methods.
                    \end{tablenotes}

            \end{threeparttable}
            \end{table}

        \vspace{1.2em}
        \noindent\text{Musical preferences}
        \vspace{0.8em}

            The last two analyses examine the association between respondents’ membership in musical taste construals and their sociodemographic characteristics and cultural practices. In Analysis~3, the dependent variable is respondents’ musical taste construal membership. In Analysis 4, the dependent variable is \textit{highbrow event attendance}, measured as an additive three-point scale capturing whether respondents have visited an art museum or gallery, attended a live ballet or dance performance, or a classical music or opera performance. These analyses include the same battery of sociodemographic and attitudinal variables employed by Goldberg (\citeyear{goldberg_2011}, p.~1421). The set of variables is shown in Table~\ref{ref:apx_music_regvars}.

            \begin{table}[htbp]
                \centering
                \footnotesize
                \begin{threeparttable}
                
                \caption{Construal analysis of musical taste: regressors}
                \label{ref:apx_music_regvars}
                
            \begin{tabularx}{\textwidth}
                {@{}L{3.0cm}L{2.3cm}cX@{}}
                \toprule
                \textbf{Variable} 
                & \textbf{Type} 
                & \textbf{Range}
                & \textbf{Maximum and Minimum values} \\
                \midrule
                \addlinespace[0.6em]
                
                Age 
                & Continuous 
                & 18--99 
                & 18: age eighteen; 99: age ninety-nine or more \\
                
                Education 
                & Categorical 
                & 0--20 
                & 0: no formal schooling; 20: eight years of college \\
                
                Occupational prestige 
                & Continuous 
                & 17--86 
                & 17: lowest HSR occupational prestige score; 86: highest prestige score \\
                
                Town size (logged) 
                & Continuous 
                & 0--15.8 
                & 0: lowest logged value; 15.8: highest logged value \\
                
                Income (logged) 
                & Categorical 
                & 6.2--10.2 
                & 6.2: log of midpoint income of lowest income category; 
                  10.2: log of midpoint income of highest income category \\
                
                Church attendance 
                & Categorical 
                & 0--8 
                & 0: never attends church; 8: attends church more than once a week \\
                
                Racism$^1$ 
                & Additive Index 
                & 0--5 
                & 0: espouses no racially discriminatory stances; 
                  5: espouses five discriminatory stances \\
                
                Gender 
                & Binary 
                & 0, 1 
                & 0: male; 1: female \\
                
                White 
                & Binary 
                & 0, 1 
                & 0: white; 1: other \\
                
                Southerner 
                & Binary 
                & 0, 1 
                & 0: Southern census region; 1: other region \\
                
                \bottomrule
                \end{tabularx}
                
                \begin{tablenotes}[flushleft]
                \footnotesize
                \item\textit{Notes:} Data drawn from the 1993 GSS.
                \item [$^1$]Additive scale of five binary question on attitudes on racial issues:
                (1) Would you yourself have any objection to sending your children to a school where more than half of the children? (2)In general, do you favor or oppose the busing of Black and White school children from one school district to another? (3-5) On average Blacks have worse jobs, income, and housing than White people. Do you think these differences are mainly due to (3) discrimination (4) because most Blacks don’t have the chance for education that it takes to rise out of poverty  discrimination? (5); because most Blacks just don’t have the motivation or will power to pull themselves
                
                \end{tablenotes}
                \end{threeparttable}
                \end{table}

            Table \ref{ref:apx_determinants_conttrad} reports the results of Analysis~3. In this analysis, we estimate logistic regression models predicting respondents’ membership in contemporary–traditional musical preference construals identified by BCA (Model~1), RCA (Model~2), CCA (Model~3) and RRCA (Model~4). The key independent variables in Models~1 through~4 are the socioeconomic and political engagement covariates listed in Table~\ref{ref:apx_politics_regvars}. Odds ratios are reported for brevity and substantive clarity; estimated coefficients and standard errors are available upon request.
            
            \begin{table}[htbp]
                \centering
                \caption{Logistic regressions of contemporary/traditional construal memberships: odds ratios} 
                \label{ref:apx_determinants_conttrad}
                
                \begin{tabular*}{\textwidth}{@{\extracolsep{\fill}}L{3.6cm}cccc}
                

                \toprule
                
                \addlinespace[0.8em]
                            & (1)   & (2)  & (3)  & (4) \\
                 \addlinespace[0.6em]
                            & DV: BCA & DV:RCA  & DV:CCA  & DV:RRCA     \\
                            & cont-trad  & cont-trad & cont-trad & cont-trad  \\
                            &  assignment &  assignment &  assignment &  assignment \\
                \addlinespace[0.6em]
                \midrule
                \addlinespace[1.0em]
                
                \text{\textit{Independent variables:}} & & & &  \\
            
                Age                     & 0.990       & 1.009      & 1.033***  & 1.070***   \\
                \addlinespace[0.4em]
                Education               & 0.930**     & 0.993      & 1.020     & 1.034      \\
                \addlinespace[0.4em]
                Occupational prestige
                                        & 1.003       & 1.013      & 1.029***  & 1.011      \\
                \addlinespace[0.4em]
                Income                  & 0.901       & 0.507***   & 0.371***  & 0.484***   \\
                \addlinespace[0.4em]
                Town size               & 0.946*      & 0.949      & 0.965     & 0.993      \\
                \addlinespace[0.4em]
                Female                  & 0.717*     & 0.745       & 0.965     & 0.875      \\
                \addlinespace[0.4em]
                Ethnicity: White        & 0.844       & 1.630      & 1.168     & 1.354      \\
                \addlinespace[0.4em]
                Residence: South        & 1.420*     & 1.240       & 1.709**   & 1.617*     \\
                \addlinespace[0.4em]
                Religiosity             & 0.964       & 1.042      & 1.069     & 1.203      \\
                \addlinespace[0.4em]
                Racism                  & 1.151*       & 1.015     & 0.995*    & 1.153      \\
                
                \addlinespace[1.2em]
                \midrule
                
                AIC                     & 720.108       & 523.157  & 454.1      & 384.211   \\
                $N$                     & 523           & 523      & 523        & 523       \\
                
                \bottomrule
                \end{tabular*}
                
                \caption*{\footnotesize
                \textit{Notes:} cont-trad: contemporary traditional.  Odds ratios reported; estimated coefficients and standard errors available upon request. * $p<0.10$, ** $p<0.05$, *** $p<0.01$.
                }
            \end{table}

            Table~\ref{ref:apx_determinants_partisanship} reports the results of Analysis~4, which estimates OLS regression models of attendance at high-brow cultural events. The key independent variables are categorical indicators of respondents’ membership in one of three substantive types of musical preference construals: contemporary–traditional (reference category), high–low-brow, and omni–univore. These categorical regressors correspond to membership in BCA construals in Model~1, RCA construals in Model~2, CCA construals in Model~3, and RRCA construals in Model~4. The reference category is contemporary–traditional membership. All models include the set of control variables listed in Table~\ref{ref:apx_music_regvars}.

            \begin{table}[tbp]
                \centering
                \footnotesize
                \caption{OLS regressions of attending to high-brow cultural events}
                \label{ref:apx_music_go2events_determinants}
                
                \begin{tabular*}{\textwidth}{@{}L{3.6cm}@{\extracolsep{\fill}}cccc@{}}
                
                    \toprule
                    \addlinespace[0.8em]
                     & (1)  & (2) & (3) & (4)  \\
                    \addlinespace[0.6em]     
                     & key regressors:  & key regressors: & key regressors: & key regressors: \\
                    \addlinespace[0.4em]
                    & BCA        & RCA        & CCA         & RRCA       \\
                    & categories & categories & categories  & categories \\                               
                    \midrule
                    
                    \addlinespace[1.2em]
                    \text{Construal membership indicators \textit{(ref: contempo/traditional)}:}  \\
                    \addlinespace[0.4em]
                    
                    high/low brow           & 0.324****     &  0.016        &  ---       &  ---         \\
                                            & (0.076)       &  (0.111)      &  ---       &  ---         \\
                    \addlinespace[0.4em]                
                    omni/univore            &  ---          & 0.050         &  -0.103    &  -0.296***   \\
                                            &  ---          & (0.102)       &  (0.105)   &  (0.114)     \\

                    \addlinespace[1.2em]
                    \text{Sociodemographic controls}: \\
                    \addlinespace[0.4em]
                    Age                     & 0.001         & 0.002         & 0.002     & -0.000      \\
                                            & (0.003)       & (0.003)       & (0.003)   & (0.003)     \\
                    \addlinespace[0.4em]
                    Education               & 0.081****     & 0.087****     & 0.087*****& 0.086****   \\
                                            & (0.016)       & (0.016)       & (0.016)   & (0.016)     \\
                    \addlinespace[0.4em]
                    Occupational prestige
                                            & 0.003         & 0.003         & 0.002     & 0.003       \\
                                            & (0.003)       & (0.003)       & (0.003)   & (0.003)     \\
                    \addlinespace[0.4em]
                    Income                  & 0.059         & 0.060         & 0.083     & 0.099       \\
                                            & (0.076)       & (0.079)       & (0.079)  & (0.078)      \\
                    \addlinespace[0.4em]
                    Town size               & 0.042***      & 0.046***      & 0.047*** & 0.047***     \\
                                            & (0.013)       & (0.013)       & (0.013)  & (0.013)      \\
                    \addlinespace[0.4em]
                    Female                  & 0.185**       & 0.209***      & 0.211*** & 0.216***     \\
                                            & (0.075)       & (0.077)       & (0.076)  & (0.076)      \\
                    \addlinespace[0.4em]
                    Ethnicity: white        & 0.110         & 0.122         & 0.120    & 0.111        \\
                                            & (0.106)       & (0.109)       & (0.108)  & (0.108)      \\
                    \addlinespace[0.4em]
                    Region: South           & 0.031         & 0.007         & -0.004   & -0.014       \\
                                            & (0.081)       & (0.083)       & (0.083)  & (0.082)      \\
                    
                    \addlinespace[1.2em]
                    \text{Attitudinal controls}: \\
                    \addlinespace[0.4em]
                    Religiosity             & 0.026*        & 0.029*        & 0.028*   & 0.023       \\
                                            & (0.015)       & (0.015)       & (0.015)  & (0.015)     \\
                    \addlinespace[0.4em]
                    Racism                  & -0.055*       & -0.067*       & -0.066** & -0.070**    \\
                                            & (0.040)       & (0.041)       & (0.041)  & (0.040)     \\
                                        
                    \addlinespace[0.4em]
                    Constant                & -1.847**      & -1.861**      & -1.932** & -1.797**    \\
                                            & (0.756)       & (0.774)       & (0.770)  & (0.766)     \\
                    
                    \midrule
                    Adj. $R^2$              & 0.171     & 0.140             & 0.161    & 0.152       \\
                    $N$                     & 521       & 521               & 521      & 521         \\
                    \bottomrule
                \end{tabular*}
                
                \caption*{\footnotesize Standard errors in parentheses.  
                \textit{Notes:}  The reader should note that groups of a same type might not be identical acros methods.  
                * $p<0.10$, ** $p<0.05$, *** $p<0.01$, **** $p<0.0001$.
                }
            \end{table}

        \begin{table}[tbp]
        \centering
        \footnotesize
        \begin{threeparttable}
        
        \caption{Construal estimations across CCMS in additional datasets$^{1}$}
        \label{ref:apx_extra_analyses}
        
        \begin{tabular}{p{4.0cm}ccc ccc ccc ccc}

            \hline
            \addlinespace[0.4em]
            & \multicolumn{3}{c}{(1) } 
            & \multicolumn{3}{c}{(2) } 
            & \multicolumn{3}{c}{(3) } 
            & \multicolumn{3}{c}{(4) } \\
            \addlinespace[0.2em]
            & \multicolumn{3}{c}{RCA} 
            & \multicolumn{3}{c}{CCA} 
            & \multicolumn{3}{c}{RRCA} 
            & \multicolumn{3}{c}{BCA} \\ 
            \addlinespace[0.2em]
            \hline

            \addlinespace[1.2em]
            \multicolumn{13}{l}{\textit{I. 1988 GSS: Attitudes on Science and Religion}} \\
            \addlinespace[0.4em]
            
                \cmidrule(lr){2-4}
                \cmidrule(lr){5-7}
                \cmidrule(lr){8-10}
                \cmidrule(lr){11-13}
                
                Number of construals:   & \multicolumn{3}{c}{3}
                                        & \multicolumn{3}{c}{4}
                                        & \multicolumn{3}{c}{3}
                                        & \multicolumn{3}{c}{3} \\

                \cmidrule(lr){2-4}
                \cmidrule(lr){5-7}
                \cmidrule(lr){8-10}
                \cmidrule(lr){11-13}
            
                \addlinespace[0.6em]
                Construal 1             & \multicolumn{3}{c}{\oldstylenums{30.20}\%}
                                        & \multicolumn{3}{c}{\oldstylenums{54.28}\%}
                                        & \multicolumn{3}{c}{\oldstylenums{42.42}\%}
                                        & \multicolumn{3}{c}{\oldstylenums{39.24}\%} \\ 
                \addlinespace[0.2em]
                Construal 2             & \multicolumn{3}{c}{\oldstylenums{27.5}\%}
                                        & \multicolumn{3}{c}{\oldstylenums{45.72}\%}
                                        & \multicolumn{3}{c}{\oldstylenums{31.29}\%}
                                        & \multicolumn{3}{c}{\oldstylenums{38.99}\%} \\
                \addlinespace[0.2em]
                Construal 3             & \multicolumn{3}{c}{\oldstylenums{25.18}\%}
                                        & \multicolumn{3}{c}{\oldstylenums{0.12}\%}
                                        & \multicolumn{3}{c}{\oldstylenums{26.28}\%}
                                        & \multicolumn{3}{c}{\oldstylenums{21.76}\%} \\

                \addlinespace[0.2em]
                Construal 4             & \multicolumn{3}{c}{\oldstylenums{16.38}\%}
                                        & \multicolumn{3}{c}{\oldstylenums{0.12}\%}
                                        & \multicolumn{3}{c}{---}
                                        & \multicolumn{3}{c}{---} \\                                                         
                Construals 5 to 11      & \multicolumn{3}{c}{\oldstylenums{0.12}\%}
                                        & \multicolumn{3}{c}{---}
                                        & \multicolumn{3}{c}{---}
                                        & \multicolumn{3}{c}{---} \\     
            \addlinespace[0.8em]

            \addlinespace[1.2em]
            \multicolumn{13}{l}{\textit{II.  TGHD on Trait Desirability for Women (UK sample, 2020})} \\
            \addlinespace[0.4em]
            
                \cmidrule(lr){2-4}
                \cmidrule(lr){5-7}
                \cmidrule(lr){8-10}
                \cmidrule(lr){11-13}
                
                Number of construals:   & \multicolumn{3}{c}{4$^{*}$}
                                        & \multicolumn{3}{c}{4}
                                        & \multicolumn{3}{c}{3}
                                        & \multicolumn{3}{c}{4} \\

                \cmidrule(lr){2-4}
                \cmidrule(lr){5-7}
                \cmidrule(lr){8-10}
                \cmidrule(lr){11-13}
            
                \addlinespace[0.6em]
                Construal 1             & \multicolumn{3}{c}{\oldstylenums{19.60}\%}
                                        & \multicolumn{3}{c}{\oldstylenums{36.90}\%}
                                        & \multicolumn{3}{c}{\oldstylenums{39.04}\%}
                                        & \multicolumn{3}{c}{\oldstylenums{68.62}\%} \\ 
                \addlinespace[0.2em]
                Construal 2             & \multicolumn{3}{c}{\oldstylenums{18.86}\%}
                                        & \multicolumn{3}{c}{\oldstylenums{31.47}\%}
                                        & \multicolumn{3}{c}{\oldstylenums{35.75}\%}
                                        & \multicolumn{3}{c}{\oldstylenums{11.94}\%} \\
                \addlinespace[0.2em]
                Construal 3             & \multicolumn{3}{c}{\oldstylenums{18.45}\%}
                                        & \multicolumn{3}{c}{\oldstylenums{24.67}\%}
                                        & \multicolumn{3}{c}{\oldstylenums{25.21}\%}
                                        & \multicolumn{3}{c}{\oldstylenums{11.70}\%}\\

                \addlinespace[0.2em]
                Construal 4             & \multicolumn{3}{c}{\oldstylenums{18.45}\%}
                                        & \multicolumn{3}{c}{\oldstylenums{6.92}\%}
                                        & \multicolumn{3}{c}{---}
                                        & \multicolumn{3}{c}{\oldstylenums{7.74}\%} \\                      
            
            \addlinespace[0.8em]

            \hline
            \hline
            \end{tabular}

            \begin{tablenotes}[flushleft]
            \footnotesize
            \item 
            \textit{Notes:} 
            \item [$^1$] Construals are numbered from largest to smallest; one-member construals respondents not shown.
            \item [$^{**}$] 18\% of population in RCA was ranked as members of 299 one-member construals.
            \end{tablenotes}
            \end{threeparttable}
            \end{table}

    \clearpage

 \begin{table}[!h]
 \centering
 \caption{Wall-clock computational time (in seconds) across datasets}
\begin{tabular}{lcccc}
        \hline
        \addlinespace[0.4em]
        & (1) & (2) & (3) & (4) \\
        \addlinespace[0.2em]
        & RCA & CCA & RRCA & BCA \\
        \addlinespace[0.2em]
        \hline

        \addlinespace[1.0em]
        \addlinespace[0.6em]

        ANES 2004 \textit{(611 respondents, 29 questions)}              & 70.69  & 0.02  & 248.96 & 0.33 \\
        GSS 1993  \textit{(1,532 respondents, 17 questions)}            & 146.18 & 0.16  & 238.36 & 0.17 \\
        GSS 1988  \textit{(1,481 respondents, 15 questions)}            & 29.90  & 0.04  & 43.50  & 0.04 \\
        TGHD      \textit{(1,214 respondents, 12 questions)}            & 52.61  & 0.15  & 83.41  & 0.10 \\

        \addlinespace[0.8em]
        \hline
        \hline
\end{tabular}
\label{tab:time_empirical}
\end{table}

\end{document}